\documentstyle[prb,aps,epsf,floats]{revtex}
\newcommand{\beq}{\begin{equation}}
\newcommand{\eeq}{\end{equation}}

\def\eqa{\begin{eqnarray}}
\def\eea{\end{eqnarray}}
\begin{document}
\draft \flushbottom \twocolumn[
\hsize\textwidth\columnwidth\hsize\csname
@twocolumnfalse\endcsname
\title{ Hamiltonian theory of gaps, masses and polarization in quantum Hall states:
 full disclosure. }
\author{    R.Shankar  }
\address{
 Department of Physics, Yale
University, New Haven CT 06520}
\date{\today}
\maketitle

\begin{abstract}

In two  short papers I had described an extension, to all length
scales, of the hamiltonian theory of composite fermions (CF)  that
Murthy and I had developed for the infrared,  and applied it to
compute finite temperature quantities for quantum Hall fractions.
I furnish details  of the extended theory and apply it to Jain
fractions $\nu = p/(2ps + 1 )$. The explicit operator description
in terms  of the CF  allows one to answer quantitative and
qualitative issues, some of which cannot even be posed otherwise.
I compute activation gaps for several potentials, exhibit their
particle hole symmetry,  the profiles of charge density in states
with a quasiparticles or hole, (all in closed form) and compare to
results from trial wavefunctions and exact diagonalization. The
Hartree-Fock approximation is used  since much of the
nonperturbative physics is built in at tree level.  I compare the
gaps to experiment and comment on the rough equality of normalized
masses near half and quarter filling. I compute the critical
fields at which the Hall system will jump from one quantized value
of polarization to another, and the polarization and relaxation
rates for half filling as a function of temperature and propose a
Korringa like law. After providing some plausibility arguments, I
explore the possibility of describing several magnetic phenomena
in dirty systems with an effective potential, by extracting  a
free parameter describing the  potential  from one data point and
then using it to predict all the others from that sample. This
works to the accuracy typical of this theory (10 -20 percent). I
explain why the CF behaves like free particle in some magnetic
experiments when it is not, what exactly  the CF is made of, what
one means by its dipole moment, and how the comparison of theory
to experiment must be modified to fit the peculiarities of the
quantized  Hall problem.

\end{abstract}
\vskip 1cm ]

\newtheorem{tabel}{Table}

\section{ Introduction}
There is a consensus  among theorists and experimentalists  that
the concept of the Composite Fermion (CF)\cite{jain} is  very
useful  in understanding the fractional quantum Hall effect
(FQHE)\cite{review1}. This  concept allows us, among other things,
to decide which fractions are robust (the Jain series), to
generate very  accurate  trial wavefunctions and gaps, and to
determine the allowed values of polarization when the spin is not
fully polarized.

The aim of this paper is to describe, in  detail, a hamiltonian
formulation which  provides a comprehensive way to describe FQHE
states qualitatively and quantitatively, at zero and nonzero
temperatures.  Recall that in the theory of superconductivity or
Fermi liquids  one always seeks a transformation relating the
original electronic variables to those of the ultimate
quasiparticles, these being the Cooper pairs and Landau's
quasiparticles respectively.  In the hamiltonian formalism used
here one passes form a description in terms of electrons to one in
terms of CF's through a sequence of transformations. One ends with
an operator description of the CF.
 Even though such a
change of basis is approximate, it  provides us with valuable
insights,  and occasionally,  quantitative information  that is
unavailable in the wavefunction approach, such as ways of coupling
to impurities and calculating unequal-time correlations.

Given that the FQHE has no small parameters, how was this passage
possible, even approximately?  The answer has two parts.
Originally  Murthy and I \cite{gmrs} developed the transformation
using a combination of the random Phase Approximation (RPA) and
the infrared limit, and  obtained the electron density operator in
the CF basis at small $ql$, $q$ being the momentum and $l$ the
magnetic length. The  RPA kept operators at different $q$'s from
mixing during the transformation. Given the density operator, the
hamiltonian, which is   just the interaction energy when electrons
are restricted to the lowest Landau level (LLL), could be written
down. For potentials that were soft at large $ql$, we could
compute objects like activation gaps\cite{gaps}.

 I then extended the infrared results to all $ql$ by appealing
to certain algebraic properties known to be true for the LLL
projected problem in any basis. The extension  consisted of taking
the small $ql$ series for charge and the   constraints, and
exponentiating them. The resulting operators obeyed the desired
algebras. Given that  the extensions were   not unique, but just
minimal and consistent, it was very satisfying that  they
embodied,  despite their  questionable pedigree, much of what was
known about the internal structure of the CF, illustrating   once
again that the tight constraints of the FQHE problem in the LLL
can actually  work in our favor. Two short
papers\cite{prl1}-\cite{prl2} described this extension and its
application to finite temperatures, $T>0$. Here I provide the
promised details and additional insights gained in the meantime.

The topics covered here fall  into two classes.

The first has to do with matters of principle. For example, can CF
be  free or nearly so? Without going into any details, we can say
no,  since it takes two very different masses $m_a$ and $m_p$, to
describe their polarization and activation phenomena, something
impossible in a free theory. Yet certain polarization phenomena at
$T=0$ seem to be very accurately fit by free particles of mass
$m_p$. The hamiltonian formalism not only allows one to compute
these distinct masses, it also  resolves the paradox posed above.
Next, CF are supposed to derive their kinetic energy from
electron-electron interactions. The present formalism provides an
explicit expression for not only this kinetic  energy but also the
CF interactions, i.e., the full CF hamiltonian. Both kinetic and
interaction terms have unusual functional forms, which are
determined uniquely by the theory. Those attempting  to fit data
to traditional forms of  energy should bear this in mind. Having a
concrete hamiltonian also eliminates questions such as which of
$m_a$ and  $m_p$ should be used at $T>0$.  A Hartree-Fock
calculation on the full CF hamiltonian gives the answer.

The second set of topics consists of  application  aimed at
showing that it is possible to compute, by analytic means and
often in closed form, numerous physical quantities pertaining to
FQHE states.  I consider here the computation of activation and
polarization masses and gaps, charge profiles of quasiparticles
and quasiholes, critical fields for magnetic transitions from one
quantized value of polarization to another for the gapped states,
all of  which are  $T=0$ quantities. I also compute  the
polarization $P$ and relaxation rate $1/T_1$ at $T>0$ for the
gapless fractions.

Given the importance and utility of the CF idea, and the potential
for misunderstandings, I have made every effort to make my
arguments accessible to as wide an audience as possible,
underscoring the various assumptions that go into the
calculations,  emphasizing not only the formalism but the physical
picture that goes with it.

In Section II, I     describe how one arrives at  the hamiltonian
in terms of CF, starting with electrons. This discussion will be
brief, given that details have already been published. It does
include recent insights on the internal structure of the CF, such
as what it is made of and what exactly its dipole moment means. In
Section III, the final equations are analyzed to gain familiarity
with their main properties. The theory is then  recast in a form
that makes it more amenable  to the Hartree Fock approximation.
 Readers not interested in
looking under the hood, may begin with the equations listed
towards the end of Section III, which   form the basis for the
subsequent calculations.

In Section IV, I  calculate the activation gaps $\Delta_a$ for
several fully polarized fractions within the HF approximation
 and compare to the results
 of Park, Meskini and   Jain, ( PMJ)\cite{pmj}
   based on trial
wavefunctions. With one exception, all calculations will be
carried out using the Zhang-Das Sarma (ZDS)\cite{zds} potential
\beq v(q) = {2 \pi e^2 \over q}e^{-ql\lambda} \eeq
 where $l$ is the magnetic length
and $l\lambda =\Lambda$ was originally introduced to describe
sample thickness, but employed here as a free parameter. I use
this potential to illustrate the method, which is instantly
adaptable to any other.

  This present calculation differs from earlier
work\cite{gaps} based on the infrared theory in that it yields
finite results even for the coulomb case, $\lambda =0$. The
numbers agree to within $10-20\%$ (and occasionally better) for
potentials that seem to describe real systems ($\lambda \simeq
1-2$). Similar results are  found when I compare to the exact
diagonalization results of Morf {\em et al}\cite{Morf} for a
similar range of the parameter $b$ that enters their potential:
\beq
   v(q) = {2 \pi e^2\over q}
{\Large e^{(qlb)^2}\ Erfc\  (qlb)}. \eeq Why bother to reproduce
numbers that are already known, to a lower accuracy? The point is
that the present approach is fully analytic, makes the underlying
physics very transparent, and furnishes an explicit operator
description of the final quasiparticles, which permits  a precise
formulation of
 many question pertaining to them that would be otherwise nebulous.
  It is also worth mentioning that
  the closed  expressions for physical quantities
  allows them to be
computed in a few seconds on a PC. The reduced precision is a
price we must pay in return.

I compare the theoretical activation gaps   to the experiments of
Du {\em et al}\cite{du} and Pan {\em et al}\cite{Pan}. In
comparing the theory to experiments, one needs to decide how to
handle disorder and LL mixing, which are suppressed in the PMJ and
Morf {\em at al}  computations. There does not exist at present an
analytical  theory for incorporating disorder.  (There does exist
numerical work demonstrating the effect of disorder,  see for
example  Ref.\cite{dis}). My approach has been to take
 experimental points and fit them to the theory with the ZDS potential
 and ask what $\lambda$ is needed. This is done solely to
  get a feeling for its size and also compare it to the
 values computed for the pure system with no LL mixing, using say,
 the  Local Density Approximation (LDA)\cite{LDA},\cite{pmj}.
 It is not assumed that the
  ZDS potential actually describes the problem at hand.
  It simply taken as  a reasonable  variant of the coulomb potential
   with a free parameter that can parameterize sample thickness and illustrate the
   hamiltonian method.
   When
  results for
 polarization phenomena are compared to experiment,
  a more ambitious approach to $\lambda$ is
 undertaken.

 I explore the question raised by Pan {\em
et al}\cite{Pan} of how the normalized  effective mass of CF near
half filling compares  with that near quarter filling. Is the
rough  equality,  observed experimentally, in accord with  theory
(in the absence of disorder)? In general the answers depend on how
the fractions are reached-- by varying the density, the field or a
combination of both. Typically these masses  lie within a factor
of two of each other and  there appears to be  no deep reason why
they should be exactly equal, a  point also  made in Ref.
\cite{j24}

I provide  the profiles of charge density in some gapped states
with one quasiparticle or quasihole and compare  to the
unpublished work of Park and Jain based on trial wavefunctions. I
explore the $\nu =1/2$ case, especially the dipole moment and what
it means,  in some detail.

Since the CF hamiltonian naturally separates into a free part
$H_0$ and an interaction $H_I$,  I explore the effect of turning
off $H_I$ and find it can change the answer by as much as a factor
of two.

I ask how well particle-hole symmetry works, i.e., to what extent
gaps at $\nu = p/(2ps+1)$ in the fully polarized case equal those
at $ 1-\nu $ and find it works very well. I   point out that this
was not a foregone conclusion since the formulae for the two cases
are quite different.

Section V is devoted to spin physics at  $T=0$, an area
investigated  in the past\cite{spin} and more recently by Park and
Jain\cite{parkjain} in the wavefunction approach. In the absence
of an overwhelming Zeeman term, one has to consider CF of both
spins. The polarization of the ground state will be decided by a
competition between ferromagnetism and antiferromagnetism. When
the energy difference (not counting the Zeeman energy) between two
ground states of different polarizations equals the corresponding
Zeeman energy difference, a transition will take place.  The
transition can be driven,  for example, if  the density and field
are varied together at fixed filling fraction,  or by tilting the
sample at fixed perpendicular field and density. The critical
fields $B^c$ at which these transitions happen are calculated. The
calculations reveal a feature   noticed by Park and Jain using
trial wavefunctions, namely that they may be fit very well to free
fermions with a constant polarization gap, $\Delta_p$. How do we
reconcile this with the fact that activation gap $\Delta_a$ is
substantially different from $\Delta_p$,   and that turning off
$H_I$ makes a sizeable difference to $\Delta_a$? I will show how
two-dimensionality and rotational invariance can conspire to mimic
free-field behavior for these polarization phenomena. For example,
in the gapless case of $\nu =1/2$, I will show that while the HF
energies ${\cal{E}}(k_{\pm F})$ of fermions on top of the spin
up/down Fermi seas are not even   quadratic functions  of the
corresponding momenta $k_{\pm F}$, (and have substantial $k_{\pm
F}^{4}$ pieces), the energy cost of {\em transferring} a particle
from the top of one sea to the top of the other (which is what
determines the polarization)  takes the free-field form $(k_{+
F}^{2}-k_{- F}^{2})/2m_p$. A similar situation exists for the
gapped fractions. These arguments should caution experimentalists
and theorists against misinterpreting the free-field fit.

My results for the critical fields $B^c$ are then compared to the
data of Kukushkin {\em et al}\cite{kuk}. In the case of magnetic
phenomena I take a different approach to   $\lambda$. First
$\lambda$ at one transition   is obtained by fitting to the
observed $B^c$. Scaling laws then determine it for the other
transitions, whose $B^c$ can be predicted to within $20\%$. In
other words, it seems to be true for magnetic transitions in the
samples considered,  that the disordered system can be described
by an effective translationally invariant potential.  I describe a
limit in which this result can be justified. However, I pursue
this approach for all magnetic phenomena, even though at present I
cannot provide similar arguments for all of them. I do so  because
it  works to within the accuracy typical of this theory, a feature
that needs to be understood.

Section VI considers magnetic phenomena  of gapless states {\em as
a function of temperature $T$}. The $T>0$ physics is  the first
instance  the present methods outperform complementary  approaches
based on exact diagonalization (limited to small systems) or trial
wavefunctions (limited the ground state and very low excitations.)
It has no finite size effects since  one works in the themodynamic
limit all along. One need not agonize over whether $m_a$ or $m_p$
should be used in computing a $T>0$ quantity such as
polarizations: given a concrete hamiltonian, a Hartree-Fock (HF)
calculation gives the results. The HF works well because most of
the right, nonperturbative physics is already built into the
hamiltonian.

A Hartree-Fock calculation gives the polarization $P$ and
relaxation rate $1/T_1$  as a function of $T$ and the potential.
These numbers are then compared to experiment, again by assuming
that the real system can be described by an effective potential,
 fitting $\lambda$ at any one data point from
each sample and explaining the rest of the data from that sample
at other fields, tilts and temperatures. Into these calculations
go the noncanonical, nonconstant,  density of states peculiar to
this hamiltonian. These results are compared to experiments of
Dementyev {\em et al} \cite{dem} who measured $P$ and $1/T_1$ at
zero and a $38.3\ {}^0$ tilt for a range of temperatures. They had
pointed out that attempts to fit all four graphs  with standard a
hamiltonian (with a mass $m$ and Stoner coupling $J$) led to four
disjoint set of values.  On the other hand if the hamiltonian
theory, with its peculiar functional form for $H$ is used,    a
good fit to all four graphs is possible with a single  $\lambda =
1.75$.

 On
comparing to the polarization data of   Melinte {\em et
al}\cite{melinte} I find the predictions work for the untilted
case but not
 at a tilt of $61 \ {}^o$. The reasons for this are discussed.

  I
 provide  an approximate expression for $1/T_1$ as a function of
 temperature. It has the Korringa form only in the critical case
 where the polarization saturates exactly at $T=0$.

Conclusions follow in Section VII and end with a discussion of a
procedure for  comparison of theory to experiment that is  tailor
made for the Hall problem.  Many details are relegated to the
Appendix, which ends with a summary of symbols for which there
does yet exist a uniform convention.

\section{ The Hamiltonian Formalism}

Let us begin by tracing the path from the  hamiltonian in terms of
electronic coordinates  to that in terms of CF,  focusing on the
spin-polarized case for fractions of the form
 \beq \nu = {p\over 2ps +1}. \eeq
 The results extend easily to
 $ \nu = {p\over 2ps -1}.$
 The treatment of old published steps, presented here for completeness,
  will be schematic.

 Consider  electrons of band mass $m$ and number   density $n$, described by the following
 first quantized hamiltonian:

\begin{eqnarray}
H_{el} &=& \sum_i {({\bf p}_i + e{\bf A})^2 \over 2m} +V\\ &\equiv
&  \sum_i {({\bf \Pi}_i )^2 \over 2m}+V\\
 &=& \sum_i {(\mbox{\boldmath $\eta $}_i )^2 \over 2ml^4}+V \\
   \mbox{\boldmath $\eta $}&=& {1 \over 2} {\bf r} + l^2 \hat{z}\times {\bf p}= l^2\hat{z}\times
 {\bf \Pi}
\\ l^2 &=& { 1 \over eB}\\ \nabla \times
{\bf A}&=&-eB
\end{eqnarray}
where $\hbar = c=1$, $\hat{z}$ the unit vector along the z-axis,
$l$ is the magnetic length, $B$ is the applied field, $V$ is the
inter-electron potential, and $\mbox{\boldmath $\eta $}$ is the
{\em cyclotron coordinate}, whose components are canonically
conjugate:
 \beq \left[ \eta_{x}\
, \eta_{y} \right] = il^2. \eeq

 Thus the  spectrum is given by Landau Levels (LL):
\begin{eqnarray}
 E &=&    \omega_0 (n + 1/2) \\
 \omega_0&=& {eB/m}.
 \end{eqnarray}
 In the lowest Landau level (LLL),
 \beq
 \langle \eta \rangle_{LLL} = l.
 \eeq

 There is a huge degeneracy of each LL due to the fact
that  the guiding center coordinate \beq {\bf R}= {1 \over 2} {\bf
r} - l^2 \hat{z}\times {\bf p} \eeq whose components obey \beq
\left[ { R}_{x}\ , { R}_{y} \right] = -il^2 \eeq  does not enter
$H$. The conjugate pair $(R_x,R_y)$ ranges over the entire sample,
whose area is its phase space, and determines the degeneracy if
one employs Bohr-Sommerfeld quantization with $l^2$ playing the
role of $\hbar$.

At the Jain fractions,  the inverse filling fraction
\begin{eqnarray}
\nu^{-1} &=& {eB \over 2\pi n} =2s+{1 \over p}\\
 &= &\mbox{flux quanta
per electron}\\ & =& \mbox{states in the LL per electron}.
\end{eqnarray}

If $\nu \le 1$, there is enough room in the LLL  to fit all the
electrons in the noninteracting case. One expects that if the
cyclotron energy $  \omega_0$ is much larger than the
interelectron potential, the ground state and low lying
excitations will be formed out of states in the LLL.

Since \beq {\bf r}= \ {\bf R} + \mbox{\boldmath $\eta $} \eeq a
natural  projection to the LLL is  \beq {\cal P}_{} \Rightarrow
{\bf r} \to {\bf R}. \eeq After this projection the two commuting
coordinates  become canonically conjugate.

Given the huge degeneracy of the LLL for $\nu <1$, the problem is
the selection of a unique ground state. Laughlin blazed one trail,
writing down inspired trial wavefunctions
 for $\nu =1/(2s+1)$. The other route is to try to start with the electronic
hamiltonian and try to reach, through a sequence of
approximations,  the final quasiparticles, which in  this work,
will be  the composite fermions.

For  Laughlin fractions,  where the wavefunction is
extraordinarily compact and  simple, one has the option of working
with Composite Bosons (CB)\cite{cb},  which have considerable
appeal of their own.

\subsection{What is a Composite Fermion?}

 What exactly is a CF composed of?
  I am
grateful to G. Murthy for some very useful discussions of this
issue.

 Laughlin showed (using
arguments involving adiabatic introduction of a flux quantum)
that at  $\nu =1/(2s+1)$ the elementary excitations have a charge
\beq e^* ={1\over (2s+1)}. \eeq

Consider the following state  \beq \psi_{vortex} = \prod_{j}(z
-z_0) \ \ \prod_{i<j} (z_i -z_j)^{2s+1}. \eeq   The prefactor
$\prod_{j}(z -z_0)$, multiplying Laughlin's ground state wave
function ( whose gaussian factor has been suppressed) is a {\em
vortex}. Due to the zero at $z_0$, there is a charge deficit near
that point, whose value, in electronic units,  can be shown to be
$-1/(2s+1)$.

{\em In CF theory the quasiparticle is believed  to be an electron
bound to $2s$ vortices.} We shall see that while  this is clearly
so for the Laughlin series, the situation for the   Jain series
 is more complex.

If we look at the Laughlin wavefunction we see a $2s+1$-fold zero
at each electron: one is due to the Pauli principle and the other
$2s$ represent the zeros due to the captured vortices. There is no
question of which vortex belongs to which electron since the
vortices are sitting {\em} on the electrons.
 The charge of the electron plus
 $2s-$fold vortex, i.e.,  CF charge is given by
 \beq 1-{2s\over 2s+1}= {1 \over 2s+1}=e^*. \eeq

 Sometimes the  vortex is incorrectly used  interchangeably with
a {\em flux tube},
 \beq \mbox{flux tube} ={\prod_{j}(z
-z_0) \over \prod_{j}|z -z_0 |} \eeq
 which has the phase of the
vortex but not the zero. In other words the CF is described as an
electron bound to $2s$ flux tubes. It was emphasized very early on
by Halperin \cite{halp} that for Laughlin fractions, electrons
like to bind to vortices due to the coulomb attraction. This was
also at the heart of  Read's work \cite{read1}, which extended the
concept  to $\nu =1/2$ where the wavefunction is obtained by
projection to the LLL (more on this shortly)  which in turn causes
the vortices to move off the electrons. In any event, electrons
are not attracted to flux tubes, which are neither charged nor low
energy excitations.

Consider  the Jain wavefunctions at  $\nu =p/(2ps+1)$:
 \beq \Psi_{Jain} = {\cal P} \prod_{i<j} (z_i
-z_j)^{2s}\chi_p(z,\bar{z}). \eeq The factor $\chi_p(z,\bar{z})$
describes $p$-filled CF Landau levels and the Jastrow factor
$J(2s)  = \prod_{i<j} (z_i -z_j)^{2s} $ describes vortices sitting
at the locations of the particles and ${\cal P}$, the LLL
projector,   replaces the $\bar{z}$'s as per \cite{gj} \beq {\cal
P} : \bar{z} \to 2l^2 {\partial \over \partial z} .\eeq

 Let us
first ignore ${\cal P}$. Then there are indeed $2s$ zeros per
particle in the Jastrow factor (located on the electrons) and one
(not necessarily analytic) zero per particle in
$\chi_p(z,\bar{z})$. (By zeros of the wavefunction, I always mean
as a function of  one coordinate, all others being held fixed.)
The $2s$-fold vortex has a charge \beq e_v = -{2ps\over (2ps+1)}
\eeq in electronic units, a result that can be deduced from just
the Hall conductance and incompressibility of the state. If we add
the vortex charge to that of the electron, we do indeed get \beq
e^* = {1 \over 2ps+1} ,\eeq
 which according to Su\cite{Su}, is the
correct, quasiparticle charge at all gapped fractions. This is
also confirmed by focusing experiments. \cite{focus}. So we may
say {\em on the basis of this unprojected wavefunction} that the
CF is the union of an electron and a $2s$-fold  vortex. Since the
vortices sit right on the electrons, there is no confusion on who
they are bound to and  all moments vanish except the total charge.

A lot of this changes upon projection by ${\cal P}$:  ${\partial
\over
\partial z}$   acting on the Jastrow factors, moves the zeros
  away from the particles and many of them vanish,   leaving $2s+1/p$ per electron, (determined by the
number of flux quanta per electron). Thus, after projection,
    vortices  cannot be
  associated with the electrons  in
   an unambiguous way.
 For example at $2/5$, there are
  2.5 zeros per electron. One  sits on the electron due to the Pauli principle,
   leaving 1.5 non-Pauli zeros per electron, which will neither lie on
   the other electrons, nor be  numerous enough  to  form two vortices per electron.
      Presumably, in
states involving projection, where the wavefunction has a very
complicated form due to the action of ${\cal P}$, there is some
nontrivial  sharing of vortices between electrons. In the limiting
case of $\nu
    =1/2$ only one non-Pauli zero per electron will remain after projection.
     We
    shall return to  $\nu
    =1/2$ later.

    These remarks do not imply
    that the CF approach to writing down trial wavefunctions based on electrons
    binding to vortices is in
    jeopardy. Thinking in terms of vortices still gives the
    unprojected wavefunction. This is all one needs, since the act of projection,
    while
    complicated, has  a definite algorithm that is
    routinely carried out, and yields wavefunctions with their
    incredible overlaps with exact results. My message is only that in the end, if
    one looks at the projected wavefunction, (which is  going to very complicated)
    one is not likely to find any simple correlation between
    electrons to vortices. These remarks apply to any projected
    wavefunction $\Psi_{LLL}$.

 It is quite remarkable that
even though we cannot  assign to each electron a $2s$-fold vortex
in $\psi_{Jain}$,  {\em $e^*$ is still given by adding the
electron's charge to that of  a charge $e_v=-2ps/(2ps+1)$ object}.
In other words, the charge we associate with the CF, being linked
to the Hall conductance and incompressibility,  is unaffected by
the projection,  although the notion of each electron having $2s$
vortices sitting on (or even near) it is no longer true. What then
is this object that pairs with the electron and how is one to
describe it theoretically ?

 The hamiltonian theory described here provides an answer. In this theory
  we   enlarge the
Hilbert space to include additional degrees of freedom,
accompanied by an equal
 number of constraints.  These new degrees of freedom
 (prevented from having any density
fluctuations by  constraints)   will turn out to have  charge
$e_v$ and  pair with electrons and change $e$ to $e^*$. {\em  I
shall refer to them as vortices, for want of a better name,  but
in view of what was said above, they are not related in any simple
way to zeros of $\Psi_{LLL}$.}

We now review the hamiltonian description, which, not
surprisingly,   is relies heavily  on earlier work.
\begin{itemize}
\item
  Lopez and Fradkin\cite{lopez}  took the first major step and
   attached  $2s$ flux tubes
 by the singular gauge transformation of
the wavefunction (due to  Leinaas and Myrrheim\cite{leinaas}),
from electrons to Chern-Simons fermions: \beq \Psi_e = \prod_{i<j}
{(z_i - z_j) ^{2s}\over |z_i -z_j|^{2s}}\Psi_{\rm
CS}\label{eq-phase} \eeq and opened up the   field theoretical
description of the Jain states. This was applied to the gapless
case $\nu=1/2$ by Kalmeyer and Zhang\cite{kalmeyer}, by Marston
{\em et al} who used bosonization\cite{brad},  and in a very
exhaustive treatment
 by Halperin, Lee and Read\cite{hlr}(HLR). ( Flux attachment had been
 done earlier to convert electrons to
  the composite boson  by Zhang, Hansson and Kivelson
 \cite{zhk} and earlier still to anyons by  Fetter, Hannah and
 Laughlin\cite{fetter}. Attachment of vortices instead of flux tubes by
 a complex gauge transformation was explored by Rajaraman and Sondhi\cite{rs}.)
\item Murthy and I introduced collective  coordinates,
 to describe long wavelength density fluctuations,
  as did Bohm and Pines \cite{BP} in their
 treatment of plasmons.
 For every extra degree
of freedom so introduced, there was a constraint on physical
states, to keep the problem same as before. The collective
coordinates corresponded to oscillators at the cyclotron scale.
Putting them in their ground states and projecting to the physical
sector using constraints, was seen to produce the zeros that
turned flux tubes into vortices, i.e., produced the Jastrow
factors.
\item To expose the low energy physics, Murthy and I introduced an additional
  unitary
transformation that decoupled the oscillators and the fermions.
This was however done  approximately:

1:  We worked at long distances. Thus if any quantity had an
expansion in powers of $ql$, we  kept just the leading term.\\
 2.  When  the density operator was encountered in a
product with other operators, we used the RPA: \beq \sum_j
e^{i{\bf (q-k)\cdot r}_j} \simeq n (2\pi )^2 \delta^2 ({\bf q-k}).
\eeq

We made the first approximation so that we could introduce a small
parameter $ql$ where there was none. The second ensured that the
operators at small $q$ like $\rho (q)$, did not mix with those at
high $q$ in the unitary transformation. These were the minimal
assumptions we had to make before we could  carry out the
decoupling  transformation. Despite these approximations a
reasonable quantitative and qualitative description emerged.

The long-distance, low-energy  theory Murthy and I derived was
given by the following set of equations:
 \begin{eqnarray}
H &=&  V  = {1 \over 2} \int{d^2q\over (2\pi )^2} \bar{\rho} ({\bf
q}) v(q) \bar{\rho} (-{\bf q})  \\ \bar{ {\rho} }(\bf{ q}) &=&
\sum_j e^{-i{\bf q \cdot r}_j}\left[ 1 - {i\l_{}^{2}\over (1+c) }
{\bf q}\times {\bf \Pi^{*}}_{j} +\cdots \right]  \label{baro} \\
\bar{\chi}(\bf{q}) &=& \sum_j  e^{-i{\bf q \cdot r}_j}\left[  1 +
{i\l_{}^{2}\over c(1+c) } {\bf q}\times {\bf \Pi^{*}}_{j} +\cdots
\right]\ \ \label{series}\\  0 &=& \bar{\chi}({\bf
q})|\mbox{Physical State}\rangle \\
 {\bf \Pi^{*}}
&=& {\bf p + eA^*}\\ {\bf A}^* &=& {{\bf A}\over 2ps+1}\\
 c^2& &= {2ps\over 2ps+1}=2\nu s
\end{eqnarray}
 Note  that physical states are to be annihilated by the
constraints $\bar{\chi}$. The magnetic moment of $e/2m$ on each
particle,  predicted by Simon, Stern and Halperin\cite{SSH},  that
arises naturally here is not shown, and neither is the
contributions to $H$ or the charge from the oscillators, which are
frozen in their ground state. The kinetic energy of the fermions
is quenched in the small $q$ sector if the number of oscillators
equals the number of particles, i.e., $Q$, the largest oscillator
momentum obeys $ Q=\sqrt{4\pi n}$. With this choice of $Q$, $H$
reduces to the electrostatic interaction between electrons written
in the new basis.

While this formalism is good only for small $ql$, it can still be
useful.  For the  Zhang-Das Sarma (ZDS) potential with  $\lambda
>1
$
 we were able to calculate\cite{gaps}
gaps that agreed well  with the results of Park and Jain. In a
collaboration with Park and Jain\cite{gapsfour} we also
established some scaling relations between fractions at the same
$p$ (number of filled CF Landau levels)  but different $2s$
(number of vortices attached)
 that seemed to work very well.

\item
While the small $ql$ theory had its share of  numerical successes,
it had some disturbing conceptual problems. For example  the
constraints did not close to form an algebra and did not commute
with the charge or the hamiltonian built out of it, except to
leading order in $q$. This implied that charge was not gauge
invariant and the physical sector  not defined: how could
constraints at two different $q$'s annihilate the physical states
but not their commutator?
    While    gauge invariance can be implemented order by order in  a
coupling constant, this is not so with respect to   $q$ which is
integrated over. (One {\em can}  use   $Q$ as a small parameter,
but some of the central  physics   gets
modified.\cite{HS1}-\cite{simon})

These problems  were resolved in my   minimal extension\cite{prl2}
of these results to all $ql$ that  is mathematically and
physically attractive. {\em Let us assume that
Eqns.(\ref{baro}-\ref{series}) represents the beginnings of two
exponential series and adopt the following expressions for charge,
constraint, and hamiltonian:}
\begin{eqnarray}
\bar{\bar{\rho}}(\bf{q}) &=&  \sum_j \exp (-i{\bf  q}  \cdot ({\bf
r}_j - {l^2\over 1+c}\hat{\bf z}\times {\bf \Pi^{*}}_j
))\label{robar}\\ &\equiv & \sum_je^{-i{\bf q\cdot R_{ej}}}\\
\bar{\bar{\chi}}(\bf{q}) &=&  \sum_j \exp (-i{\bf  q} \cdot  ({\bf
r}_j + {l^2\over c(1+c)}\hat{\bf z}\times {\bf \Pi^{*}}_j ))\\
 &\equiv &  \sum_j
e^{-i{\bf q\cdot R_{vj}}}\label{chibar}\\ H &=& {1 \over 2}
\int{d^2q\over (2\pi )^2} \bar{\bar{\rho}}({\bf q}) \
v(q)e^{-(ql)^2/2}\ \bar{\bar{\rho}}(-{\bf q})\label{Ham}
\end{eqnarray}
 Note that  ${\bf R}_e\ \mbox{and}\  {\bf R}_{v} $ were fully determined by
the two terms we did  derive. The gaussian in Eqn. (\ref{Ham})
will be explained shortly.
\end{itemize}

  This is my final answer.

\section{Analysis of the CF  hamiltonian}
Readers who either skipped the derivation or were troubled by the
approximations, are invited to take Eqns. (\ref{robar}-\ref{Ham})
as an effective field theory of CF for which the author can
provide a plausible lineage going back to the electronic
hamiltonian.

To understand what these equations imply, let us begin with the
coordinate appearing in the exponential in the expression for
$\bar{\bar{\rho}}(q)$:
\begin{eqnarray}
{\bf R}_e &=& {\bf r} -{l^2\over (1+c)}\hat{\bf z}\times {\bf
\Pi^{*} }\label{re}.
\end{eqnarray}
Its components   obey  \begin{eqnarray} \left[ R_{ex}\ , R_{ey}
\right] &=& - il^2,
\end{eqnarray}
the commutation rules of the guiding center of a unit charge
object.  This, together with the fact that it enters the electron
density, tells us it describes, in the CF basis, the guiding
center coordinates of the  electron.

Next consider the  coordinates appearing in $\bar{\bar{\chi}}(q)$
\begin{eqnarray}
{\bf R}_v &=& {\bf r} +{l^2\over c(1+c)}\hat{\bf z}\times {\bf
\Pi^{*} }\label{rv}\\ \left[ R_{vx}\ , R_{vy} \right] &=& il^2/c^2
\end{eqnarray}
These   describe, in the CF basis,  the guiding center coordinates
of a particle whose charge is $-c^2= e_v = -(2ps)/(2ps+1)$.

 This is exactly charge of
the $2s$-fold vortex. It will be seen to pair with the electron
and reduce the charge down to $e^*$. For these reasons  we shall
refer to it as the vortex, although the  nomenclature is not
ideal. For one thing, we know that this object does not correspond
to the 2s-fold zeros of the LLL wavefunction, which  does not
generally have $2s$ non-Pauli zeros per electron. In addition, the
assignment of physical meaning to objects that appear in an
enlarged Hilbert space is at best schematic. Recall the
oscillators, which when put in the ground state and projected to
the physical sector, produced the Jastrow factor with its
vortices. We could, for this reason, call them vortices. But we
must not forget that prior to projection, neither the oscillators,
nor  their wavefunction had any meaning in electronic language,
and that conversely,   in the electronic Hilbert space, there were
no independent  degrees of freedom corresponding to vortices,
which  are really made up of electrons.
 The whole idea of going to an enlarged space, as in
the case of  Bohm and Pines, is to be able to handle, in
intermediate stages,  collective variables   as canonical
coordinates independent  of electrons.

Finally \beq \left[ {\bf R}_e\ , {\bf R}_{v} \right] = 0.
\label{evcomm} \eeq Thus the four dimensional fermionic phase
space has yielded two independent sets of canonical coordinates,
${\bf R}_e$ and ${\bf R}_v$. Now this    is twice as many
coordinates per particle as in the {\em electronic LLL problem. }
But the constraints Eqn. (\ref{chibar})  tell us the density
formed out of ${\bf R}_v$ has no fluctuations, so that the number
of independent coordinates matches the LLL. This  is reminiscent
of Bohm-Pines theory\cite{BP}, where, once plasmons are introduced
at small $q$'s, the fermions are not allowed collective density
oscillations at these $q$'s.

\begin{figure}
\epsfxsize=4in \centerline{\epsffile{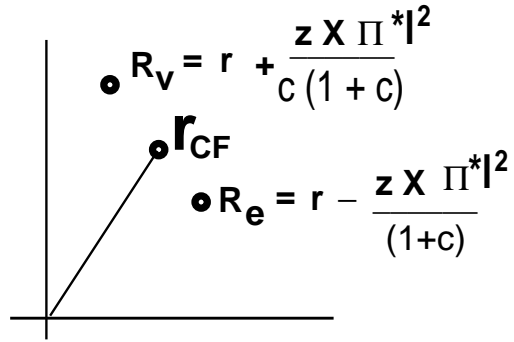}}
\vspace*{-2in} \caption{Anatomy of the CF: note that a CF at a
point ${\bf r}$ is flanked by the vortex and electron. They are
bound by terms in $H$ that grow with $\Pi^{*}$.  }
\label{cfcoordinates}
\end{figure}

Consider   Eqn. (\ref{re} and  \ref{rv}). They show that  a CF
  at ${\bf r}$ with
kinetic momentum ${\bf \Pi^{*}}$,  is flanked by the electron and
"vortex" within a distance of order $l^2 \Pi^{*}$. (See Figure
\ref{cfcoordinates}). Its total charge is their sum $e^* =
1/(2ps+1)$. Its dipole moment (in the frame ${\bf r}=0$) is $d^* =
-l^2 \hat{\bf z}\times {\bf \Pi^{*}}$. Its size $l^2 \Pi^{*}
\simeq l$ near the Fermi surface,  making it a well defined object
in this energy range.  The vortices are bound to the electrons,
since as their separation (proportional to $\Pi^{*}$) grows, so
does the energy since $H$ will be seen to have  terms that grow
with $\Pi^{*}$. Thus we have managed to reach one of the goals of
any theory of the FQHE: explain how CF get their
 kinetic  energy from the electrostatic
 energy of the
electrons. Indeed the entire CF hamiltonian is just the
electrostatic interaction of electrons written in the CF basis.
While these features are interesting, they are still heuristic.
First, the  discussions involving operators are semiclassical.
Next, ${\bf \Pi^{*}}$ is not a constant of motion except at $\nu
=1/2$ when it equals ${\bf p}$. Finally the first quantized
expressions do not include the effects antisymmetrization.
However, as we proceed, we will find an operator expression of
these ideas that is robust and survives in a second quantized
theory  fermions.

Given the commutation relations of ${\bf R}_e$ and ${\bf R}_v$ we
easily deduce those of $\bar{\bar{\rho}}$ and $\bar{\bar{\chi}}$:
\begin{eqnarray}
 \left[ \bar{\bar{\rho}}({\bf q}) , \bar{\bar{\rho}}({\bf q'}) \right] &=& 2i \sin
\left[ { ({\bf q\times q'})\ l^2 \over 2}\right] \bar{\bar{\rho}
}({\bf q+q'})\label{GMP} \\ \left[ \bar{\bar{\chi}}({\bf q}) ,
\bar{\bar{\chi}}({\bf q'}) \right] &=& -2i \sin \left[ {({\bf
q\times q'})  l^{2}) \over 2c^2}\right] \bar{\bar{\chi}} ({\bf
 q+q'}) \\ \left[ \bar{\bar{\chi}}({\bf q}) ,
\bar{\bar{\rho}}({\bf q'}) \right] &=& 0 .
 \end{eqnarray}
  One sees  that $\bar{\bar{\rho}}$ is not only algebraically closed, but obeys the
  Girvin-MacDonald-Platzman  GMP algebra\cite{GMP} for
  magnetic translations. I will keep referring to it as the projected charge
   density,  from which it  differs  by a factor $e^{-(ql)^2/4}$.
   In the hamiltonian Eqn. (\ref{Ham}), which  is just the electrostatic interaction written in
the CF basis, the   factor $e^{-(ql)^2/2}$ takes this difference
into account.

Note that $\bar{\bar{\rho}}$ constitutes a specific realization or
representation of the GMP algebra in terms of the final,
canonical, CF variables, a feature that allows one
 to apply standard many-body methods to  $H$.

Finally the constraint commutes with the projected charge and
hence the low energy hamiltonian which is quadratic in it. As in
the  Yang-Mills case,
 the constraints form a
nonabelian algebra and commute with $H$.

How does the small $q$ - RPA treatment manage to generate these
 coordinates $R_e$ and $R_v$ with their nice commutation
 relations?
  This is my current understanding. The first two terms in $\bar{\bar{\rho}}$
   and $\bar{\bar{\chi}}$, which fully determine these coordinates,
    could be derived in another theory, equivalent
    to ours for small $ql$ but not all $ql$: the theory explored by
     Stern {\em et al}\cite{simon}.
     In this theory   $Q$, the upper cut-off for the oscillators is, assumed to be
      vanishingly small. This means that we can safely  assume that every
       $q$ in the problem, including those integrated over,
        are small, being bounded by $Q$. Next, RPA becomes
         exact in this case since non RPA terms involve multiple
          $q$ integrals  which are suppressed by higher powers of
          $Q$ in the unitary transformation. In such a derivation one gets
           exactly the same first two terms.
           These encode the  charge and dipole moment of the CF,
            which characterize the CF in the infrared. While my  extension to all $q$
             is  mathematically satisfactory, it need not
              be numerically exact     down
               to arbitrarily small length scales. We should have
               been prepared for this since one cannot
               have at the same time a scheme that is analytically
               tractable and  numerically
               exact, unless we are  dealing with exact solutions. We
                managed to make the
               nonperturbative passage from electrons to CF
               by the exponentiation, which in turn
               was guided  by the  LLL algebras.

   Recall that all proofs of nonzero static compressibility at $\nu=1/2$
    relied on a careful
implementation of the constraints or     gauge invariance
  \cite{HS1}, \cite{read2}, \cite{DH}. We may now understand
this as a follows: {\em $\bar{\bar{\chi}}=0$ means  that  only the
electrons in  the CF respond  to the  static potential, exhibiting
nonzero static compressibility of unit charge objects.}

\subsection{How to solve $H$?}

Now we ask how we are to handle Eqns. (\ref{robar}-\ref{Ham}). As
shown in Appendix 2, $p$-filled LL's and  particle and hole
excitations on top of it,
 are HF states of our $H$. (The proof relies on the
rotational invariance of the hamiltonian.) One possibility is to
ignore the constraints altogether and proceed with the HF
approximation. This will however lead to the following fatal flaw:
transition matrix elements of $\bar{\bar{\rho}}$ will start out as
$q$ and the structure factor $S(q)$ will go as $q^2$, in violation
of Kohn's theorem. (The the $q^2$ sum rule is saturated by the
oscillators that were decoupled.)

We must therefore bring in the constraints and hope they will bail
us out. A standard
 way to incorporate first class constraints is to introduce  them into
 the path integral with a Lagrange multiplier and try to solve the theory in
  an approximation that respects the gauge  symmetry i.e., the constraints.
   We will discuss it shortly, but
in the present case I will use a solution that is essentially what
Murthy and I used in the small $q$ theory. \cite{gmrs}: replace
$\bar{\bar{\rho}}$ by  the {\em preferred combination} for charge
density \beq \bar{\bar{\rho}}^p = \bar{\bar{\rho }} - c^2
 f  \bar{\bar{\chi}} \eeq
 where \beq
 f = e^{-q^2l^2/8ps}\ \ \ \ \mbox{Vortex Form Factor}\eeq
 This combination is   equivalent to
$\bar{\bar{\rho }}$ in the physical sector. The factor $e^{-q^2\
l^2/8ps}$ (absent in the earlier work and unimportant for any
fraction other than $1/3$) takes into account the fact that since
the vortex and the electron have different magnetic lengths, to
convert the vortex magnetic translation operator to the magnetic
number density we need a different gaussian from the one we
absorbed into $H$ in Eqn. (\ref{Ham}).  Note that
$\bar{\bar{\rho}}^p$ is  {\em weakly gauge invariant}:

\beq \left[ \bar{\bar{\chi}} \ , \bar{\bar{\rho}}^p\right] \simeq
\bar{\bar{\chi}}. \eeq Clearly so is the  $H(\bar{\bar{\rho}}^p)
$\ that I shall use. Weak gauge invariance is enough to keep
physical and unphysical states from mixing.

Consider the series expansion of  $\bar{\bar{\rho}}^p$: \beq
\bar{\bar{\rho}}^p = \sum_j e^{-i{\bf q \cdot r}_j}\!\! \left( \!
{1 \over 2ps+1}\!  -  {i\l_{}^{2}  }{\bf  q}\times {\bf
\Pi^{*}}_{j}  \! +{0} \cdot \left( {\bf q}\times {\bf \Pi^{*}}_{j}
\right)^2 \! + ..\! \right) \label{rostar} \eeq

If we expand  $e^{-i{\bf q \cdot r}_j}$ to first order in $q$ we
can verify that the term linear in $q$ contains only the guiding
center   coordinate of the CF ($ {\bf r} - l^{*2} \hat{z} \times
{\bf \Pi}^*$) with no admixture  of the cyclotron coordinate. Thus
it does not contribute to the order $q$ transition matrix element.
This is the unique multiple of the constraint we can add to
$\bar{\bar{\rho}}$, with this property.

With the constraint implemented this way, we are in compliance
with
 Kohn's theorem. But there is more.
Consider the series Eqn. (\ref{rostar}). The first term,
proportional to CF density has the coefficient $e^*$. The next
term has the dipole moment given by Read using wavefunction
arguments. (This was done for $\nu =1/2$ and is expected for the
whole series, as suggested by   Figure 1.) The vanishing of the
third order term explains the success of the small $q$ theory.

It can also shown that  if only terms linear in $q$ are kept in
 $\bar{\bar{\rho}}^p$,
  the algebra closes  with
\beq \sin ({\bf q \times q'}\ l^2 /2) \to
 {\bf q \times q'}\ l^2/2
\eeq in the structure constant of the GMP algebra Eqn.
(\ref{GMP}). The significance of this is not known.

It is remarkable that a single guiding principle, Kohn's theorem,
leads to a combination with all these properties. Since the
internal structure of the CF is built in  at tree level, we expect
that vertex corrections (due to the constraints) must vanish as $q
\to 0$. We shall not refer to the constraint any further.

Recall the cautionary note about ascribing meaning to objects in
an enlarged Hilbert space. Why should we give any significance to
the terms in Eqn. (\ref{rostar})? After all, we could have added
any multiple of the constraint $ \bar{\bar{\chi}}$  to $
\bar{\bar{\rho}}$ in the physical sector and each would have had a
different monopole and dipole term in the series expansion. This
is indeed true and in an exact calculation it would not have
mattered  how much of the constraint we added, the final $e^*$ and
$d^*$ would have been the same. However in our  HF calculation we
have included the constraint in an unusual way, guided by Kohn's
theorem. That is, we have implemented the constraint to the extent
we ever will. Since this led to a unique preferred combination (at
small $q$), we assume that any corrections due to constraints will
affect only the higher order moments.

In the operator approach there is no problem with how to assign
this or that vortex to an electron. All one claims  is that the
density operator that obeys Kohn' theorem in our HF calculation
couples to any external potential with a charge $e^*$ and a dipole
moment $d^*=l^2\ {\bf q}\times {\bf \Pi^{*}}$.

 There are no problems with antisymmetrization among particles:
we simply express this first quantized density operator in second
quantized form  with Fermi operators.

Let us return to the standard
 way of incorporating first class constraints  in
  an approximation that respects the gauge  symmetry i.e., the constraints.
  While this has not
 been done for general fractions, it has been done by Read\cite{read2} for
 bosons at $\nu=1$, which turn into fermions in zero field  upon {\em
 single}
 vortex attachment {\em a la} Pasquier and Haldane\cite{PH}. (The lessons learnt
 from this exercise are directly applicable to us since the only
 difference is in the coupling constant of the gauge and matter
 fields.)
 Implementing  the constraint in
a conserving approximation leads to a propagating gauge field
whose longitudinal part $a_L$ screens the charge fully,  leaving
behind dipoles of  moment $d^*$ which then interact via the
transverse gauge field $a_T$.   The propagator of $a_T$,  just as
in HLR, favors
 the region $i\omega \simeq q^3$.
 (The  field $a_T$,
  also produces mass divergences at the Fermi
energy as in HLR and restores compressibility.)
 {\em However, away from the
ultra-low  frequency region, the  answer is given by the
correlation function  of independent   objects of charge $e^*=0$
and dipole moment $d^*=l^2\ {\bf q}\times {\bf p}$. }

For  gapped fractions (not too close to $\nu =1/2$) and/or at
$T>0$,  a description in terms of independent particles with  the
right $e^*$ and $d^*$  is likewise expected to be a good
approximation, likely for all $\omega$, since the gap and/or $T$
will cut-off the low frequency end where  $a_T$ raises its head,
and the major effects of $a_L$ are already encoded in $e^*$ and
$d^*$. Rather than reach this description  using the conserving
approximation (which is  very difficult away from $\nu =1/2$ due
to LL structure,\cite{GMMP}) I use a scheme Murthy and I devised
for small $ql$. (Away from $\nu=1/2$, $a_T$  could affect the
statistics of the quasiparticles.)

{\em  For the benefit of the readers who just joined in,
  I display the  equations will be used in the subsequent calculations}:
\begin{eqnarray}
H^p &=&  {1 \over 2} \int{d^2q\over (2\pi )^2} \bar{\bar{\rho}}^p
\ ({\bf q}) \ v(q)e^{-(ql)^2/2} \ \bar{\bar{\rho}}^p \ (-{\bf q})
\\ \bar{\bar{\rho}}^p (\bf{q}) &=&  \bar{\bar{\rho }}({\bf q}) -
c^2 e^{-q^2l^2/8ps}
 \bar{\bar{\chi}}(\bf{q})\\
\bar{\bar{\rho}}(\bf{q}) &=& \! \sum_j \exp (-i{\bf  q} \! \cdot
\! ({\bf r}_j\! - \! {l^2\over 1+c}\hat{\bf z}\times {\bf
\Pi^{*}}_j ))\label{robar2}\\
 \bar{\bar{\chi}}(\bf{q}) &=& \! \sum_j \exp (-i{\bf
q}\! \cdot  \!({\bf  r}_j \!+ \!{l^2\over c(1+c)}\hat{\bf z}\times
{\bf \Pi^{*}}_j ))\label{chibar2}\\
   \left[ H^p , \bar{\bar{\chi}} \right] &\simeq   &
\bar{\bar{\chi}}  \\ 0 &=&\bar{\bar{\chi}}|\mbox{ physical
state}\rangle   \label{preferred}
\end{eqnarray}
Many  leading long wavelength  effects of the constraints have
been built into $\bar{\bar{\rho}}^p$;   they will be ignored in
the subsequent  HF calculations, as will be the superscript on
$H^p$ since we shall always use this  expression in terms of the
preferred charge $\bar{\bar{\rho}}^p$.

 We separate  $H$ into free and interacting parts
 \beq
 H=H_0 +H_I
 \eeq
wherein the two pieces $H_0$ and $H_I$  correspond to diagonal
($i=j$) and off-diagonal ($i\ne j)$ terms when the double sum over
particles is expanded. In the simplest case
 $\nu =1/2$
 we have
 \beq
 H_0 \ (\nu =1/2) = \sum_i 2  \int { d^2q\over 4\pi^2} \sin^2 \left[ {{\bf q\times
 k}_i
 l^2\over 2} \right] \bar{v}(q) e^{-q^2l^2/2}\label{freeh}
 \eeq
If we expand the $\sin$ in a series and keep the lowest term, we
get an expression quadratic in momentum from which we can define
an effective mass $1/m^* $. Thus we have managed to generate the
CF  kinetic energy,{\em in operator form}, in terms of the
electron-electron interaction. But we have more. First, there are
more powers of momentum in the kinetic energy. (We shall however
see that only quartic term is important  in HF.) Next, there is
also $H_I$, which can modify all the numbers. Thus $1/m^*$ depends
on the momentum and
 will be defined at the
Fermi surface. For $\nu =1$ bosons,  Haldane and Pasquier\cite{PH}
obtained  the same $H_0$ by algebraic methods aimed at a direct
LLL formalism. In the present case we have $H$ for the entire Jain
sequence. Away from ${1 \over 2}$ or ${1 \over 4}$, $H$ is even
more complicated, but of a very definite, known,  functional form.

\section{The Activation Gaps of Fully Polarized States}
Here we use  the hamiltonian theory  to compute activation gaps
for
 fractions $\nu = p/(2ps+1)$ in a field so strong that
the system is fully polarized. We will   probe the theory in the
following ways:
\begin{itemize}
\item We will compare the gaps to those obtained by Park, Meskini
and Jain (PMJ)\cite{pmj} using trial wavefunctions for the
Zhang-Das Sarma (ZDS) potential \beq v(q) = {2 \pi e^2 \over
q}e^{-ql\lambda} \eeq for $p=1,2,3$ and $4$ and $s=1$. These
serve as a benchmark, at least for fractions not too close to $\nu
=1/2$.
\item For the benefit of other users, a fit to the gaps in the
experimentally significant region $1\le \lambda \le 2$ will be
given for both $s=1$ and $s=2$. The gaps are also expressed in
terms of an effective activation mass $m_a$.
\item We will compare the theory to the experiments of Du {\em et
al}\cite{du} and Pan {\em et al}\cite{Pan}.
\item We will examine the charge density profiles in states with a
quasiparticle or quasihole.
\item The gaps are computed for a gaussian potential and
compared to PMJ for $p=1,2,3,4$ and $s=1$.
\item
 The effect of turning off the interaction $H_I$ in $H=H_0+H_I$, will
be explored.
\item We will compare the results to the exact diagonalization
results of Morf {\em et al}\cite{Morf}.
\item It will be seen how well particle-hole (PH) symmetry works.
For example, is the activation gap for $2/5$ same as that for
$3/5$ in the fully polarized case?

\end{itemize}

\subsection{Comparison of gaps in the HF approximation to PMJ}

 We use the HF approximation.
As shown in  Appendix 2,  $p$-filled CF LL and particle-hole
excitations thereof are HF states of our $H$.
 The HF  ground state is given by CF
filling $p$ LL's, which we will denote by $|{\bf p}\rangle$. I
will use a  boldface symbol such as ${\bf p}$  to label a Slater
determinant with $p$  occupied Landau Levels. Nonboldface symbols
will label  single particle states.  Note also that the actual LL
index $n$ for the state labeled by $p$ is $n=p-1$ since the LLL
has index $n=0$.

 The gap is defined by \beq \Delta = \langle {\bf p} +
PH | H|{\bf p}+ PH \rangle -\langle {\bf p} | H| {\bf p} \rangle
\eeq where $PH$ stands for a widely separated particle-hole pair.
Now that
 this is exactly how gaps are computed in PMJ's wavefunction approach\cite{pmj}.
 There is however one big difference hidden in the notations.  There
 too the hamiltonian is just the interaction, but {\em written in the
  electronic basis} (with $\rho ({\bf q}) = \sum_j \exp (i{\bf q\cdot r}_j)$)
   while the states (which carry the same label) are these simple wavefunction,
    multiplied by the Jastrow factor   and then projected to the LLL.
    Projection leads to a very complicated expression
    for the wavefunctions. In the present approach we have tried to incorporate these effects by
     going in the reverse direction, from electrons to CF's,  and obtaining complicated expressions
      for the charge and other operators, but with  simple expressions for the wavefunctions.
       While  these operator  expressions are unusual in form, they are
      still simple enough in appearance and amenable to exact analytical
      treatment, because of the approximations that went into the
      derivation. Thus
       does not expect the present results  to match those of the Jain
       approach in their accuracy.
       This is indeed the case, unless the potential is fairly soft ($\lambda$ is
        larger than, say unity). The aim of the present approach is
        to provide a $10 -20 \% $ theory for soft potentials
        (which do seem relevant to experiment) in which the
        calculations can be performed analytically and the physics
        of the quasiparticles is transparent.

Rather than work with a widely separated particle-hole (PH) pair,
I first find the energy
 in a state with just the particle and add to it the energy of a
  state with just the hole and subtract
  double the ground state energy. While the details are relegated
  to
   Appendix 4, here is the central idea.

   One begins with the second quantized expression for the
   preferred charge operator $\bar{\bar { \rho}}({\bf q})$:
   \beq
 \bar{\bar{\rho}}^p({\bf q}) =
 \sum_{m_2n_2;m_1n_1}d^{\dag}_{m_2n_2}d_{m_1n_1}\rho_{m_2n_2;m_1n_1}
 \eeq
 where $d^{\dag}_{m_2n_2}$ creates a particle in the state
 $|m_2\ n_2\rangle$ where $m$ is the angular momentum and $n$
 is the LL index of  CF in the weakened  field $A^*=A/(2ps+1)$ with a magnetic length
\beq l^*=l\sqrt{2ps+1}. \eeq The key ingredient in the HF
calculation is the
 matrix element $\rho_{m_2n_2;m_1n_1}$. It is shown in Appendix 1
 that this matrix element factorizes:
 \beq
 \rho_{m_2n_2;m_1n_1}=\rho_{m_2m_1}^{m}\otimes \rho_{n_2n_1}^{n}
 \eeq

The gaps depend only on $\rho_{n_2n_1}^{n}$, the superscript on
which will be generally dropped.

 As shown in Appendix
4,
\begin{eqnarray}
\Delta_a (p,s,\lambda )&=& \int {d^2q\over 4\pi^2} \exp
(-x/(2ps+1)) v(q) G(p,x,s)\label{gaps1}\\ x&=& {q^2l^{*2}\over 2}
= {q^2l^{2}(2ps+1)\over 2}\label{gapstart}\\ G(p,x,s)&=&
G_0(p,x,s)+\Phi (p,x) \label{freeplusint}\\ G_0(p,x,s)&=& c^2
e^{-x\tau /2} \exp \left({-x\over (4ps)(2ps+1)}\right) \nonumber
\\ &\times & (L_{p-1}(x\tau ) -L_p(x\tau ))\\ \Phi (p,x) &=&
\sum_{m=0}^{p-1}(R\  (p-1,m,x)-R\  (p,m,x))\\ R\  (a,b,x) &=&
x^{a-b}\left( {a!\over b!}\right) \left[
c^{a-b}e^{-xc^2/2}L_{b}^{a-b}(xc^2) \right. \nonumber
\\
  & & \!\!\!\!\!\!\! \!\!\!\!\!\!\! \left. - c^{2-a-b}\exp \left({-x\over 4ps(2ps+1)}\right)
L_{b}^{a-b}(x/c^2) \right]^2\\
 c^2 &=& {2ps\over
2ps+1}\\ \tau &=& (c-1/c)\label{gaps2}
\end{eqnarray}
where $L_m^n$ is the associated Laguerre polynomial.

Often we will use the {\em dimensionless gap} $\delta    $ defined
by \beq \Delta_a = {e^2\over \varepsilon l}\  \delta_a .
\label{reduced}\eeq

It is useful to know  that  \beq {\Delta_a \over k_B}= {e^2\
\delta_a\over \varepsilon lk_B}\   \simeq 50\sqrt{B(T)}\delta_a \
{}^oK \eeq I shall  use the values given in  Table
(\ref{numbers}):

\vspace*{.2in}
\begin{large}
\begin{tabular}[c]{|c|c|}\hline \hline
 $ $  & $ $ \\
  ${e  B\over m_ek_B}$ & 1.34 $ B(T){}^oK$ \\
  $ $  & $ $ \\
 $ {e^2 \over \varepsilon
lk_B}$ & 50 $\sqrt{B(T)} {}^{o}K$\\ $ $  & $ $ \\
 $ {\varepsilon \over e^2l}$ & $.026\ m_e \sqrt{B(T)}$\\
$ $  & $ $ \\ \hline \end{tabular}
\end{large}
\vspace*{.2in}

\begin{tabel}\label {numbers}
  Approximate numbers used in this
 paper,
with  $k_B$ the  Boltzmann's constant, $B(T)$  the field in Tesla.
\end{tabel}
\vspace*{.2in}

\begin{figure}
\epsfxsize=3in \centerline{\epsffile{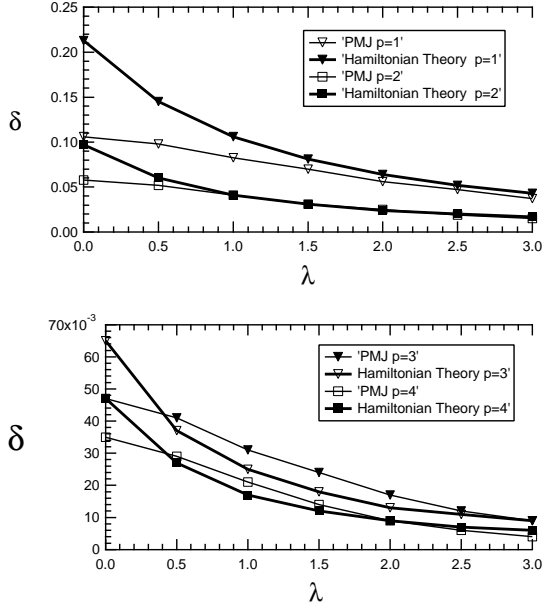}} \vskip 0.15in
\caption{Comparison of dimensionless  activation gaps $\delta_a$
to the work of Park, Meskini and Jain {\em et al}  for the
fractions $1/3,2/5,3/7,4/9$ ($p=1,2,3,4$, \mbox{and} \ s=1) as a
function of $\lambda$,
  the thickness parameter in the ZDS potential.}
\label{gaps}
\end{figure}

Figure \ref{gaps}  shows the gaps computed for $1/3,2/5,3/7$ and $
4/9$ for the ZDS potential
 and compared to the work of PMJ \cite{pmj}
  in the region $0\le \lambda \le 3$.
The following features are noteworthy.
\begin{itemize}
\item At $\lambda =0$, the coulomb case, the gaps are finite in contrast to
the small $q$ theory\cite{gaps}. This is due to the gaussian
factor $e^{-q^2l^2/2}$ which was  absent there. The slope of the
graphs in the present theory is nonzero at this point. It is
readily
 verified that $d\Delta /d\lambda $ at $\lambda =0$ is the gap due to a
  delta-function potential, and  should vanish for spinless fermions.
  It does happen  for PMJ, whose electronic wavefunction is explicitly antisymmetric.
The present theory of CF does not give good answers for potentials
as short ranged
 as the delta function. Indeed even the coulomb interaction is too
 singular, and
  the theory begins to work well
 only beyond $\lambda \simeq 1$.
\item Beyond $\lambda \simeq 1$ the agreement is quite fair in general and very good for $2/5$.
\item The gaps do not vanish for any fraction and any finite $\lambda$.
\end{itemize}

\subsection{Activation masses}
 I have computed gaps for many other fractions, including for
 $s=2$, when four vortices are attached to form CF's.

Rather than show more plots, I will now analyze the theory in
terms of $m_a$ the activation mass defined by

 \beq
 \Delta_a = {eB^*\over m_a}={eB\over(2ps+1)  m_a}.\label{acdef}
 \eeq

Comparison to Eqn. (\ref{reduced}) shows that

\beq {1 \over m_a} = {e^2 l\over \varepsilon}\delta_a (2ps+1)
\equiv {e^2 l\over \varepsilon} C_a . \eeq Thus  \beq C_a =
\delta_a (2ps+1). \eeq

The significance of $C_a$ is that it approaches a limit as we
approach $\nu =1/2$ or $1/4$, as first emphasized by HLR
\cite{hlr}. What we will see now is the $C_a$ does indeed have a
nice limit, but this limit depends on $\lambda$, a parameter that
was set equal to zero (coulomb case) in HLR. We will focus on the
$s=1$ series, which converges to $\nu =1/2$ as $p \to \infty$. To
remind us of this fact a superscript $2=2s$ will be appended.
Consider the Table (\ref{cscale}),  of  gaps fitted as a function
of $\lambda$ in the interval $1\le \lambda \le 2$, although the
fit will work for modest excursions on either side.

\vspace*{.2in}

\begin{tabular}[t]{|c|c|c|c|}
\hline \hline
 p  & ${\Delta_{a}^{(2)} / k_B}=50 \sqrt{B(T)}\delta_{a}^{(2)}$ & $\delta_{a}^{(2)}$ &$
 C_{a}^{(2)}$
 \\ \hline
  1 & $ {5.31 \sqrt{B(T)}/ \lambda}$ & ${.106/ \lambda}$ & ${.32/ \lambda}$ \\
  2 & ${2.08 \sqrt{B(T)}/ \lambda} $& ${.042/ \lambda}$ &${ .21/ \lambda}$ \\
  3 &$ {1.23 \sqrt{B(T)}/ \lambda} $& ${.025/ \lambda}$ & ${.17/ \lambda}$ \\
  4 & ${0.87 \sqrt{B(T)}/ \lambda}$ & ${.017/\lambda}$ &$ {.16/ \lambda} $\\ \hline
\end{tabular}

\begin{tabel}
\label{cscale} Activation gaps as a function of $\lambda$ for
$1\le \lambda \le 2$ according to the hamiltonian theory. Note the
convergence of $C_{a}^{(2)}$ as $p \to \infty$.
\end{tabel}

\vspace*{.2in}

Based on Table (\ref{cscale}), and a similar one for states near
$\nu =1/4$, I find that we may write, {\em near these states},

 \begin{eqnarray}
C_{a}^{(2)} &=& {.160\over \lambda}\label{ca}\\ C_{a}^{(4)} &=&
{.148\over \lambda^{5/4}}\label{ca2}
\end{eqnarray}
where the expression, including the  exponents ($1,{5 \over 4}$)
are approximate.

 Consider the  normalized  mass defined by Pan {\em et
 al}\cite{Pan}
 \beq
 m^{nor}_{a} = {m_{a}\over m_e\sqrt{B(T)}}
 \eeq
where $m_e$ is the electron mass and $B(T)$ is the field in Tesla.

  In terms of  $C$, \beq
m^{nor}_{a}={.026\over  C_{a}^{(2s)}}\label{norma} \eeq

Combining Eqns. (\ref{ca}-\ref{norma})
 \begin{eqnarray}
  m^{nor}_{a}
  &=& .163 \lambda \ \ (s=1)\label{nor1}\\
  &=& .175 \lambda^{5/4}\ \  (s=2)\label{nor2}
  \end{eqnarray}

  We find that the suitably scaled masses $m_{nor}$ are  comparable for
  $s=1$ and $s=2$ but not exactly  equal. No fundamental arguments
  exist for their equality since the answer depends on the
  potential, parameterized  by $\lambda$.

\subsection{Comparison to data}
 In comparing  these gaps to   experiments, I will limit myself
  to  $\nu \le 1/2$. States like $3/4$ are related by
    particle-hole symmetry if full polarization is assumed, and
    states with $\nu >1$ require assumptions about filled
    electronic LL's which I do not want to make.

Consider the experiments of Du et al\cite{du}, who have extensive
data on activation gaps. Given that the experiments, unlike PMJ,
   have an unknown contribution from   LL mixing and impurities,
    it is not clear how to
   apply the theory. There is no {\em ab initio } calculation that
   includes these effects. (There is however reliable evidence that LL mixing is a very small
   effect at the values of $\lambda$ under consideration.\cite{LLmix}). I will therefore compute gaps using
 the ZDS potential with  $\lambda  $ as a free parameter,
     and ask what $\lambda$ fits the data, just to get a feel for its
     size.
   The results are summarized in
Table \ref{duac}.

\vspace*{.2in}

\begin{tabular}{|c|c|c|c|c|} \hline \hline
$ \nu \ \ \  $  &$B(T))$& $\Delta_{a}^{exp} ({}^oK)$  &
$\Delta_{a}^{theo} ({}^oK)$ & $\lambda$\\ \hline
  1/3  &13.9 & 8.2 &$5.3 \sqrt{B(T)}/\lambda$ & 2.4  \\
 2/5 & 11.6&3 & $2.08 \sqrt{B(T)}/\lambda$ &2.4  \\
3/7  &10.8 &2 &$1.23 \sqrt{B(T)}/\lambda$ &2.0 \\
 \hline \hline
\end{tabular}

\vspace*{.2in}

\begin{tabel}\label{duac}
 Comparison of activation masses to Du
{\em et al}, sample A, which has a density $n=1.12\cdot 10^{11}
cm^{-2}$. The last column gives the best fit to $\lambda$.
\end{tabel}

\vspace*{.2in}

Is it possible to describe the disordered sample by some effective
$\lambda$? It does not seem likely, given these three data points:
the line through them (plotted against $B^*=B-9.27$ Tesla) has a
negative intercept, while the pure system calculations give a gap
that never vanishes for any finite $\lambda$. Furthermore, the
negative intercept is between $1{}^o-2{}^oK$, while the gaps are
at best  $6-8{}^oK$. Thus the effects of disorder appear to be
quite significant.   Therefore no attempt will be made to find an
effective $\lambda$. If one day we get samples for which the
disorder broadening is much smaller, we can attempt this. In the
meantime, LDA and exact diagonalization calculations suggest that
the answers for the pure system differs by roughly a factor of two
from the data\cite{pmj},\cite{Morf}.

 Consider now the results of Pan\cite{Pan} {\em et
al}. Rather than attempt to fit their gaps to the theory,  I
consider the  following issue they raise about the normalized
activation mass \beq m^{nor}_{a}= {m_a\over m_e \sqrt{B(T)}}. \eeq
 They observe that these masses are in the range $.25-.35$ near
$\nu =1/2$ and $1/4$. How does this rough equality of masses of
fermions with two and four vortices fit in the present theory? If
we compare their results  to Eqns. (\ref{nor1}-\ref{nor2}) we
extract the  range of values  for $\lambda$ listed in Table
\ref{Pan}. They seem to be spread over a range that is more or
 less equal. The theoretical prediction for this case is sensitive
 to how exactly the two fractions are reached. For the present case,
 wherein $n$ is fixed and $B$ is altered, we expect
$\lambda^{(4)}/\lambda^{(2)}=\sqrt{2}$. I suspect that the
experimental $\lambda$'s do not show this factor of $\sqrt{2}$
because the effects of disorder are most pronounced near the
gapless states.

\vspace{.2in}
\begin{tabular}{|c|c|}\hline \hline
  Theoretical value & $ \lambda$ implied by  data \\ \hline
$ m^{nor}_{a}=.163 \lambda^{(2)}\  (s=1)$& $\lambda^{(2)} = 1.5-2$
\\
 $ m^{nor}_{a}=.175 (\lambda^{(4)})^{5/4} \ (s=2)$ & $\lambda^{(4)} = 1.3-1.75$ \\ \hline
\end{tabular}
\vspace{.2in}
\begin{tabel}
\label{Pan}  Given that normalized masses for $2$ and $4$ vortex
attachment are in the range $.25-.35$ the table asks what the
corresponding values of $\lambda$ are.Theory predicts that
$\lambda^{(4)}= \sqrt{2}\lambda^{(2)}s$.
\end{tabel}

\subsection{A study of particle and hole profiles.}

Now we consider the charge densities in a state with  either a
widely separated particle-hole pair in one of the gapped fractions
or just a CF at $\nu =1/2$. The calculations are detailed in
Appendix 3.

\begin{figure}[t]
\epsfxsize=3.5in \centerline{\epsffile{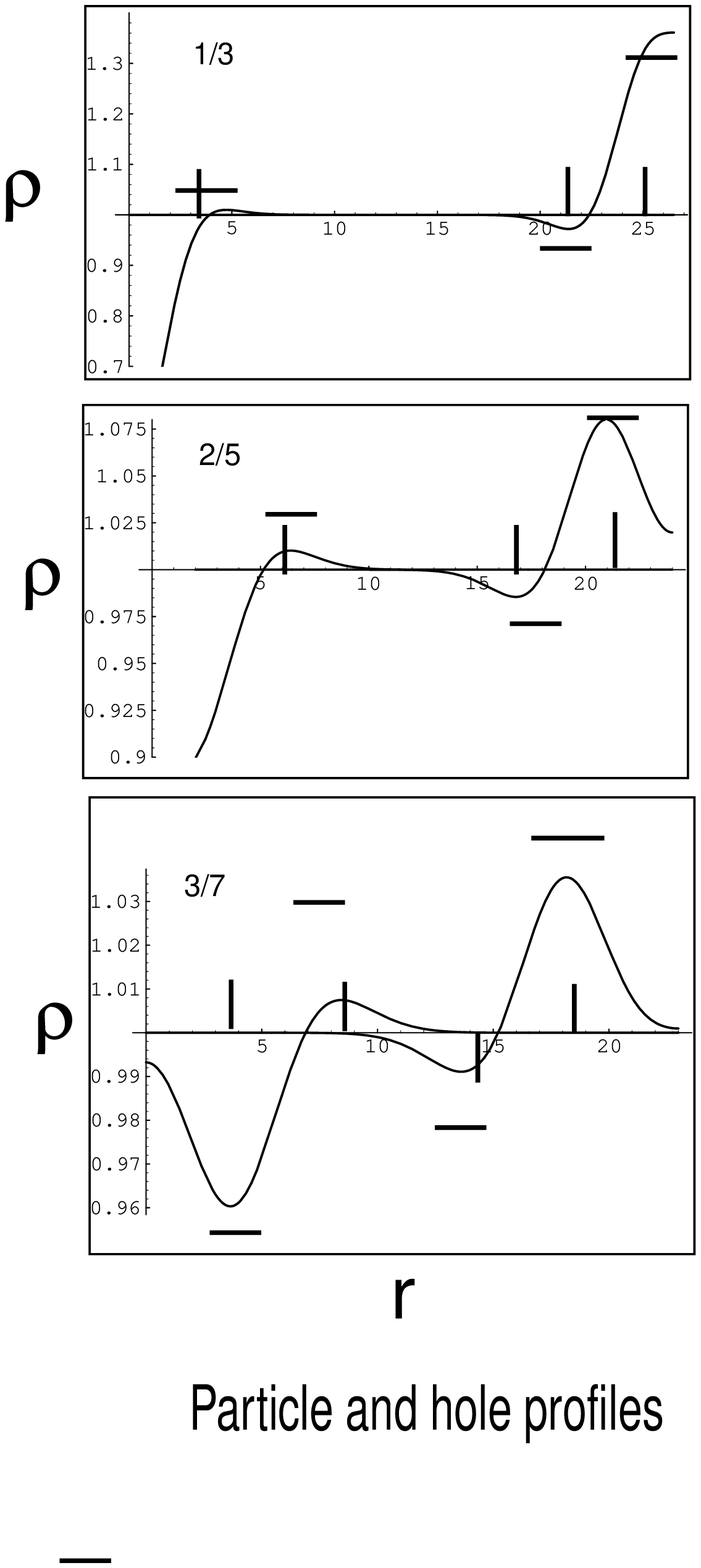}} \vskip 0.15in
\caption{Particle-hole profiles, in units of the ambient density,
at $1/3,2/5,3/7$. The results are compared to the unpublished work
of Park and Jain by placing them at antipodal points of a sphere,
with the hole at the north pole $r=0$. The coordinate   $r$ is the
distance on the sphere measured in units of $l$. The solid lines
are my predictions, the horizontal (vertical) lines show the
heights (locations) of  the maxima and minima observed by Park and
Jain. }
 \label{profilesgapped}
\end{figure}
  Figure
(\ref{profilesgapped}) which shows the distribution of charge in
units of the ambient density, for  three gapped fractions with a
widely separated particle and hole. To compare to the unpublished
work of Park and Jain who placed them
 at antipodal points of a sphere, I have done the same.
 The hole is placed at the north pole $r=0$  and $r$ is the distance along
  a great circle in units of $l$. The solid lines refer to my
results. The small horizontal lines indicate the values of the
maxima and minima of Park and Jain results, while the small
vertical lines show their locations.
 Three features are noteworthy.

  First,  the density is best predicted at $2/5$. It
is not surprising that the gap follows PMJ best  in this case.

  Next, in
the case of $3/7$,  the particle and hole actually overlap and get
entangled  in the
 Monte Carlo work on a finite sphere. In the present infinite
volume approach,  we have no trouble distinguishing them apart
since they were computed individually and superposed to draw the
figure.

Finally, consider the case of $1/3$, where one  expects to find
the best results for gaps  (due to a large gap), but  does not.
Since there is just one filled CF level, if you make a hole at
$r=0$ the density must vanish. This feature found in Park and
Jain, is absent in the present approximation where the function
drops to about half  the ambient density. This breakdown of the
present picture at short distances is generic. Note also that
while our unitary transformations attempt to go from electrons to
CF
 while Jain's wavefunction approach  goes from
 CF to the electrons, the two are not of equal accuracy.
Given the simplicity and tractability  of the final operators in
the present model (which cannot possibly carry the complexity of
the projected Jain wavefunctions in electron coordinates) this is
to be expected.

\subsection{A closer look at $\nu=1/2$}

At $\nu =1/2$ the CF of momentum ${\bf k}$  is predicted  to have
a dipole moment $l^2 \hat{z}\times {\bf k}$ \cite{read1}. Let us
understand this statement. It stems from examining the trial state

\beq \Psi = {\cal P} \prod_{i<j}(z_i-z_j)^2 Det \left| e^{i{\bf
k}_i\cdot {\bf r}_j}\right| .\eeq Before projection, the vortices
are on the electrons. The CF charge is $e^*$ but the dipole moment
is zero. To project, one begins with
 \beq e^{i{\bf k\cdot r}} = \exp {i \over 2}
({k\bar{z}+ \bar{k}z})\eeq (where $k= k_x+ik_y$, )  recalls that
$\bar{z} \to 2l^2 {\partial /\partial z}$, and sees that
 the derivatives in the exponential will move $z$ to $z+ik$. This
 suggests that the vortex will move off the particle by an amount
 $ikl^2$, which explains  the origin of the dipole and its moment.
This analysis in terms of  just one
  particle corresponds to
 what was inferred from  Figure 1 or the analysis of ${\bf R}_e$ and ${\bf R}_v$.

 While this is true, {\em it does mean that   the spatial distribution of zeros in
 the LLL wavefunction (as a function of one coordinate, the others being frozen)
  will  resemble the momenta that fill
 the Fermi sea.} There are several reasons for this.
 \begin{itemize}
 \item Besides the one exponential we considered,
 there are other  exponentials that
 move {\em every} other coordinate by the corresponding $k$, so that in the
 end
 $(z_i-z_j)^2\to (z_i+il^2k_i-z_j-il^2k_j)^2$.  This dependence on the {\em
 difference} of momenta is in accord with the symmetry first noted by  Haldane,
 and dubbed  K-invariance by Stern {\em et al}\cite{simon}.
  \item All of the above  refers to just one term in the expansion of the determinant.
  We must still antisymmetrize
 over all possible pairing of
 momenta with coordinates.
 \item Many of these  zeros will
 have to disappear upon projection, leaving just a total of two per
 particle. It is far from   clear how the zeros will  be
 distributed in the end. All we know is that
  one has to lie on the electron by the Pauli principle, and the other
   forms a single
  vortex, as in D.H. Lee's picture\cite{DH}. The latter
  must be at a  distance $2kl^2$ from the electron to preserve the dipole
 moment.
Since the "size" of the CF at the Fermi energy, $2l$,
 is close to the
  interparticle spacing of $2\sqrt{\pi }l\simeq 3.5 l$,
 there is
 definitely going to be some ambiguity in pairing zeros with electrons.

 \end{itemize}

These complications notwithstanding, the dipole makes a clear
appearance in the operator approach.  It is   not tied to zeros of
the wavefunctions and emerges as follows: if the expression for
the preferred charge density $\bar{\bar{\rho}}^p$, (which obeys
Kohn's theorem), is coupled to an external potential, the second
term in the series Eqn. (\ref{rostar}) is precisely that of a
dipole of
 strength $l^2 \hat{z}\times {\bf k}$.
  This operator definition is unchanged by
antisymmetrization (we simply write Eqn. (\ref{rostar}) in terms
of fermion operators in second quantization) and crystallizes the
notion of the dipole inspired by an analysis of the trial
wavefunction. The dipoles also emerge  in the high frequency
density-density response computed in the conserving
approximation\cite{read2}, as stated earlier. The  zeros of trial
wavefunctions\cite{han} do not seem to be  the likely place to
look for evidence of dipoles.

\begin{figure}
\epsfxsize=3in \centerline{\epsffile{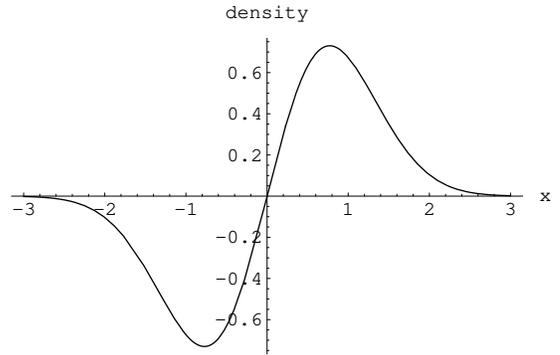}} \vskip 0.15in
\caption{The charge density of a CF showing its dipole moment. The
CF is localized in $x$ but spread out uniformly in $y$. Luckily
this does not smear out the dipole moment. } \label{profiledipole}
\end{figure}

 I will now illustrate these points  in a many-body state
with an extra CF. It however takes a bit of work.  Suppose we
create a CF in a state of definite momentum ${\bf p}$ above the
Fermi surface. If we evaluate $\bar{\bar{\rho}}^p ({\bf r})$ in
this state, we get zero. This is because the CF has an equal
amplitude to be at all places and the dipolar density gets smeared
out completely. If we localize it at the origin, we use all
momenta, and hence all values of the dipole moment which again
averages  to zero charge density. Consider however the following
state \beq |p_0\rangle = \int_{-\infty}^{\infty}{dp_x}d^{\dag}
(p_x,p_o)|FS\rangle \eeq
 where $|FS\rangle$ is the Fermi sea, and
$p_0>p_F$. This is a superposition of  states of all values of
$p_x$, localizing the CF at  $x=0$ but with fixed $p_y =p_0$. (As
$p_0>p_F$, all these states lie on a line outside the FS.) The
particle is thus spread out completely in the $y$-direction.
However, its dipole moment is fixed at $l^2 p_0$ along the
$x$-axis. A calculation done in Appendix 3 yields the following
result in compliance with these expectations: \beq \rho ({\bf r})
\simeq \left[ \exp \left[{-(x-{1\over 2}p_ol^2)^2\over l^2}\right]
- \exp\left[{-(x+{1 \over 2}p_ol^2)^2\over l^2}\right]\right] \eeq
depicted in Figure (\ref{profiledipole}).

If one evaluates the dipole moment of this charge distribution,
one finds it equals $p_ol^2$. This is the sense in which the
dipole moment appears in the  the operator approach.

I conclude with one significant difference between $e^*$ and
$d^*$. The CF charge $e^*$ is robust under projection, while $d^*$
is not.  In the unprojected wavefunction at $\nu=1/2$, the dipole
moment of the CF is zero since the vortices are on top of the
electrons. The charge is zero as well. Consider a   system with an
interaction comparable to the cyclotron gap, such  that   the best
wavefunction is the above  unprojected one. Based on this
wavefunction we would assign to the  CF a charge $e^*$, but no
dipole moment. Consider now a problem where the projected state is
the best. Instead of applying ${\cal P}$ in one shot, imagine
slowly reducing the non-LLL component of the wavefunction to zero.
Along the way, the zeros get ripped off the electrons and some of
them disappear. Through it all $e^*$ is invariant, since it
depends on the Hall conductance, which is constant. But the dipole
moment, which describes the internal structure of the CF, changes
from zero to some value in the LLL. This value seems to be $l^2 \
{\bf z \times k}$, in the wave function based arguments of Read,
in the operator series and the density-density response. I do not
know if the dipole moment is robust within the LLL, but suspect it
is, since Kohn's theorem (which limits the matrix element in the
LLL) gives a unique answer in the operator approach.

Some caution must be used in looking for this moment. Even if we
know how the CF couples to an external potential, the response
function will depend on the Hamiltonian as well. In particular, we
know that at $\omega =0, \ q\to
 0$, the compressibility will not vanish as
$q^2$ but as a constant, because of the special symmetries of $H$,
a point made by Halperin and Stern and discussed at length in
Refs. \cite{HS1},  \cite{simon} and \cite{read2}. Only at high
frequencies will the dipoles behave classically.

\begin{figure}
\epsfxsize=3in \centerline{\epsffile{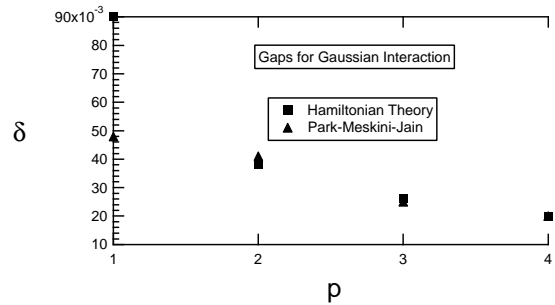}} \vskip 0.15in
\caption{Comparison to the work of Park, Meskini and Jain {\em et
al} for the gaussian potential $v(q) = {2\pi e^2  l} \
e^{-q^2l^2/2}$ for $p=1,2,3,4$} \label{fig4}
\end{figure}

\subsection{Other potentials}

Figure \ref{fig4} shows a comparison to the PMJ results for a
gaussian potential \beq v(q) = {2\pi e^2  l} \ e^{-q^2l^2/2} \eeq
Note that except for $\nu = 1/3$ the agreement is exceptional.
This is the kind of potential for which the present theory works
best.

On the other hand for a potential, \beq v(r) = {e^{-\kappa r}\over
r} \eeq the agreement is worse than for the coulomb case since
this potential is
 just as bad as $r\to 0$ and does not give the large $r$ values a chance.
 Likewise $1/r^2$ fares worse than $1/r$.

From playing with these potentials and using the PMJ results as a
benchmark
 we can thus learn when the present model can be trusted.

\subsection{Effects of CF interactions}

\begin{figure}
\epsfxsize=3in \centerline{\epsffile{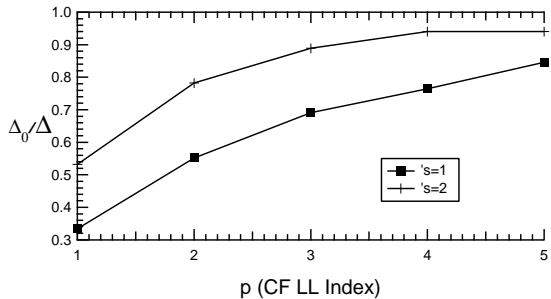}} \vskip 0.15in
\caption{Ratio of $\Delta_0$, the gap computed without $H_I$ to
the to $\Delta$, the full HF gap, for $p=1,2,3,4$ and $s=1,2$.}
\label{fig3}
\end{figure}

Figure \ref{fig3} shows what happens to the gaps if the CF
interactions are turned off.
 These correspond to the contribution from the $G_0$ term in Eqn. (\ref{freeplusint}).
 Note that interactions seem less important for $\nu =1/4$ and
 systematically get less important as $p$ increases.

 There is some freedom in defining the  measure of  interactions,
 which I took to mean the effect
 of $H_I$ on $\Delta_a$. But the effect of $H_I$ is two fold: it renormalizes
  the self-energy of the particles and also mediates interactions
 between them. One can  envisage a situation in which
 individual energies get strongly renormalized by $H_I$ but the dressed
  particles are barely
 interacting, i.e., their energy barely varies with separation.
 In this case one could argue that CF are weakly
 interacting. Such a separation, between the quasiparticle and quasihole,  was set to infinity in  our
 gap calculations.  This energy, as a  function of separation,
 is contained in the
  $q$-dependence  of
 the magnetoexciton \cite{magp},\cite{GMMP} spectrum. Since the variation is
 typically comparable to the gap,
CF seem to be
  quite strongly
 interacting by this measure as well.

\subsection{Comparison to the work of Morf {\em et al}}
\begin{figure}
\epsfxsize=3in \centerline{\epsffile{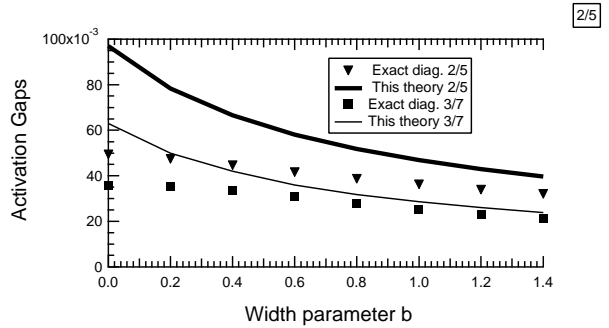}} \vskip 0.15in
\caption{Comparison of hamiltonian theory to the exact
diagonalization results of Morf {\em et al} for $p=2 \mbox{and} 3$
}
  \label{Morf}
\end{figure}

Morf {\em et al}\cite{Morf} have calculated the activation gaps
for $1/3$, $2/5$ and $3/7$ by exact diagonalization of finite
systems, paying great care in extrapolating to the thermodynamic
limit. The potential they use is \beq v(q) = {2 \pi e^2\over q}
{\Large e^{(qlb)^2)}\ Erfc\  (qlb)} \eeq where $b$ is the analog
of $\lambda$. Our numbers are compared in Figure \ref{Morf}.

The following features are worthy of note.
\begin{itemize}
\item The calculated gaps always lie above the exact
diagonalization results for the two fractions shown (as well as
for the $1/3$ case, not shown). This  result agrees with the
general belief that HF always overestimates the gaps by neglecting
fluctuations. Compare this to the case of PMJ where for $3/7$ and
$4/9$ the calculated gaps were sometimes lower and sometimes
higher.

The fact that the theory predicts gaps that always exceed the
exact diagonalization results suggests the possibility that the
problem may lie not in $H^p$ but in the HF approximation. If $H^p$
were solved by a more accurate method than HF, the agreement might
have  been better. This however merely remains a possibility till
someone solves it by, say,  exact diagonalization. This will be
hard since  the theory has constraints and is formulated in the
full fermionic Hilbert space.

\item The general agreement is worse for this potential than for
the ZDS case. This is because at large $q$ this potential goes as
$1/q$ while the ZDS potential falls exponentially.
\end{itemize}

\subsection{Particle-hole symmetry}
For the fully polarized case one expects that within the LLL,
$\Delta_a$ for  $\nu $ and $ 1-\nu$ will be equal. Let us focus on
just the case $s=1$, which corresponds to two-vortex attachment.
The fractions related are now $\nu$ and \beq 1-\nu = 1-{p \over
2p+1} =  {p+1\over 2(p+1)-1} \eeq Thus to find the gap at $1-\nu $
corresponding to a certain $\nu = p/(2p+1)$ we must
\begin{itemize}
\item Replace $p$ by $p'=p+1$ and consider fractions of the form
$p'/(2p'-1)$. The flux quantum per CF is now $-1/p'$ which means
the double-vortices have overturned the applied field and changed
its direction and $p'$ levels are filled in this weakened and
reversed field. In Eqns. (\ref{gaps1}-\ref{gaps2}) we must make
the obvious changes, $p\to p'= p+1$ and   $(2p+1)\to (2p'-1)$, in
all the denominators, and the not so obvious change $\exp
(-x/(4p(2p+1))\to \exp (x/(4p'(2p'-1))$ in the double-vortex form
factor which reflects the fact that the vortex charge is bigger
than that of the electron.
\item If we are working at $\Lambda = l \lambda \ne 0$, we have to
keep $\Lambda$ constant in the comparison. Thus we need to verify
that \beq \Delta ( {\nu}, \lambda )=\Delta \left(1-\nu ,
\lambda\sqrt{{\nu \over 1-\nu}} \right) \eeq or for the
dimensionless gap $\delta_a = \Delta_a/(e^2/\varepsilon l))$, \beq
\delta_a (\nu , \lambda ) = \sqrt{{\nu \over 1-\nu }}\delta_a ({
1-\nu},\lambda\sqrt{{\nu \over 1-\nu }}  )\eeq

\end{itemize}
My results are summarized in Table \ref{phsym}.

\vspace*{.2in}

\begin{tabular}{|c|c|c|c|c|} \hline \hline
$ \nu \ \ \  $ &  $\delta_a (\nu ,0)$ &${\delta_a ( 1-\nu,0)\over
\sqrt{(1-\nu )/\nu}}$ & $\delta_a (\nu,1)$ & ${\delta_a (1-\nu,
\sqrt{{\nu \over 1-\nu }})\over \sqrt{(1-\nu )/\nu}}$
\\ \hline
  1/3 & .213 &.203 &.106 & .101  \\
 2/5 &.097 & .098 &.042 & .043 \\
3/7 &.063 &.065 &.025 &.026 \\
 4/9 &.047 &.048 & .017& .018 \\
5/11 & .038 &.039 &.013& .014\\
  \hline \hline
\end{tabular}
\begin{tabel}
\label{phsym} Particle-hole symmetry in the LLL for the polarized
cases with $\nu =p/(2p+1)$ requires that the dimensionless gap
$\delta_a (\nu ,\lambda )= \sqrt{1-\nu \over \nu}\delta_a
\left(1-\nu ,\lambda \sqrt{{\nu \over 1-\nu }}\right)$.  The Table
considers $\lambda =0 \ \mbox{and} \ 1$ and indicates that the
theoretical numbers obey this symmetry.
\end{tabel}
\vspace*{.2in}

 Note that particle-hole symmetry holds
very well even at $\lambda =0$ when the absolute value of the  gap
is not close to the benchmark value set by PMJ. If one looks at
the expressions for the gap, one sees that this agreement is very
nontrivial since the two problems are very different in the CF
formalism: they have different number of filled LL, and the matrix
elements involved are quite different.

\section{Magnetic transitions at $T=0$.} Now we turn to the
behavior of the spin of the system, which was assumed to be frozen
along the applied field. This topic has explored within the CF
approach rather extensively by Park and Jain\cite{parkjain}.

 The coupling of electron spin to the applied field is given
a Zeeman term
 \beq H_Z
=-g \left( {e \over 2m_e}\right){S\over 2} B \eeq where $g=.44$,
$m_e$ is the electron mass in free space, $S$ is
  given by \beq
S=n\ P \eeq where $n$ is the density and $P$ is the polarization,
{\em to which each electron contributes $\pm 1$.}

\subsection{Magnetic transitions in gapped fractions}
 If $H_Z$ dominates, we expect the system to be fully
polarized ($P=1$). As we lower $H_Z$, we may expect $P$ to drop.
In CF theory for gapped fractions there is a discrete set of
allowed values of $P$. At $\nu =p/(2ps+1)$, these correspond to
states of the form $|{\bf p-r,r}\rangle$. These stand for
many-body states in which  $p-r$ LL are occupied by up spins and
$r$ LL by down spins. In Jain's approach, the actual wavefunction
will be such a state times the Jastrow factor, followed by
projection to the LLL. In the present approach, $|{\bf
p-r,r}\rangle$ is literally the state, but the operators for
charge and spin  are obtained by canonical transformations. For
the interested reader I mention that flux attachment and canonical
transformations are  same for both spins.

Since the uniform external field couples to  the $q=0$ component
of the spin density which  is unaffected by the canonical
transformations, $H_Z$ will have the same form in the final CF
representation.

It is important to note that even though the states are labeled by
free particles (in a weakened field $A^*$), the problem is not
really free: in Jain's version the free states  turn to  highly
correlated wavefunctions for electrons, and our case, the states
may look free, but  $H$ is not.

In any event, the allowed values of polarization are given  by
 \beq
P= {p-2r\over p} . \eeq
 Thus for example, when $p=4$, the allowed
values are $P=1,.5,\mbox{and}\  0$ corresponding to $|{\bf
4,0}\rangle \ $,$|{\bf 3,1}\rangle \ $ and $|\bf{2,2}\rangle$.

Our goal is to calculate the critical fields at which the system
will jump from one value of $r$ to the next as $H_Z$ is varied.
Let
 \beq E (p-r,r)   = \langle {\bf p-r,r}|H|{\bf p-r,r}\rangle  \eeq
 {\em where $H$ does not contain the energy due to $H_Z$}.
 This will be case for the single-particle and ground state energies,
 with one exception which will be clearly pointed out. Since $H_Z$  is diagonal in
 the
HF states  which   have definite spin, its effects can be
 trivially incorporated.

The HF calculation of $E(p-r,r)$, detailed in the Appendix 5,
gives
\begin{eqnarray*}
 \lefteqn{E(p-r,r)=}\\
 &&{n\over p}\int_q \left[ \sum_{n_1=0}^{p-r-1}\langle n_1|
 {\rho} (q)\rho (-q)|n_1\rangle -\sum_{n_1=0}^{p-r-1}\sum_{n_2=0}^{p-r-1}
 |\rho_{n_1n_2}|^2\right. \\
 &&+\left. \sum_{n_1=0}^{r-1}\langle n_1|
 {\rho} (q)\rho (-q)|n_1\rangle -\sum_{n_1=0}^{r-1}\sum_{n_2=0}^{r-1}
 |\rho_{n_1n_2}|^2\right] \\
 \end{eqnarray*}
 where $\rho_{n_1n_2}$ was  introduced earlier and discussed in
 Appendix 1 and
 \beq
\langle n|
 {\rho} (q)\rho (-q)|n\rangle = \sum_{n'=o}^{\infty}
 |{\rho} (q)_{nn'}|^2
 \eeq

 The critical field $B^c$ for the transition from  $r$ to $r+1$ is
given by : \beq E(p-r,r)-E(p-r-1,r+1) = g{ e  B^c\over
2m_e}{n\over p} \label{criticalB}\eeq where the right hand side
denotes the Zeeman cost of flipping  the $n/p$  spins in  the  LL
that switched its spin. This discussion assumes that $B$ is
perpendicular to the sample. If there is a tilt $\theta$,  we
write
 \beq E(p-r,r)-E(p-r-1,r+1) = g{ e  B^{c}_{\perp}\over
2m_e \cos \theta }{n\over p} \label{criticalB2}\eeq

When these $B^c$'s were calculated, I noticed the same  remarkable
regularity first noted by  Park and Jain\cite{parkjain}, namely
that they
  could be fit by a theory of free fermions of
mass $m_p$  (the polarization mass)  that  occupy LL with a gap
$\Delta_p ={e B^*/ m_p}$. In this case we would have \beq  E(p\!
-\! r,r)-E(p\! -\! r\! -\!1,r\! +\! 1)=   {n(p-2r-1 )\over p}
\Delta_p \label{gapdefine} \eeq
 since    $(n/p)$ spin-up fermions of energy $(p-r-1+{1\over
2})\Delta_p$ drop to the  spin-down level with energy  $
(r+{1\over 2}) \Delta_p$.

Suppose we evaluate the left-hand-side of Eqn. (\ref{gapdefine})
in the   HF approximation to  $H$ and {\em define}

\beq \Delta_{p}(r)^{def} = {p\over n} { E(p-r,r)-E(p-r-1,r+1)\over
p-2r -1}.\eeq

Given that  $H$ is not free, there is no reason why
$\Delta_p(r)^{def}$ should be  $r$-independent. But  it is very
nearly so. For example at $p=6, \lambda =1$,

\beq \Delta_p(0,1,2)^{def} = {e^2\over \varepsilon l}(0.00660,
0.00649,  0.00641) \eeq which describe $|{\bf 6,0}\rangle \to
|{\bf 5,1}\rangle$, $|{\bf 5,1}\rangle \to |{\bf 4,2}\rangle$, and
$|{\bf 4,2}\rangle \to |{\bf 3,3}\rangle$. This $r$-independence
of the gaps was true for every fraction and every  value of
$\lambda$ I looked at. Yet I knew that $H$ was definitely not free
since the activation gap $\Delta_a$ to make a widely separated
particle-hole pair differs from $\Delta_p$ by factors like $2$ or
$4$ (depending on $\lambda$) and turning  off $H_I$ makes a
substantial difference, as demonstrated earlier.

 I will place this result in perspective shortly, after noting
that it has a counterpart in the gapless case as well. But first I
present the HF results for (the $r$-independent)  $m_p$ and
$\Delta_p = eB^*/m_p$.  At and near $\nu = {1\over 2} \ \mbox{and}
\ {1\over 4}$, for
  $.75 < \lambda < 2$, $m_p$  may be approximated by
\begin{eqnarray}
{1 \over m^{(2)}_{p}} &=& { e^2 l \over \varepsilon }C_{p}^{(2)}
(\lambda )\ \ \ \
 \ \ \ \  C_{p}^{(2)} (\lambda )  = {.087 \over \lambda^{7/4}} \label{mp2}\\
{1 \over m^{(4)}_{p}} &=& { e^2 l \over \varepsilon }C_{p}^{(4)}
(\lambda ) \ \ \ \ \ \ \ \
   C_{p}^{(4)} (\lambda )={.120 \over \lambda^{7/4}}\label{mp4}
\end{eqnarray}
 For fractions like $2/5$, not too close to $1/2$, I will use the actual $m_p$
  in comparing to experiment.

The transition $|{\bf p-r,r}\rangle \to ||{\bf p-r-1,r+1}\rangle $
occurs when \beq
 g{e\over 2m_e} {B_{\perp}^{c}  \over \cos \theta}
 =(p-2r-1)\Delta_p .\label{tran}
             \eeq

\subsection{Magnetic transitions of gapless fractions}
Let us now turn to the gapless fractions $1/2$ and $1/4$. The
discrete labels $p-r$ and $r$ of the HF states  that count the
spin-up and down LL's are now replaced by continuous variables
$k_{\pm F}$ which label the Fermi momenta of the spin-up and down
seas. These momenta are such that the total number of particles
equals $n$:  \beq k_{+F}^{2}+k_{-F}^{2} = k_{F}^{2}=4\pi n \eeq
where $k_F$ denotes the Fermi momentum of a fully polarized sea.

In the gapped case there were several critical fields $B^c$, each
corresponding to one more CF -LL  flipping its spin, each
describing one more jump  in the allowed values of $P$. In the
gapless case the situation is different. For very large Zeeman
energy, the sea will be fully polarized. It will not be worth
including even one fermion of the opposite spin since the  Zeeman
energy cost alone will exceed the Fermi energy of the polarized
sea. As we lower the Zeeman term, we will reach a critical field
at which it will be worth introducing one fermion of the other
spin with zero kinetic energy. At this point the energy of a
particle on top of the spin-up sea obeys \beq {\cal E}_+(k_{+ F})
=g{e\over 2m_e} {B_{\perp}^{c} \over \cos \theta}. \eeq If we
lower the Zeeman term further, the polarization will fall
continuously and be determined by ${\cal E}_{\pm}(k_{\pm F})$, the
energies of the particles on top of these two seas according to
\beq {\cal E}_+(k_{+F}) -{\cal E}_-(k_{-F}) = g{e\over 2m_e}
{B_{\perp} \over \cos \theta} \eeq since this equation states that
the system is indifferent to the transfer of a particle from one
sea to another, i.e., has minimized its energy with respect to
polarization.

Using the fact that at these fraction when the mean magnetic field
vanishes, we deal with a very simple expression: \beq
\bar{\bar{\rho}}^p({\bf q})=\int {d^2k\over 4\pi^2}(-2i) \sin
({{\bf q \times k }\ l^2\over 2}) d^{\dag}_{{\bf k - q}}d_{{\bf
k}} \eeq it is easy to do a HF calculation and obtain
\begin{eqnarray*}
\lefteqn{{\cal E}_{\pm}(k))=}\\ && 2\int{d^2q \over
4\pi^2}\check{v}(q)\sin^2 \left[{{\bf k \times q}l^2\over
2}\right] \\ && -4\int{d^2k' \over
4\pi^2}n^{F}_{\pm}(|k'|)\check{v}(|{\bf k -k'}|)\sin^2 \left[{{\bf
k' \times k}\ l^2\over 2}\right]\\ &&\equiv {\cal E}_0+{\cal E}_I
\end{eqnarray*}
where the Zeeman energy is not included,  $n^{F}_{\pm}$ is the
(step)  Fermi function for the two species, ${\cal E}_0$ and
${\cal E}_I$ represent single particle energy (due to what was
called $H_0$ earlier) and the energy of interaction of this
particle at the Fermi surface with those inside the sea,  and \beq
\check{v}(k)=v(k)e^{-k^2l^2/2}. \eeq

When this result is used to compute ${\cal E_+}(k_{+F}) -{\cal
E}_-(k_{-F})$, I find once again  that the numbers fit a free
theory in the following sense. Imagine that CF were free and had a
mass $m_p$. We would then have \beq {\cal E}_+(k_{+F}) -{\cal
E}_{-}(k_{-F})= {k_{+F}^{2}-k_{-F}^{2}\over 2m_p} \eeq What I find
is that the HF number for ${\cal E}(k_{+F}) -{\cal E}(k_{-F})$ may
be fit to the above form {\em with an $m_p$ that is essentially
constant as we vary $k_{\pm F}$ i.e., the relative sizes of the up
and down seas (which is analogous to an $m_p$ that does not depend
on the index $r$ in the gapped case) and that this $m_p$ matches
smoothly with that defined for the nearby gapped fraction.}

This result is surprising because  we know the CF are not free
from a variety of reasons. Indeed the HF energies do not have a
quadratic dispersion: for example at $\nu ={1\over 2}$ and
$\lambda =1$
 \beq
{{\cal E}(k_{\pm F}) \over (e^2/\varepsilon l)} = a \left({k_{\pm
F} \over k_F}\right)^2 +b \left({k_{\pm F} \over k_F}\right)^4
\label{dispersion}\eeq where $a=.075$,    $b=-.030$.

The proper interpretation of this free-field behaviour will be
taken up shortly. For now I assume this feature of the results and
define  $m_p$  by \beq {1\over m_p} =2{{\cal E}_+(k_{+F}) -{\cal
E}_-(k_{-F})\over k_{+F}^{2}-k_{-F}^{2}}.\eeq As mentioned above,
$m_p$ for the gapless cases merges smoothly with the $m_p$ for the
nearby gapped states. The results are thus given by  Eqns.
(\ref{mp2}-\ref{mp4}) for the range $.75 \le \lambda \le 2$ for
the ZDS potential.
\subsection{Why the free-field behavior?}
The fact that magnetic phenomena at $T=0$ can be described (to
excellent accuracy)  by free fermions of mass $m_p$  needs to be
properly understood and interpreted. For example one must resist
the thought that perhaps by some further change of variables one
could take the present hamiltonian and convert it to a free one.
This is because if there were really an underlying free theory, it
would be able to predict an activation mass $m_a$  and this  would
have to  coincide with $m_p$. We know within this theory, within
Jain's approach, or from experiment, that these masses differ by
at least a factor of two.

I will now show that a single assumption about the form of the
ground state energy, an assumption that is not equivalent to the
free-field assumption or even to a quadratic dispersion relation
in the gapless cases,  will explain this behavior for gapped and
gapless fractions. Consider $E(S)$, the ground state energy as a
function of $S=nP$. By rotational invariance it must have only
even powers of $S$ in its series. Assume the series is dominated
by the first two terms: \beq E(S) = E(0)+{\alpha \over 2}S^2\eeq
where $\alpha$ is the inverse linear static susceptibility.

Consider first  the gapless case.
  When  $dn$  particles go from  spin-down to spin-up,
 \begin{eqnarray}
 dE &=&{\alpha   } \ S\  dS = {\alpha   }\  S \ ( 2 \ dn)\\
 &=& \alpha {k_{+F}^{2}- k_{-F}^{2}\over 4\pi} (2 \   dn)
 \end{eqnarray}
using  the volumes of the Fermi seas.
  We see that $dE$  has precisely the form of the kinetic
 energy difference of particles of mass  $m_p$ given by
\beq
 {1 \over m_p} = {\alpha \over \pi}.
  \eeq
 Thus $m_p$ is essentially  the static susceptibility,
  which happens to have dimensions of mass in
 $d=2$. The statement that $m_p$ has no $r$-dependence in the gapped
  case or no spin dependence in the gapless case is the same
 as saying that the full nonlinear susceptibility does not depend on the
  spin
 $S$, which in turn means $E(S)$ is quadratic in $S$.

  Note that the free-field form of $dE$
  comes from  $E \simeq S^2$   {\em and}
   $d=2$: in $d=3$, we would have $dE/dn \simeq  S
 \simeq (k_{+F}^{3}- k_{-F}^{3}) $ which no one would interpret   as a
 difference of  kinetic energies.

This  general argument notwithstanding, it is worth explicitly
considering the case in hand.

 First, one must not think that
$E(S)$ being quadratic in $S$ as
 equivalent to assuming that the single-particle HF energies are
 quadratic in momenta.  Consider the HF energies
 quoted earlier

\beq {{\cal E}(k_{\pm F}) \over (e^2/\varepsilon l)} = a
\left({k_{\pm F} \over k_F}\right)^2 +b \left({k_{\pm F} \over
k_F}\right)^4 \label{dispersion2}\eeq

The quartic terms miraculously drop out in the energy cost of {\em
transferring} a particle from the top of the spin-down sea to the
top of the spin-up sea:

\begin{eqnarray}
{dE \over (e^2/\varepsilon l)}&=&a{k_{+F}^{2}-k_{-F}^{2}\over
k_{F}^{2}}+b{k_{+F}^{4}-k_{-F}^{4}\over k_{F}^{4}}\\ &=&
a{k_{+F}^{2}-k_{-F}^{2}\over
k_{F}^{2}}+b{(k_{+F}^{2}-k_{-F}^{2})(k_{+F}^{2}+k_{-F}^{2})\over
k_{F}^{4}}\\ &=&{(a+b)\over k_{F}^{2}}(k_{+F}^{2}-k_{-F}^{2})
\end{eqnarray}
using \beq k_{+F}^{2}+k_{-F}^{2}=k_{F}^{2}. \eeq Note how $d=2$
was essential to this argument: in $d=3$ we would have
$k_{+F}^{3}+k_{-F}^{3}=k_{F}^{3}$.

{\em Thus the $k^4$ terms in ${\cal E}(k_{\pm})$ are not the cause
of the
 $S^4$ term.} However,    a small $ k^6$ term
 in ${\cal E}(k_{\pm})$, corresponds to  small quartic terms in
 $E(S)$.

To understand  why the $k^6$ term is so small, we turn to Eqn.
(\ref{freeh}) for $H_0$. Expanding the $\sin^2$ in a series, we
find the $k^6$ term is down by a
 factor of at least 15 (50) relative to the $k^2$ term, at $\lambda
=0$  ($\lambda =1$), all the way up to $k=k_F$. Presumably this
feature
 (and its counterpart in the gapped case)
persists in the HF approximation to  $H$ and keeps $E(S)$
essentially quadratic, which in turn mimics free-field behavior.

To really drive home the point, consider a problem where particles
are free and  have a dispersion relation \beq {\cal{E}}(k)= ak^2 +
bk^4 . \eeq Let $yn$ be density of spin-down particles  and
$(1-y)n$ particles that of  spin-up particles. Thus \beq S=(1-2y)n
.\eeq The total energy, as a function of $y$ is
 \begin{eqnarray} E(y) &\simeq & \int_{0}^{4\pi
ny}dk^2 (ak^2+bk^4)+\int_{0}^{4\pi n(1-y)}dk^2
(ak^2+bk^4)\nonumber \\ &=& a'(y^2+(1-y)^2) +b'(y^3+(1-y)^3)
\end{eqnarray}
Note that the cubic terms in $y$ cancel. Since $y$ is linearly
related to $S$, it follows $E(S)$ is also quadratic in $S$.
However a $k^6$ term would have led to $S^4$ terms in $E(S)$.

We wrap up this topic with one thought: even if CF are free or
nearly so, there is no reason their kinetic energy should be
quadratic in momentum. These particles owe their kinetic energy to
electron-electron interactions,   and given this fact, all we can
say is that their energy must be an even function of $k$, starting
out as  $k^2$  at  small $k$. What constitutes small $k$ is an
open question that is answered unambiguously here: our expression
of the energy has substantial $k^4$ terms for momenta of interest.

Let us now turn to the gapped case and verify that   $E(S) = E(0)+
{\alpha \over 2}S^2$ implies that $\Delta (r)$ will be
$r$-independent. First note that \beq S=nP=n(1-{2r\over
p})={n\over p}(p-2r). \eeq It then follows that
\begin{eqnarray*}
\lefteqn{E(p-r,r)-E(p-r-1,r+1)=}\\ && {\alpha \over
2}(S^2(p-r,r)-S^2(p-r-1,r+1))\\  &=& {\alpha \over
2}\left[{n^2\over p^2}\right] (p-2r+p-(2r+2))(p-2r-p+2r+2)\\
&=&{\alpha }\left[{n^2\over p^2}\right](2p-4r+2)\\ &\equiv
&{n\over p} (p-2r-1)\Delta_p(r)^{def}
\end{eqnarray*}

We find that the $r$-dependence of $\Delta_p(r)^{def}$ drops out
and gives \beq \Delta ={2n\alpha \over p} .\label{Delta}\eeq

If we write
\begin{eqnarray}
\Delta_p &=& {eB^*\over m_p}={eB\over m_p(2ps+1)}\\ &=& {2\pi n
\over \nu m_p}{1 \over 2ps+1} = {2\pi n \over m_p p}
\end{eqnarray} it implies, upon comparing to Eqn.(\ref{Delta})
that, as in the gapless case \beq {1 \over m_p} ={\alpha \over
\pi}.\eeq

Analogously to the   gapless case,  we can show that if the HF
energies vary with the LL index $n$ as ${\cal E}(n) = an+bn^2$,
$E(S)$ will be quadratic in $S$. Thus the CF-LL's do not have to
be uniformly spaced for them to behave as if they were (with a
spacing $\Delta_p$) in  $T=0$ magnetic transitions.

\subsection{ Effective potentials for dirty systems}

Here I ask if it is possible that a ZDS potential with some
effective
 $\lambda$ can describe the dirty system. First of all, I realize
 this cannot be true with respect to all observables, if at all it
 is true for any. For example, if one were considering
 conductance, one knows the electron in a disordered potential will
 typically
 get localized whereas no ZDS interaction will predict this. As
 for transport gaps, the present day  samples, with a disorder
 broadening of the same order as the gaps, again preclude this
 possibility. Magnetic transitions, on the other hand, are
 controlled by total energies and one may expect that disorder will have a rather innocuous
 effect and  can
 be represented in an average way by some translationally
 invariant interaction. I will show below that at least in some
 limiting case this is a reasonable approximation. I will
 show that even in cases where I cannot provide a similar argument,
 it  seem to work. This needs to be  fully understood.

 As seen above, the critical fields are controlled by  ground
 state energies in states of different polarization. Let us write
 the ground state energy as
 \beq
E_0=  \langle \Omega |H|\Omega \rangle \eeq where $|\Omega
\rangle$ is the ground state, and  \beq H = {1 \over 2}
\int{d^2q\over (2\pi )^2} \bar{\bar{\rho}}^p \ ({\bf q}) \
v(q)e^{-(ql)^2/2} \ \bar{\bar{\rho}}^p \ (-{\bf q}). \eeq Now add
on a perturbation that couples the system to an external impurity
potential $\Phi (q)$. To second order we find, upon disorder
averaging,
 \beq E=E_0 +\int{d^2q\over (2\pi )^2}e^{-q^2l^2/2}{|\langle N
 |\bar{\bar{\rho}}^p|\Omega \rangle |^2 \Phi^{2}_{0} (q) \over E_0-E_N}
 \eeq
 where $N$ is any intermediate state and $\Phi_{0}^{2}$ is the average of $\Phi (q)\Phi (-q)$ over
 realizations. We will now replace $E_N-E_0$
 by $\Delta$,  the smallest  excitation gap  produced by $\bar{\bar{\rho}}^p$.
 In our HF states, this is the gap to the next CF LL, ignoring the
  $q$ dependence of the magnetoexciton, or the roton minimum if one includes it.
 This replacement overestimates the
 second order contribution. Since matrix elements to more distant LL
 are accompanied by higher powers of $q$, the error will be  small if the
 disorder potential (due to faraway impurities) has only long
 wavelength components. If we now use
 completeness, we see that the second order energy can be  found
 by sandwiching a hamiltonian, once again quadratic in $\bar{\bar{\rho}}^p$, but
 with
  an effective potential
 \beq
 v_{eff}(q)= v(q) - {\Phi_{0}^{2}(q)\over \Delta}\label{veff}
 \eeq
  While the second term, due to disorder,
  need not be of   the original form, what is important is that it is
  translationally invariant and makes a negative definite contribution
  to $v_{eff}$.   We
 could incorporate its effect by increasing $\lambda$ in the original ZDS term
 by a
 suitable amount. While such a replacement cannot reproduce all the effects of $v_{eff}$
 in detail, let us note that  the energies depend on
  the potential $v(q) $  principally  via the corresponding
  Haldane pseudopotential\cite{pseudo} $V_1$. (This is for $2s=2$. For $2s=4$, it is
  $V_3$. In any event, one only dominates.)   We may thus choose the effective
 $\lambda$,
 so that it reproduces the dominant  $V$ corresponding to $v_{eff}$.

 We see that at $T=0$, in a state with a robust  gap, to
 second order in the impurities, there is an effective $\lambda$,
  if we ignore higher CF LL's.
  These are a lot of restrictions. Of these, only use of second
   order perturbation theory may by
 justifiable since the CF has a charge $e^*$ and not $e$, and
 correlations, which are taken into account from the outset,  reduce the
 coupling to disorder by a factor $1/(2ps+1)^2$ at small $q$. The
 other restrictions cannot be justified at this point.
I will however pursue the notion of an effective $\lambda$ for
all fractions and at $T>0$.

Specifically, $\lambda$ will be
      extracted from one data point and used to explain the rest
      of the data from that sample. If the other data points differ
      only in the temperature $T$, the same $\lambda$ will be
      used. If it differs in $B$ or $n$ or $\nu$, the following
      scaling argument will be used.\cite{scaling}

In a heterojunction, the donors of density $n$ produce a confining
linear potential of slope that goes as $n$. If one considers a
variational  wavefunction of the  Frank-Howard \cite{FH}  form
$\psi (z) = A(w) z \exp (-z/w)$ in the transverse direction, then
the optimal $w$ (to which $\Lambda$ must be  proportional), varies
as $w \simeq n^{-1/3}$. Consequently $\lambda = \Lambda / l$
varies as \beq \lambda  \simeq n^{-1/3}B^{1/2}\simeq
B^{1/6}\nu^{-1/3}\simeq n^{1/6}\nu^{-1/2}.\label{lamscale}\eeq

\subsection{Comparison to data of  Kukushkin {\em et
al .}} Kukushkin {\em et al}\cite{kuk} vary both $n$ and $B$ and
drive the system through various transitions at $T=0$ (by
extrapolation). The field $B$ is always perpendicular to the
sample. We will compare the hamiltonian  theory to these
experiments by calculating the critical fields at which the $\nu
=1/2$ and $\nu =1/4$ systems  saturate ($P=1$) and the gapped
fractions undergo transitions from one quantized value of $P$ to
the next.

Let us recall that as far as these transitions go, the systems
behave like free femions of  mass $m_p$ which is independent of
the index $r$ which labels how many LL's have reversed their spins
in the gapped case or in the gapless cases, the size of the up and
down Fermi circles. At and near the gapless states $m_p$ may be
fit by the expressions
\begin{eqnarray}
{1 \over m^{(2)}_{p}} &=& { e^2 l \over \varepsilon }C_{p}^{(2)}
(\lambda )\ \ \ \
 \ \ \ \  C_{p}^{(2)} (\lambda )  = {.087 \over \lambda^{7/4}} \label{mp22}\\
{1 \over m^{(4)}_{p}} &=& { e^2 l \over \varepsilon }C_{p}^{(4)}
(\lambda ) \ \ \ \ \ \ \ \
   C_{p}^{(4)} (\lambda )={.120 \over \lambda^{7/4}}\label{mp44}
\end{eqnarray}
where the superscripts on $C$ refer to the number of vortices
attached.

I consider $B^c$'s at which the systems at $1/4,2/5,3/7,4/9, $ and
$1/2$ lose full polarization ($r=0$ for gapped cases, saturation
for the gapless cases) and, for $4/9$, also  the $r=1$ transition,
$|\bf{3,1}\rangle \to |\bf{2,2}\rangle$.

 I fit $\lambda$ to the
$\nu =3/7$ transition $|\bf{3,0}\rangle \to |\bf{2,1}\rangle$ at
$B^c=4.5 T$.

Two points need to be mentioned in connection with the experiment.
First, each of these transitions seems to take place via a narrow
intermediate  step with a polarization half-way between the ones
allowed by CF theory based on spatially homogeneous states. Murthy
has suggested\cite{gmstep} that these correspond to spatially
inhomogeneous states. I use the center of these narrow steps as
the transition points for comparison to the present theory.
Secondly, I use the actual values for $C^{(2)}_{p}$ for these
fractions, rather than the asymptotic value in  Eqns.
(\ref{mp22}).

I then   obtain $\lambda_{3/7}=1.42$ on solving Eqn.({\ref{tran})
which takes the following specific form here :

\beq g{e \over 2m_e}{  B^c\over (e^2/\varepsilon l)} = {2 \Delta_p
\left[ (3,0)\to(2,1)\right]\over (e^2/\varepsilon l)}={2  (
.0117)\over \lambda^{7/4}_{(3/7)}} \eeq

For  transitions  at other   $B_{\perp}$ and $n$,  I need the
corresponding $\lambda $'s. One can argue, as per  Eqn.
(\ref{lamscale}) that $\lambda \simeq \nu^{-1/3} B^{1/6} $, from
which it follows that \beq \lambda_{\nu} = \lambda_{3/7} \left[ {B
\over 4.5 }\right]^{1/6}\left[ {3 \over 7 \nu}\right]^{1/3} = .83
\ {B^{1/6}\over \nu^{1/3}}.\label{lambda} \eeq

Given $\lambda$ one  finds $B^c$ using  Eqn. (\ref{tran}) for
gapped cases.

 For the gapless cases, there are two equivalent
approaches. First, at the critical field  the Fermi energy of the
up spins equals the Zeeman energy of the down spins:
  \begin{eqnarray}
g\left[ {e  B^c \over 2m_e}\right]&=&{k_{F}^{2}\over 2m_p} \\ &=&
{2\pi n\over m_p}={eB \nu \over m_p}\\ &=&{e^2\over \varepsilon
l}{.087\over 2 \lambda^{7/4}} \ \ \  \mbox{ $\nu={1\over
2}$}\label{bchalf}\\ &=& {e^2\over \varepsilon l}{.120 \over 4
\lambda^{7/4}} \ \ \ \mbox{ $\nu={1\over 4}$}\label{bcquarter}
\end{eqnarray}
Using Eqn.(\ref{lambda}) one solves for $B^c$ and obtains the
values given in Table \ref{critical}.

Equivalently we can write for the total ground state energy
density $E^Z(S)$, (where the superscript indicates that the Zeeman
energy is included),
 \beq
 E^Z(S)={\alpha \over 2 } S^2 -g {e\over 2m_e} {B_{\perp} S \over \cos \theta}
\eeq where  $\alpha  ={\pi / m_p}$. This expression  is minimized
(for$P\le 1$) to give
 $P$:
\begin{eqnarray}
 P &=&  {.13 \sqrt{B_{\perp} }\lambda^{7/4}\over \cos
\theta}\label{22} \ \ \ \ \ \ \ \ \nu = {1 \over 2},\  \mbox{ B in
Tesla} \label{27}\\ &=&\label{28} {.19
\sqrt{B_{\perp}}\lambda^{7/4}\over \cos \theta}\ \ \  \ \ \ \ \
\nu = {1 \over 4},\ \mbox{ B in Tesla}
\end{eqnarray}
Setting $P=1$ gives the critical fields.

\vspace*{.2in}

\begin{tabular}{|c|c|c|c|c|} \hline \hline
$ \nu \ \ \  $ & comment &  \  $B^{c}$ (exp) &\  $B^{c}$ (theo) &\
\ $\nu B^c$ (exp)
\\ \hline
  4/9 & $(3,1)\to (2,2)$ & \ \ 2.7 T & \ \ 1.6 T & \ \  1.2\\
 2/5 & $(2,0)\to (1,1)$ &\ \ 3 T & \ \   2.65 T & \ \  1.2\\
1/4 & saturation &\ \   5.2 T &\ \  4.4 T &\ \  1.3\\ 3/7 &
$(3,0)\to (2,1)$ &\ \ 4.5 T&\ \   4.5 T & \ \    1.93\\ 4/9 &
$(4,0)\to (3,1)$ & \ \ 5.9 T & \ \  5.9 T & \ \   2.62\\ 1/2 &
saturation&\ \  9.3 T&\ \ 11.8 T&\ \  4.65 \\ \hline
 \end{tabular}

\vspace*{.2in}
\begin{tabel}
  Critical fields based on a fit at $3/7$.The rows are ordered by
the last column which measures density.\label{critical}
\end{tabel}

\vspace*{.2in}

Note that in rows  above (below)  $3/7$, where I fit $\lambda$,
the predicted $B^c$'s are   lower (higher) than the observed
values, i.e.,  the actual $\lambda$'s are less (more) than what
Eqn. (\ref{lambda}) gives.
  This is consistent with the
expectation that
 interactions
   will increase
 the effective thickness with increased  density.
If I fit  to the $2/5$ point, I obtain similar numbers, with the
agreement worsening as we move  off in density from 2/5. Thus
$3/7$ was chosen as the fitting point since its density was
somewhere in the middle of all the densities considered.

An alternate  approach is to attempt to calculate $\lambda$  {\em
ab initio} using, for example, the Local Density Approximation
(LDA) as done by of PMJ and numerous predecessors\cite{LDA}. This
method will however not include the effect of disorder. It
typically gives a $\lambda$ that is half as big. We cannot
attribute the entire difference to disorder. For one thing, the HF
approximation tends to inflate $\lambda$. Next LDA , as the name
suggests, is an approximation. Lastly, LL mixing can account for
some of the difference, though not much at these values of
$\lambda$\cite{LLmix}. It is possible, that due to all these,
 the effects of disorder are not that  significant, for magnetic phenomena. The
present strategy of determining $\lambda$ phenomenologically
stands or falls depending on how well the fit to any one data
point allows us to make predictions for other measurements made on
the same sample. Table (\ref{critical}) suggests it is quite a
useful point of view.

\section{Physics at nonzero temperatures $T>0$.}

So far we have seen the hamiltonian theory may be used to compute
quantities such as gaps, particle-hole profiles, critical fields
for magnetic transitions and so on. All such quantities have been
readily computed using trial wavefunctions, giving numbers that
are superior to ours. The main idea of this paper so far has been
to get a 10-20$\%$ theory in which we see the underlying physics
as transparently as possible and to resolve questions such as why
CF behave like particles on some occasions.

We turn to physics at finite $T$ where the method has few rivals.
Exact diagonalization is limited to very small systems and trial
wavefunctions typically cover the ground state and very low
excitations. The hamiltonian approach is able to yield, in the HF
approximation, the polarization $P$ and the relaxation rate
$1/T_1$ for the gapless states as a function of temperature. If
$\lambda$ is treated as before (fit to one data point per sample)
we will see it is possible to give a very satisfactory account of
experiments up to about  $1{}^oK$, which is of the order of the
Fermi energy.
\begin{figure}
\epsfxsize=3.0in \centerline{\epsffile{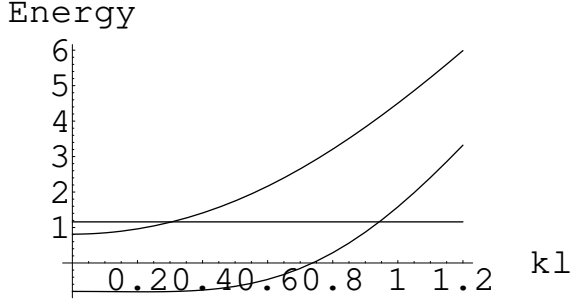}} \vskip 0.15in
\caption{Hartree Fock energies at $\nu =1/2$  for up and down
spins (upper and lower curves) at $T=.3{}^oK$ at $B=5.52 \ $ and
zero tilt. Note that they are not simply quadratic in momenta, and
that at the chemical potential, indicated by the horizontal line,
the two graphs have very different slopes, i.e., density of
states. } \label{HFE}
\end{figure}

The HF energy of a particle {\em including the Zeeman energy} is
the self-consistent solution to
\begin{eqnarray*}
\lefteqn{ {\cal E}_{\pm}^{Z}(k)=}\\ &&\mp {1 \over 2}g \left[ { e
B\over 2m}\right] + 2\int{d^2q \over 4\pi^2}\check{v}(q)\sin^2
\left[{{\bf k \times q}l^2\over 2}\right] \\ && -4\int{d^2k' \over
4\pi^2}n^{F}_{\pm}(|k'|)\check{v}(|{\bf k -k'}|)\sin^2 \left[{{\bf
k' \times k}l^2\over 2}\right]
\end{eqnarray*}
where the superscript on ${\cal E}_{\pm}^{Z}$ reminds us it is the
total energy including the Zeeman part, the Fermi functions \beq
n^{F}_{\pm}(|k|)={1 \over \exp \left[ ({\cal E}_{\pm}^{Z}(k)-\mu
)/kT\right] + 1 }\eeq depend on the energies ${\cal
E}_{\pm}^{Z}(k)$ and the chemical potential $\mu$. At each $T$,
one must choose a $\mu$, solve for ${\cal E}_{\pm}^{Z}(k)$ till a
self-consistent answer with the right total particle density $n$
is obtained. From this one may obtain the polarization by taking
the difference of up and down densities. As usual we use the ZDS
potential for which \beq \check{v}(q) = e^{-q^2l^2/2}{2\pi e^2
e^{-ql\lambda} \over q} .\eeq

The computation of $1/T_1$ is more involved. The question we ask
is the following. The fermions  are in a quantum well, with their
density varying across the width. So  the nuclear relaxation rate
will be a function of position.  Consider a nucleus at the center
of the quantum well, (as well as the $x-y$plane) where the density
is the largest. Let us call this point the origin and  let $1/T_1$
be the relaxation rate here. The theory predicts
\begin{eqnarray}
1 \over T_{1} &=& 4\pi k_BT\left( {K^{max}_{\nu}\over
n}\right)^2\nonumber \\ &&\! \! \! \! \! \! \! \times
 \int_{E_0}^{\infty} dE \left(
{dn^F(E)\over dE}\right)
\rho_{+}(E)\rho_{-}(E)F(k_+,k_-)\label{oneovert}\\
F&=&e^{-(k_{+}^{2}+k_{-}^{2})l^2/2}I_0(k_+k_-l^2)\\
 \rho_{\pm}(E) &=&\int{kdk\over
2\pi}\delta (E-{\cal E}_{\pm}^{Z}(k))\label{oneovert1}
\end{eqnarray}
where $E_0$  is the lowest possible energy for up spin fermions,
and   $K^{max}_{\nu}$ is the measured maximum Knight shift for the
fraction $\nu =1/2\  \mbox{or}
 \  1/4$.

Here is a rough description of the derivation, the details of
which may be found in the Appendix 6. Suppose for a moment we were
dealing with electrons and not CF's. The Knight shift at the
chosen point, the origin, will be determined by the spin density
there. The same parameter enters the $1/T_1$ calculation
quadratically. This is why $K^{max}_{\nu}$ enters the answer. The
idea is that $K^{max}_{\nu}$ is not calculated {\em ab initio} but
taken from the same experiment.  The density of states and Fermi
factor are standard. The only new feature here is the presence of
$F(k_+,k_-)$ which  reflects the fact that the spin density has to
be projected into the LLL when going to the CF basis. The effect
of this factor (which is none other than the $e^{-q^2l^2/2}$ which
appeared on the projected charge density) is to suppress processes
with momenta much larger than $1/l$, as these have no place within
the LLL.

\subsection{Comparison to experiment}
\begin{figure}
\epsfxsize=3in \centerline{\epsffile{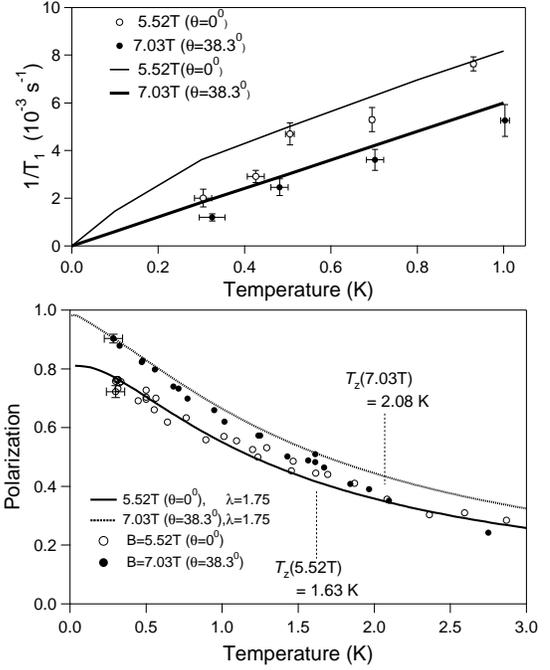}} \vskip 0.15in
\caption{Comparison to the work of Dementyev {\em et al}.
  The value of $\lambda$ is fit to $P$ at $300 \ mK,\  B_{\perp}=5.52\ T$
   and the  rest follows from the theory.
Notice the correlation between the curvature of
 $1/T_1$ and the limit of $P$ as
$T\to 0^{0}K$. } \label{fig1}
\end{figure}
We now compare to some experiments at $\nu=1/2$ and $T>0$.
 Consider first Dementyev
{\em et al} \cite{dem}.  From their data point  $P=.75$ for
$B=B_{\perp}= 5.52 T$ at $300 \ mK$ I deduce \beq \lambda =1.75.
\eeq   I have once again  chosen instead to match my HF results
with the above data point, (which gave $\lambda =1.75$) and see to
what extent a {\em sole} parameter $\lambda$, can
 describe  $P$ and $1/T_1$ for the
given sample at a given $B_{\perp}$, but various temperatures and
tilts.

 Since there does not exist a model, including disorder,  that describes
   how $\lambda$
should vary with tilt  I include no such variation.

 Dementyev {\em et al} find
$K_{1/3}^{max}= 4.856 \cdot 10^{-7}{}^{o}K$, which is believed to
describe a saturated system  at $P=1$. They estimate that
$K_{1/2}^{max}=.953 K_{1/3}^{max}$, which is what we need here. It
is assumed that the nuclear wavefunction $u_E(0)$ (defined in Eqn.
(\ref{200}) Appendix 6) does not vary between $\nu =1/3$ and $\nu
=1/2$. Given this information, $1/T_1$ follows.

 The top and  bottom halves of  Figure \ref{fig1}  compare the   HF calculation of
 $1/T_1$ and  $P$ respectively,  to the
  data. (The graphs for $1/T_1$ differ slightly from those in
   Ref.\cite{prl2} since the present calculation
    treats the spin of the CF more carefully. The $1/T_1$ graph
    at $5.52 \ T$
      appears a little jagged since it was computed at just six points
      which were then connected.This is not apparent in the tilted case since the points lie on a straight line.)

  Dementyev {\em et al} had pointed out that a two parameter fit
  (using a mass $m$ and interaction  $J$), led to  disjoint pairs of values
for these  curves. Given that $H$ is neither free nor of the
standard form ($p^2/2m +V (x) $) this is to be expected. By
contrast, a single $\lambda$  is able to describe the data here
rather well  since  $H$ has the right functional form. Given how
the theory fits the polarization data up to  the Fermi energy of
$\simeq 1^o K$, it is clear that changing the data point used to
fix $\lambda$ will be inconsequential.

  If $P$ were
computed from the LDA value  $\lambda \simeq 1$, it would be down
by 15-50 $\%$ as $T$ drops from $1^o$K to $0^o$K. The present work
establishes a phenomenological, nontrivial  and  nonobvious fact
that a single $\lambda$ parameter, (like $g$ or $\varepsilon$)
 determined from one data point, can describe both $P$ and $1/T_1$
 for the given sample under a variety of conditions.
  That the fitted $\lambda$ is larger than the LDA value makes sense, as
  both disorder and LL mixing will
 lower the gap and  raise $\lambda$. As mentioned earlier, it is
 not clear exactly how strong the disorder corrections are since
 there is LL mixing and some intrinsic errors in the LDA.

 Consider next sample M280 of  Melinte {\em et al}
\cite{melinte} which had $P=.76 $ at $.06^{o}K$ and $B=B_{\perp} =
7.1 T $, from which I deduced $\lambda =1.6$. Figure (\ref{fig2})
compares my $T$ -dependence with data. The initial rise of $P$
with temperature
  was also  seen by  Chakroborthy and Pietlianean\cite{cp}.

\begin{figure}
\epsfxsize=2.4in \centerline{\epsffile{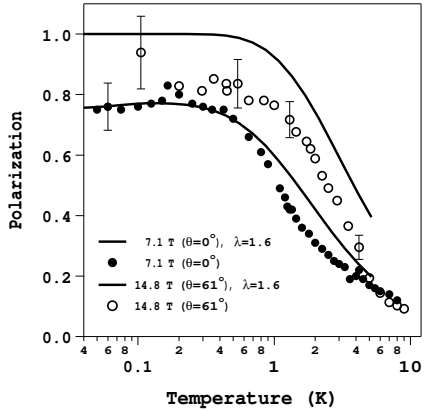}} \vskip
0.15in \caption{Comparison to  Melinte {\em et al}. with $\lambda$
fit to $P$ at $60 \ mK,\  B_{\perp}=7.1\ T$. Some typical error
bars are  shown.} \label{fig2}
\end{figure}
Note that agreement is quite poor for the tilted case. There are
confusing aspects of both the theory an experiment. In the
experiment, one may ask why the polarization does not increase
with increased tilt and hence increased Zeeman coupling. Of course
this will happen if $P=1$ to begin with, but it is not, it is
clearly below unity, say $80\%$. On the other hand, the theory for
tilted fields in not in good shape either. First of all, orbital
effects have to be considered due to the tilt. The thickness
parameter $\Lambda$ can be affected by it. As pointed out by
Jungwirth in a private communication, once there is an in-plane
component of $B$, the problem is no longer rotationally invariant.
This means that our states are no longer HF states and get
scattered into each other by the potential. Presumably these
effects were  negligible in the case of Dementyev {\em et al} but
not at nearly twice the tilt of  $61{}^o$. No attempt is made here
to take into account all the effects of the tilt. Instead I
include just the increased Zeeman coupling and hope for the best.

For the benefit of others who measure $1/T_1$ at $\nu =1/2$ in the
future on similar samples, I give some  very approximate formulae
(to be used for zero or small tilts). From Figure (\ref{fig1}), we
note that in general, the graphs of $1/T_1$ become linear and
parallel for temperatures above $.3\ {}^oK$. In this region we can
write \beq {d(1/T_1)\over dT}\simeq 3 \left[ {\bar{K}\over
\bar{n}}\right]^2 \cdot 10^{-3} s^{-1}\left[ {}^o K\right]^{-1} \
\ \ \  \mbox{for}\  T>.3 {}^o K\eeq with ${\bar K}$ the Knight
shift in KHz and \beq \bar{n} = {n \over 10^{10}/cm^2}\eeq

(In this approximate formula, I ignore the $\lambda$ dependence of
 Eqns.
  (\ref{oneovert}-\ref{oneovert1}), and the distinction between the average and
maximum Knight shift.)

The graphs do not generally obey the Korringa-like law because as
$T\to 0$  they are sublinear (superlinear)for saturated
(unsaturated) cases. Only the critical case with $P(0)\to 1$ as
$T\to 0$ is linear. For $T>.3 \ {}^o \ K$, (which in general must
be replaced by either the energy gap or energy overlap between the
up and down Fermi energies)
\begin{eqnarray} {1\over T_1} &=& \left[ 3 \  T({}^oK) + C\right]
\left[ {\bar{K}\over \bar{n}}\right]^2 \cdot 10^{-3} s^{-1} \\
C&=& 0 \ \ \  \ \ \mbox{(critical)}\\
 &=& >0\ \  \ \  \mbox{(unsaturated)}\\
  &=& <0 \ \ \ \ \mbox({saturated)}
  \end{eqnarray}

 For the critical case (only),  we have Korringa law
  \beq
{1\over T_1 \ T} = \left[ 3 \  T({}^oK) \right] \left[
{\bar{K}\over \bar{n}}\right]^2 \cdot 10^{-3} s^{-1} ({}^o\
K)^{-1} \eeq

  For Dementyev {\em et al}   $\ C\simeq 1$. This value may be used as
  a first approximation, for example, if Melinte {\em et al} measure $1/T_1$ on sample M280.
  For more accurate results they must solve Eqns.
  (\ref{oneovert}-\ref{oneovert1}) with $\lambda =1.6$.

The present formalism has been applied by G. Murthy to calculate
the $T$-dependence of polarization in the  $1/3$ and $2/5$
states.\cite{gmpol}

\section {Conclusion}
This paper  described in detail a formalism  for describing  FQHE
states, in which  the hamiltonian and various operators were
expressed in terms of the quasiparticles, the Composite Fermions.
It was shown
 that the formalism  could be used to calculate
 a variety of
quantities at zero and nonzero temperatures to some reasonable
accuracy, and to resolve   matters of principle (some  of which
could not even be posed otherwise),  such as the internal
structure of the CF, the reason it sometimes appears to be free
when it is not, and how theory is to be compared to experiment in
the quantum Hall problem.

First a review of the wavefunction approach to CF was given. It
was seen that  whereas in the Laughlin fractions, CF was evidently
an electron bound to 2s vortices, the situation was more
complicated for the Jain series  due to the action of the LLL
projector ${\cal P}$. This caused the  vortices to move off the
electrons and also got rid of many of them. It was not clear how
the vortices paired off with the electrons, say at $\nu =2/5$,
when there were 1.5 (non-Pauli) vortices per electron in the
projected state. Since the charge of the CF was the same
($e^*=1/(2ps+1)$) before and after projection, whatever paired
with the electron still had the charge of $2s$ vortices, though it
could not be associated with zeros of the wavefunction in any
simple way. What is this object and how is one to incorporate it
in the theory?

The hamiltonian theory was seen to provides the answer. In this
theory
  we   enlarged the
Hilbert space to include additional degrees of freedom,
accompanied by  constraints.  These new degrees of freedom
 (prevented from having any density
fluctuations by  constraints)   turned  out to have  charge
$e_v=-2ps/(2ps+1)$ and paired with electrons.  They were referred
to as  vortices, for want of a better name, but in view of what
was said above, they are not related in any simple way to zeros of
$\Psi_{LLL}$,  the electronic wavefunction in the LLL.

To solve the theory with constraints, the HF approximation was
used. Free particle and hole states of CF's in a reduced field
$B^*=B/(2ps+1)$  were seen to be HF states. All matrix elements
were evaluated in closed form
 and usually a single numerical integral gave the
numbers. As for the constraints,  the method proposed with Murthy
was used: the charge density operator was written as a judicious
combination of the transformed and LLL projected electronic charge
density and the constraint, which is just the vortex charge
density, the judicious  combination being the unique one (as $q\to
0$)  that  obeyed Kohn's theorem. When this combination was
expanded in $q$, the monopole term was $e^*$ and the next term
corresponded to a dipole moment $l^2 \hat{z}\times {\bf k}$. The
operator approach thus gave a precise meaning to the CF dipole
moment,  in terms of how it coupled  to an external potential.
This coupling did not change when second quantized with Fermi
fields. Since this procedure took into account the most important
effects of the constraints (away from $T=0$ and ultralow
frequencies $w\simeq q^3$), constraints were neglected thereafter.

In Section IV activation gaps for fully polarized states were
calculated for the Zhang Das Sarma (ZDS) potential as a function
of the parameter $\lambda$. Analytic expression were derived for
all fractions of the form $p/(2ps+1)$ and compared to the
Monte-Carlo work of PMJ based on trial wavefunctions for $s=1,
p=1,2,3,4$. It was found that the numbers were within 10-20$\%$
for  $\lambda
>1$.

A comparison was made to experiments of Du {\em et al}\cite{du}.
Rather than try to compute $\lambda$ ab initio, it was fitted to
the three gaps at $1/3, \ 2/5, \ /3/7$ and was seen to take the
values $2.4,2.4,\ \mbox{and}\  2.0$.

It was found that normalized activation masses (scaled by
$1/\sqrt{B}$) were not too different for $s=1,s=2$, i.e., near
$\nu =1/2$ and $\nu =1/4$, in reasonable accord with the
experiments of Pan {\em et al}\cite{Pan}.It was pointed out that
no deep reasons existed for their exact equality.

For the gaussian potential the numbers were in excellent agreement
with PMJ except for $\nu =1/3$. In was clear that although the
extended theory was defined for all length scales, it gave good
numbers only for soft potentials, ones that were smooth within a
magnetic length.

A comparison to the exact diagonalization work of Morf {\em et al}
was made for fractions $2/5$ and $3/7$. The HF results of this
theory lay consistently above their numbers. The differences were
somewhat larger than in the case of PMJ.

In the hamiltonian formalism we can make precise the question of
whether or not CF are interacting since $H$ naturally separates
into a free  and interacting parts $H_0$ and $H_I$. It was seen
that turning off $H_I$ made a sizeable difference to activation
gaps.

The theory was used to compute the profiles of charge densities in
states with a single particle or hole and compared to the
unpublished work of Park and Jain. It was seen that while the
salient features were reproduced with ups and downs at the right
places, the amplitudes were not as pronounced. The best fraction
was $2/5$, and not surprisingly, this was also the one where the
gaps came out best. The case $\nu =1/2$ and the dipole moment of
the CF were analyzed in some depth.

It was found that the activation gaps for the fully polarized
states obeyed particle-hole symmetry to an excellent
approximation. This was a nontrivial result since the
corresponding CF  states were very different as were the
expressions for gaps.

Polarization phenomena at $T=0$ were the subject of Section V. By
varying the density and field at fixed filling, or by placing the
sample in a tilted field it is possible to increase the Zeeman
coupling and  drive the system though many magnetic transitions.
For the fraction $p/(2ps+1)$, CF theory has states with  $p-r$ LL
of spin up and $r$ levels with spin down, $r=0$ being the fully
polarized case. Critical fields $B^c$ at which the system would
jump from one value of $r$ to the next were computed.  It was
found,  as Park and Jain did, that it  was possible to fit all the
numbers very well by assuming that CF were free and occupied LL
with a polarization gap $\Delta_p$ or  a corresponding
polarization mass $m_p$. Yet we know CF cannot be free, given that
the activation mass and gap are substantially different from these
values. It was shown  that rotational invariance and $d=2$
conspired to create this impression of free fermions. In
particular, for the gapless case,  while the individual HF
energies of spin up and down fermions had sizeable $k^4$  terms in
the formula for ${\cal E}_{\pm} (k)$, the energy cost of {\em
transferring } a fermion from the top of one sea to the top of the
other equaled that of free fermions with a quadratic dispersion
$k^2/2m_p$.

The theory was compared to  the experiments of Kukushkin {\em et
al}\cite{kuk} who varied both $n$ and $B_{\perp}$ to drive the
magnetic transitions. Some plausibility arguments were given for
why an effective, translationally invariant potential might be
able to describe magnetic phenomena under restricted conditions.
It however used to describe all magnetic phenomena:  an effective
$\lambda$ was deduced from one data point (here one transition),
and used to predict all the others, using scaling laws. The
numbers agreed to within $10-20 \%$.  While the arguments for an
effective potential description  were trustworthy only in a
limited region, the procedure seemed to work in a wider region,
including gapless states and $T>0$. This deserves to be
understood.

The most important results of this theory,  not limited by finite
size, involve  the computation of relaxation rates and
polarizations as a function of temperature. These were discussed
in Section VI. Finite $T$ HF equations were derive analytically
and
 solved numerically to yield these quantities. They were compared
to the work of Dementyev {\em et al} and Melinte {\em et al}. In
the first case,   the measured polarization at 300 mK for a
perpendicular field of $5.52 T$ was used to fix  $\lambda =1.75$.
Using this value  the polarization and relaxation rate $1/T_1$
were computed for a range of temperatures going up to $2^{o}K$,
and $1^{o}K$ respectively, where the latter
 is roughly  the Fermi temperature. The agreement was
very satisfactory. For a tilt of $38.3^{o}$, the  agreement was
again good till about $1^{o}K$.  In their paper Dementyev {\em at
al} had pointed out that fitting these four data sets  to a model
with a mass term $m$ and Stoner enhancement $J$ lead to four
disjoint islands in the parameter space. In the present case a
single $\lambda =1.75$ seemed to describe all four sets quite
well. It was claimed  this was because the $H$ used was of the
right functional form, with an unusual kinetic and interaction
terms chock full of momenta and currents. For any one graph it may
be mimicked by a hamiltonian of the $(m,J)$ form, but these
numbers would then vary from set to set. In particular, to get the
right relaxation rates, with just one parameter, $\lambda$,  one
needs the unusual dispersion relations (far from quadratic in
momenta) and the corresponding unusual density of states that
arise naturally here.

As for the polarization data of Melinte {\em et al}\cite{melinte},
there was reasonable agreement for the untilted sample, but not
the tilted (by over $60{}^o$) case. The latter could be attributed
to theory, which is not designed to handle such large tilts,  or
to the data which exhibit some unusual features discussed in the
text, or both.

An approximate Korringa-like law for states that are just fully
polarized at $T=0$ and the slope of $1/T_1$ versus $T$ for the
general case for $T> .3\  {}^o K$ were provided to allow future
experimenters (working on  samples similar to that used by
Dementyev {\em et al}) to make a quick and approximate comparison
to this theory without having to solve the finite-T HF equations.

 In
summary it was shown that the hamiltonian theory of CF provides a
comprehensive and analytical scheme for describing the low energy
physics of the FQHE states, clarifying concepts regarding the
internal structure of CF's and how they are to be coupled to
external potentials, and computing a variety of quantities at zero
and nonzero temperatures to an accuracy of $10-20\%$ under typical
conditions, and sometimes considerably better. The deviations
could be due to the HF approximation or to the hamiltonian itself.
In comparing to magnetic experiments, if one extracted a single
parameter $\lambda$ that characterized electron-electron
interaction in a given sample from one data point, the theory gave
a reasonable account of other data from that sample.

\subsection{General philosophy}
I conclude with some remarks  that put the present approach in
perspective and  underscore the differences between the FQHE
problem and others. Let us begin with the fact that in the FQHE
restricted to the LLL, {\em and in the absence of disorder}, every
physical quantity is a functional of the electrostatic interaction
between electrons, since the kinetic energy is quenched. I will
address the important question of disorder shortly.  If this
interaction is pure coulomb, it is a zero-parameter theory. If it
is modeled by the ZDS potential as done here, all observables --
$m_a, m_p, B^c, P(T), 1/T_1$ -- are functions of $\lambda$. Once a
single data point  is known, $\lambda$ may be determined, and from
it, all others calculated (using scaling laws if needed). While I
chose the ZDS potential to make this point, one could use another,
say the one used by Morf {\em et al}\cite{Morf}. The the numbers
in question  are predominantly  controlled by one of Haldane's
$V$'s, and these are different ways of varying it.

Why do we not do this all the time? Consider for example a Fermi
liquid.  Why do we bother with Landau's $F$ functions? Why do we
not start with the coulomb interaction and calculate everything
with zero parameters?  Or if we wanted to model it with a short
range force, why don't we employ the Hubbard model, calculate any
one observable, fit it to data, extract the Hubbard $U$ and
predict everything else?  The reason is well known:  we do not
know how to go  from the model to the physical quantities except
in perturbation theory. Let us restate this in Landau's language.
Suppose we turn off $U$. The states, at least the low lying ones,
are labeled by free field theory. We now turn on the interaction
and the states and energies evolve to those of the interacting
theory. We finally should end up with the quasiparticle basis, at
least near the Fermi surface. Unfortunately this change of basis
from free to interacting theory cannot be carried out  in practice
except perturbatively.

The same thing happens in S-matrix theory, which appears to have
guided  Landau's intuition. The $in$ and $out$ states carry free
particle labels and their dot product gives the elements of the
$S$-matrix. They are obtained by taking particles that are
infinitely separated (and noninteracting) in the distant past or
future and evolving them (forwards or backwards) in time to $t=0$.
Unfortunately this evolution too can be done only perturbatively.

 In the FQHE problem
the situation is a lot better, though it looks bad to start with.
The usual idea of starting with free electrons and turning on
interactions is doomed from the start since for $\nu<1$, the free
electrons do not have a unique ground state. But lurking in the
background are again some  free states or wavefunctions. These are
the CF states with $p$ filled LL and  some low energy excitations.
They get mapped to the electronic wavefunctions in Jain's approach
by a transformation that consists of attaching $2s$-fold vortices
with the Jastrow factor $J(2s)$ and projecting with ${\cal P}$.
The resulting wavefunctions are known to be excellent, with nearly
unit overlap with the results of exact diagonalization.  If we
know the electron-electron potential (in the absence of disorder)
we can calculate anything related to the ground state and low
energy excitations using these. In the hamiltonian approach we go
the other way: from electrons to CF. We first do $2s$-fold flux
attachment by the CS transformations and this becomes vortex
attachment following the additional transformations  Murthy and I
developed. The states not only carry free labels, they are simply
free (in the HF calculations);  it is the operators that become
complicated. They are however simple enough to work with
analytically, and  lead to less accurate numbers than Jains's
approach.

All this is possible because of the   one major difference
compared to the Fermi liquid and the $S$-matrix problems: the
transformation here is discrete and given by the attachment of
$2s$ vortices (before projection, which leads to a complicated
wavefunction, but is fully implementable). The CF know only about
$p$, the transformation about $2s$, and the electrons know about
$p$ and $2s$. It is this discreteness, (absent in the Fermi liquid
or $S$-matrix problem), that eventually works in our favor, that
sufficiently constrain the problem and   lead us to the right
answer. For example, in Jain's approach analyticity and Fermi
statistics fix the vortex number to be $2s$ and in the hamiltonian
approach the same is true as well, and additionally, the LLL
algebra pointed to the exponentiation which was the
nonperturbative step that had to exist in any transformation
linking electron to CF.

There is a mistaken belief that CF have to be free to be useful.
Note that  Jain is able to calculate a slew of quantities with his
wavefunctions and to a lesser accuracy, so can the hamiltonian
method. Neither method requires CF to be free. {\em  What is free
about the CF is the label for the states}. While this is true in
S-matrix theory and in Landau theory, what is special here, as
explained above, is that the transformation from the free states
to the interacting counterparts  is  known to a great accuracy.

I believe  this happy situation, of being able to go from the free
to the interacting states  in the FQHE, deserves further
exploitation in comparing to experiment. If there were no
disorder, what we should try to extract from the data is the
electron-electron interaction, $v(q)$, which is a c-number
function that multiplies the operator quadratic in
$\bar{\bar{\rho}}^p$ and defines $H$. By extracting $v(q)$ I  mean
some parameter like $\lambda$ of the ZDS potential or $b$ in the
Morf {\em et al} case. As mentioned earlier, the physics is
dominated by one Haldane pseudopotential, ($V_1$ for $2s=2$) and
these parameters just control it. It does not seem fruitful to fit
the complicated CF hamiltonian to a string of standard or
nonstandard {\em operators}. The hamiltonian for the CF's is not
of the canonical form with a quadratic kinetic energy and some
interaction terms of the usual (density-density) form. There is no
reason such should be the case, given the internal structure of
CF's. Rather,  the free and interacting parts of the CF
hamiltonian form a
 monolith, of unusual functional form,
 fully determined by the density operator
written in the CF basis.  While it is possible to fit any one
experiment  with one set of standard interaction parameters, (like
$m$ and $J$ that Dementyev {\em et al} used) the fact that the
functional form is wrong will manifest itself, as they found out,
in the need for many  disjoint sets of parameter for different
measurements. (This is analogous to the fact that if we try to fit
the kinetic energy of a relativistic particle as a function of its
velocity to a nonrelativistic form, we will need a velocity
dependent mass, while if we fit it to the correct form we will
extract  a fixed rest mass.)  Likewise, if one  extracts $\lambda$
(or its counterpart for another potential)  from the data, a
single simple result is more likely to emerge. This $\lambda$ will
of course have the usual variance of $10-20 \%$ characteristic of
this theory.

{\em Now it is time to wake up and smell the disorder. } Does it
completely destroy the approach presented here? Is it possible
that even the disordered system can be modeled by a pure potential
with some effective $\lambda$ or $b$? The answer is clearly
negative if one wants to describe everything. For example no
translationally invariant  interaction will predict localization.
As for activation gaps, the data suggest that disorder effects are
quite strong: the disorder broadening (measured by the negative
intercept of the gaps versus effective field $B^*$) is comparable
to the transport gaps. As $B^* \to 0$, the pure system gaps never
vanish while in experiments they do, implying no effective
potential exists. If at some future date, we get samples with even
lower disorder, we may be able describe these gaps with an
effective potential, even for small $B_{eff}$.

For the present, for  magnetic transitions at $T=0$, which depend
only on total energies, I have tried to ask if an effective
potential exists which can subsume the effects of disorder. I gave
crude arguments   that showed that in the limit of weak and smooth
disorder one could get the disorder averaged ground state energy
to second order by replacing $v(q)$ with a $v_{eff}(q)$. While
$v_{eff}$ did not have the same functional form as $v$, it was
always weaker, which meant the effective $\lambda$ (which produced
the corresponding dominant pseudopotential $V$) is  larger. {\em I
have explored the possibility, of determining an effective
$\lambda$ from one data point and using it (combined with a
scaling law if necessary) to predict the other data on that
sample, for all magnetic phenomena.} While the plausibility
arguments were given for $T=0$ and that too under a variety of
restrictions, I have tried it for all magnetic phenomena. What I
find  that if one is prepared to work with a $10-20\%$ theory,
such a program actually works.
 If such an effective  potential is
found from static or equal time data, it can be imported to the
hamiltonian scheme to do other things such as $\omega$ dependent
response functions and $T>0$ physics.

\section{Acknowledgements}
In writing this long paper, I have profited enormously from the
generous help of Ganpathy Murthy. I welcome this opportunity to
acknowledge this debt. I thank Bert Halperin for insisting on
clarity in spelling out  what assumptions were being made  in this
presentation, Jim Sauls for a critical discussion of  my treatment
of CF spin, Tomas Jungwirth for a discussion of the tilted case.
The paper went through some  revisions following a talk at the
Aspen Center for Physics. I am pleased to acknowledge the center,
as well as many participants for their input, especially, J.
Einsentein,  J.K. Jain, L. Levy,  R. Morf,  S. Das Sarma, B.
Schklovskii, S. Girvin and  D. Thouless. I thank  the National
Science Foundation for Grant DMR98-00626.

\appendix
\vspace*{.2in}
\begin{center}
APPENDIX
\end{center}
\vspace*{.2in}
 \subsection{Matrix elements}
\label{ME}
 Many of the calculations performed in this paper deal with the
 preferred density $\bar{\bar{\rho}}^p$. In second quantization we
 write it as
 \beq
 \bar{\bar{\rho}}^p({\bf q}) =
 \sum_{m_2n_2;m_1n_1}d^{\dag}_{m_2n_2}d_{m_1n_1}\rho_{m_2n_2;m_1n_1}
 \eeq
 where $d^{\dag}_{m_2n_2}$ creates a particle in the state
 $|m_2\ n_2\rangle$ where $m$ is the angular momentum and $n$
 is the LL index. They are related to the CF cyclotron and guiding
 center coordinates, ${\bf R}$ and $\mbox{\boldmath $\eta $}$  as follows. Let
 \beq
b = {R_x -iR_y\over \sqrt{2l^{*2}}} \ \ \ b^{\dag} = {R_x
+iR_y\over \sqrt{2l^{*2}}} \eeq where $l^* = l /\sqrt{1-c^2}$ is
the CF magnetic length.  These obey the oscillator algebra
 \beq
 \left[ b \ ,b^{\dag} \right] =1
  \eeq
  given
  \beq
  \left[ R_x , R_y \right] =-il^{*2}.
   \eeq
   Similarly we define, in terms of the cyclotron coordinates,
 \beq
a = {\eta_x +i\eta_y\over \sqrt{2l^{*2}}} \ \ \ a^{\dag} = {\eta
_x -i\eta_y\over \sqrt{2l^{*2}}} \eeq which obey the oscillator
algebra
 \beq
 \left[ a \ ,a^{\dag} \right] =1
  \eeq
  given
  \beq
  \left[ \eta_x , \eta_y \right] =il^{*2}.
   \eeq

   The states $|m n\rangle$ are just the tensor products
   \beq
|m n\rangle = {(b^{\dag})^m \over \sqrt{m!}}{(a^{\dag})^n \over
\sqrt{n!}}|00\rangle \eeq where $|00\rangle$ is annihilated by
both $a$ and $b$.

I will now evaluate matrix elements of  $e^{-i{\bf q \cdot R}}$
and $e^{-i{\bf q }\cdot \mbox{\boldmath $\eta $}}$ and show that

\begin{eqnarray} \langle m_2 |e^{-i{\bf q \cdot R}} |m_1\rangle
&=& \sqrt{m_2!\over m_1!} e^{-x/2} \left( {-iq_+l^*\over
\sqrt{2}}\right)^{m_1-m_2}\nonumber \\ & &\times \
L_{m_2}^{m_1-m_2}(x) \label{Rmat} \end{eqnarray}
 where
 \beq
 x= q^2l^{*2}/2,\ \ \ \ \ \    q_{\pm} = q_x \pm  iq_y
 \eeq
  $L$ is the associated Laguerre  polynomial, and  $m_1\ge
 m_2$.
 If $m_1<m_2$ one may invoke  the relation
 \beq
\langle m_2 |e^{-i{\bf q \cdot R}} |m_1\rangle = \langle m_1
|e^{+i{\bf q \cdot R}} |m_2\rangle^* .\eeq

To establish Eqn. (\ref{Rmat}), consider the coherent states
 \beq
 |z\rangle = e^{b^{\dag}z}|0\rangle = \sum_{m=0}^{\infty}
 {|m\rangle \over \sqrt{m!}}z^m
 \eeq
 with the inner product
 \beq
 \langle \bar{z}|z \rangle = e^{\bar{z}z}
 \eeq
 First we write from the definitions given above
 \begin{eqnarray}
 \langle \bar{z}|e^{-i{\bf q \cdot R}}|z\rangle \nonumber\\
  &=& \sum_{m_1=0}^{\infty}\sum_{m_2=0}^{\infty}{\bar{z}^{m_2} \over \sqrt{m_2!}}{z^{m_1} \over
  \sqrt{m_1!}}\langle m_2 |e^{-i{\bf q \cdot R}} |m_1\rangle \label{R1} \\
  &\equiv & R(\bar{z},z,{\bf q}).
  \end{eqnarray}
On the other hand
\begin{eqnarray}
\langle \bar{z}|e^{-i{\bf q \cdot R}}|z\rangle &= & \langle
\bar{z}|\exp ( -{il^*\over \sqrt{2}}(q_+b^{\dag}+q_-b))|z\rangle
\\
 &=& \langle \bar{z} - {il^*\over \sqrt{2}}q_+|{z} - {il^*\over
\sqrt{2}}q_-\rangle e^{q^2l^{*2}/4}\\ &=& \exp \left[ \bar{z} z -
{il^*\over
\sqrt{2}}(\bar{z}q_-+q_+z)\right]e^{-q^2l^{*2}/4}\label{R2}
\\
 &\equiv &
R(\bar{z},z,{\bf q})
\end{eqnarray}
Comparing Eqns. (\ref{R1}-\ref{R2}) and matching powers of
$\bar{z}^{a}z^b$ we obtain Eqn. (\ref{Rmat}) if we recall\beq
L_{m_2}^{m_1-m_2} (x) = \sum_{t=0}^{m_2} {m_1! \over (m_2- t)!
(m_1-m_2+t)!}{(-1)^t\over t!} x^t \eeq

To establish
 \beq \langle n_2 |e^{-i{\bf q}
\cdot \mbox{\boldmath $\eta $}} |n_1\rangle = \sqrt{n_2!\over
n_1!} e^{-x/2} \left( {-iq_-l^*\over
\sqrt{2}}\right)^{n_1-n_2}L_{n_2}^{n_1-n_2}(x)\label{etamat}\eeq
(again for $n_1\ge n_2$) we just  need to remember that the
commutation rules of the components of  $\mbox{\boldmath $\eta $}$
have a minus sign relative to those of ${\bf R}$, which  exchanges
the roles of creation and destruction operators and hence $q_+ $
and $q_-$.

Now we consider matrix elements of {$\  $ $\bar{\bar{\rho}},
\bar{\bar{\chi}},\ \bar{\bar{\rho}}^p \ $}. As a first step, let
us express the operators ${\bf R}_e$ and ${\bf R}_v$ in terms of
CF guiding center and vortex coordinates ${\bf R}$ and
$\mbox{\boldmath $\eta $}$.
 We have seen that  in the CF  representation
  \begin{eqnarray}
   {\bf
R}_e &=& {\bf r}-l^2{\hat{z}\times {\bf \Pi^*} \over 1+c}={\bf
R}+\mbox{\boldmath $\eta $} c
\end{eqnarray} if we recall $l^2 = l^{*2}(1-c^2)$.

It can similarly be shown that
  \begin{eqnarray}
   \mbox{\boldmath $\eta $}_e &=& {\bf r}+l^2{\hat{z}\times {\bf \Pi^*} \over
c(1+c)}={\bf R}+\mbox{\boldmath $\eta $} /c.
\end{eqnarray}
Thus in first quantization\begin{eqnarray}
 \bar{\bar{\rho}}^p &=& \bar{\bar{\rho}}- c^2 f \bar{\bar{\chi}}\\
 \bar{\bar{\rho}}&=& \sum_i  \exp (- i {\bf q \cdot
  R}_i) \exp (- i {\bf q} \cdot
\mbox{\boldmath $\eta $}_ic)\\ \bar{\bar{\chi}}&=& \sum_i \exp (-
i {\bf q \cdot
  R}_i) \exp (- i {\bf q \cdot} \mbox{\boldmath $\eta $}_i/c )  \\
  f &=& \exp \left({-q^2l^{*2}\over 8ps(2ps+1)}\right)\ \ \
  \mbox{(Vortex form factor)}
 \end{eqnarray}
 Armed with Eqns. (\ref{Rmat} and \ref{etamat}   ) we may finally write
 \begin{eqnarray*}
\lefteqn{ \bar{\bar{\rho}}_{ m_2 n_2;m_1  n_1}=} \\
 & &\sqrt{m_2!\over m_1!} e^{-x/2} \left(
{-iq_+l^*\over \sqrt{2}}\right)^{m_1-m_2}L_{m_2}^{m_1-m_2}(x)
 \\
 & &\bigotimes \left[ \sqrt{n_2!\over n_1!}
\left( {-icq_-l^* \over
 \sqrt{2}}\right)^{n_1-n_2}e^{-x c^2/2}L_{n_2}^{n_1-n_2}(xc^2)\right. \\
 & &\left. -c^2 \cdot  f\cdot
\left({-iq_-l^*\over \sqrt{2}c}\right)^{n_1-n_2}
e^{-x/2c^2}L_{n_2}^{n_1-n_2}(x/c^2)\right]\\
 & &\equiv
\rho_{m_2m_1}^{m}\bigotimes \rho_{n_2n_1}^{n}
 \end{eqnarray*}
 Superscripts on $\rho_{m_2m_1}^{m}$ and $\rho_{n_2n_1}^{n}
 $ which will be
   apparent from the subscripts, will usually  be suppressed.

\subsection{Proof of Hartree-Fock nature of trial states}
Consider \beq \langle f|H|i\rangle = \langle {\bf
p}|d_fHd^{\dag}_{i}|{\bf p}\rangle \eeq where $|{\bf p}\rangle$
stands for the (ground) state with $p$-filled LL, and $i,f$ label
single-particle excitations on top of this ground state. We want
to show that this matrix element vanishes if $i\ne f$, i.e., the
hamiltonian does not mix these putative  HF particle states. (This
result was established for the small $q$ theory by G.
Murthy.\cite{GMMP}.) The proof, which relies on just the
rotational invariance of the potential, applies with trivial
modifications to the hole states, i.e., to \beq \langle {\bf
p}|d_{f}^{\dag}Hd_{i}|{\bf p}\rangle .\eeq

This matrix element in question  takes the schematic  form \beq
 \langle f|H|i\rangle = \int_q \langle {\bf p}|d_f
 d_{1}^{\dag}d_2d_{3}^{\dag}d_4 d_{i}^{\dag}|{\bf p} \rangle
 \rho_{12}({\bf q})\rho_{34}(-{\bf q})
 \eeq
 where $1$ stands for $m_1n_1$ and so on, and   $\int_q$ stands
 for an integral over {\em a rotationally
 invariant measure}:
 \beq
 \int_q ={1\over2} \int {d^2q\over 4\pi^2}v(q) e^{-q^2l^2/2}.
 \eeq

 Now we use Wick's theorem and perform pairwise contractions on
 the vacuum expectation value,
 bearing mind that
\begin{itemize}
  \item We cannot contract the indices $1$ and $2$ or $3$ and $4$
  since this will require that  $q=0$ at which point the measure
  (which contains the potential) vanishes.

  \item If we contract $i$ and $f$ we already have the desired result.
\end{itemize}

Here is a representative of the contractions we can get:
\begin{equation} \rho_{12}\rho_{34}
\delta_{f1}(1-n_{1}^{F})\delta_{23}(1-n_{2}^{F})\delta_{4i}(1-n_{4}^{F}).
\eeq where $n_{1}^{F}$ is the Fermi function for the LL labeled by
$n_1$ \beq n_{1}^{F}=\theta (p-1-n_1) \eeq and so on. Since $f=1$
and $i=4$, the $(1-n_{2}^{F})(1-n_{4}^{F})=1. $
 The integrand assumes  the form
\begin{eqnarray*}
\lefteqn{\sum_{m_2=0}^{\infty}\sum_{n_2=p}^{\infty}\rho_{f2}({\bf
q}) \rho_{2i}(-{\bf q})=}\\ & & \left[
\sum_{m_2=0}^{\infty}\rho_{m_fm_2}({\bf q})\rho_{m_2m_i}(-{\bf
q})\right] \left[ \sum_{n_2=p}^{\infty}\rho_{n_fn_2}({\bf
q})\rho_{n_2n_i}(-{\bf q})\right]
\\ &=&
\delta_{m_fm_i}\sum_{n_2}q_{-}^{n_i-n_f}F(|q|)
\end{eqnarray*}
where we have also used the fact that $e^{-i{\bf q \cdot R}}\cdot
e^{-i{\bf q \cdot R}}=I$ in doing the sum over $m_2$, and $F(|q|)$
is some rotationally invariant function.  It follows that every
term in  the sum over $n_2$ vanishes unless $n_f=n_i$ due to the
angular integral in ${\bf q}$.

\subsection{Particle-hole profiles} Here we deal with both gapped
and gapless states.
\subsubsection{Gapped case}
         Let us create a particle in the $p+1$-th LL and ask what the corresponding
         charge density looks like. In $q$ space this is given by
         \begin{eqnarray}
         \langle \bar{\bar{\rho}}^p ({\bf q}) \rangle
         & =& \sum_{1,2}\langle {\bf p}|d_{\mu}d^{\dag}_{1}d_2d^{\dag}_{\mu}|{\bf p}\rangle
         \rho_{12}({\bf q}) e^{-q^2l^2/4}
         \end{eqnarray}
         where $\mu$ labels both $n=p$ and $m=0$ of the created particle,
         and $1$ and $2$  stand for the double labels summed over in the definition
         of $\bar{\bar{\rho}}^p$.

         Now we do the contractions bearing in mind
         that we should  not contract $1$ with $2$ since this gives the background charge.
         This gives
         \begin{eqnarray}
         \langle \bar{\bar{\rho}}^p({\bf q})\rangle &=& \sum_{1,2}\langle d_{\mu}d^{\dag}_{1}\rangle
         \langle d_2 d^{\dag}_{\mu}\rangle \rho_{12}\\
         &=& \sum_{1,2}\delta_{\mu  1}(1-n_{1}^{F})\delta_{\mu
         2}(1-n_{2}^{F})\rho_{12}\\
         &=& \rho_{\mu   \mu}\\
         &=& \langle e^{-i{\bf q \cdot R}}\rangle_{00}
         \langle \left( e^{-i{\bf q \cdot \mbox{\boldmath $\eta $} }c}-c^2 fe^{-i{\bf q \cdot \mbox{\boldmath $\eta $} }/c }\right)
         \rangle_{p p}\\
         &=& e^{-q^2l^{*2}/4}\left( e^{-q^2c^2l^{*2}/4}L_{p}\left({q^2l^{*2}c^2\over
         2}\right)\right.\\
         && \left.-c^2 f e^{-q^2l^{*2}/4c^2}L_p \left({q^2l^{*2}\over
         2c^2}\right)\right)
         \end{eqnarray}
         It is straightforward to go to real space by Fourier transform.
         The figures depict the charge so obtained plus the background charge
         $\nu /(2\pi l^2 )$ in units of the background charge. The result is
         \beq
        { \bar{\bar{\rho}}^p (r/l) \over (\nu /2\pi l^2 )}=1 + {1 \over \nu}
         \int_{0}^{\infty}y dye^{-y^2/4} J_0(yr/l)F(y)
         \eeq
         where
         \begin{eqnarray*}
         \lefteqn{ F(y)=}                          \\
         && e^{-y^2(2ps+1)/4}\left( e^{-y^2c^2(2ps+1)/4}L_{p}\left[{y^2c^2(2ps+1)\over
         2}\right] \right. \\
        &&\left. -c^2 f e^{-y^2(2ps+1)/4c^2}L_p\left[ {y^2(2ps+1)\over 2c^2}\right]\right)
         \end{eqnarray*}

The computation of the hole charge is analogous. To facilitate
comparison
         to Park and Jain I have placed the particle and hole at antipodal points
         on a sphere of the same radius as the one they used.

\subsubsection{The  CF at $\nu =1/2$}
As in the gapped case, we now place a fermion just above the Fermi
sea. As explained in the main body of the text, we need a
superposition of states of fixed $p_y =p_0$ and a sum over all
values of $p_x$ to localize the CF at $x=0$. Since the dipole
always has a nonzero $x$-component in this superposition, it does
not get washed out.

For this case, with $A^*=0$, the LL expression for charge simplify
greatly : \beq \bar{\bar{\rho}}^p({\bf q})=\int {d^2k\over
4\pi^2}(-2i) \sin \left({{\bf q \times k}\ l^2\over 2}\right)
d^{\dag}_{{\bf k - q}}d_{{\bf k}} \eeq
 In other words, the pair of labels $(m,n)$ is replaced by a momentum
 vector ${\bf q}$. The density operator (for each particle)
 \beq
 \bar{\bar{\rho}}^p({\bf q})= e^{-i({\bf q\cdot r+q \times p}\ l^2/2)} -
 e^{-i({\bf q\cdot r-q \times p}\ l^2/2)}
 \eeq
 only connects states differing by ${\bf q}$. Carrying out the
 Wick contractions we get to the following result (dropping irrelevant constants)
 \begin{eqnarray}
 \langle \rho (x,y)\rangle &\simeq & \int dp_x dq_xe^{iq_xx}e^{-q_{x}^{2}l^2/4}
 \left( e^{-iq_xp_0l^2/2}-e^{iq_xp_0l^2/2}\right)   \nonumber \\
 &\simeq & \exp -\left[\left[{ x-p_0l^2/2\over l^2}\right]^2 \right]
  - \left[ p_o \to -p_o \right]
 \end{eqnarray}

Similar methods may be employed to calculate the structure factor
$S(q)$.

\subsection{Activation gaps}
           Now we need to find the energy cost of producing a widely
           separated particle-hole (PH) pair. This will be done by evaluating
   \begin{eqnarray}
   \Delta_a &=& \langle {\bf p} + P|H|{\bf p} + P\rangle + \langle {\bf p}+H|H|{\bf p} + H\rangle
  \nonumber \\
  & & -2\langle {\bf p} |H|{\bf p} \rangle\\
   &=& \int_q E(P)+E(H).
   \end{eqnarray}
   where $P$ denotes a particle added to the state labeled $\mu =
   (n=p,m=0)$ and $H$ denotes a state in which a hole has been
   made in the state $\mu = (n=p-1,m=0)$.
   Let us consider
   \beq
   E(P) =  \langle {\bf p}|d_{\mu}
 d_{1}^{\dag}d_2d_{3}^{\dag}d_4 d^{\dag}_{\mu}|{\bf p} \rangle \rho_{12}\rho_{34}
 \eeq
 In performing the contractions we
\begin{itemize}
  \item Do not make any contractions within $H$. This gets rid of
  $E_0 =\langle {\bf p} |H|{\bf p} \rangle $, the ground state energy.
  \item Do not contract $1$ with  $2$ or $3$ with  $4$ since $v(0)=0$.
\end{itemize}
We end up with
\begin{eqnarray*}
\lefteqn{\int_q \left[ \delta_{\mu
1}\delta_{23}\delta_{4\mu}(1-n_{1}^{F})(1-n_{2}^{F})(1-n_{4}^{F})\right.
} \\ & & -\left. \delta_{\mu
3}\delta_{14}\delta_{2\mu}(1-n_{3}^{F})(n_{4}^{F})(1-n_{2}^{F})
\right]\rho_{12}\rho_{34} \end{eqnarray*}

 Since $4=\mu =1$ in the
first term, we can drop $(1-n_{1}^{F})(1-n_{4}^{F})$ and for
similar reasons $(1-n_{3}^{F})(1-n_{2}^{F})$ in the second giving
us

\begin{eqnarray}
 E(P)& =&  \sum_{m_2=0}^{\infty}\sum_{n_2=p}^{\infty}
 \rho_{\mu 2}({\bf q})\rho_{2\mu}(-{\bf q}) \\ & & -
\sum_{m_2=0}^{\infty}\sum_{n_2=0}^{p-1}
 \rho_{2\mu }({\bf q})
\rho_{\mu 2}(-{\bf q}) .
\end{eqnarray}

Since the sum over $m_2$ is unrestricted, we can use completeness
and
 $e^{-i{\bf q \cdot R}}\cdot
e^{-i{\bf q \cdot R}}=I$ to get rid of the $m$-index altogether.
Thus we end up with \begin{eqnarray} E(P)&=&  \left(
\sum_{n=p}^{\infty}|\rho_{pn}|^2
-\sum_{n=0}^{p-1}|\rho_{pn}|^2\right)\\ & =&  \left[ \langle
p|\rho (q)\rho (-q) |p\rangle -
2\sum_{n=0}^{p-1}|\rho_{pn}|^2\right]
\end{eqnarray}

A similar calculation for the hole state gives (upon dropping the
ground state energy as usual) \beq E(H)=  \left[ -\langle p-1
|\rho (q)\rho (-q) |p-1 \rangle +
2\sum_{n=0}^{p-1}|\rho_{p-1,n}|^2\right]
\end{equation}
where
 \beq
\langle n|
 {\rho} (q)\rho (-q)|n\rangle = \sum_{n'=o}^{\infty}
 |{\rho} (q)_{nn'}|^2.
 \eeq
 Putting all the  pieces together, and recalling the various matrix elements, we obtain
Eqns.{\ref{gaps1}-\ref{gaps2}).

\subsection{Critical fields for magnetic transitions}
We need to calculate \beq E(p-r,r) = \langle {\bf p-r,\ r}|H|\
{\bf p-r,\ r}\rangle \eeq the energy in a state with $p-r$ spin-up
LL's and $r$ spin-down LL's. Since the HF calculation for the
spinless case is very similar, this treatment will be brief. We
write \beq H = \sum_{1234}\int_q d_{1}^{\dag}d_2d_{3}^{\dag}d_4
\rho_{12}\rho_{34} \eeq with the understanding that a label like
$1$ stands for the triplet $(n_1,m_1,s_1)$, $s$ being the spin.
The matrix elements $\rho_{ij}$ are defined by
\begin{eqnarray*}\lefteqn{ \rho_{12}=}\\ && \langle 1|e^{-i{\bf q
\cdot R}}\left[ e^{-i{\bf q} \cdot \mbox{\boldmath $\eta $}c}-c^2
fe^{-i{\bf q \cdot \mbox{\boldmath $\eta $} }/c}\right]|2\rangle
\\ &=&
 \rho_{m_1m_2}\otimes
\rho_{n_1n_2}\otimes \delta_{s_1s_2}\end{eqnarray*}
 and as a result
 \begin{eqnarray*}
 \lefteqn{E(p-r,r)=}
 \\ &&
 \int_q\sum_{n_1n_2s}n_{1}^{F}(s)(1-n_{2}^{F}(s))|\rho_{n_1n_2}|^2\underbrace{\sum_{m}\langle
 m|
 I| m\rangle }_{=n/p}
 \end{eqnarray*}
 where we acknowledge the fact that the occupation factors  $n_{1}^{F}$ and $n_{2}^{F}$ can depend on
 the spin. We have also used the fact that the sum over all values
 of $m$ is the degeneracy of each CF-LL, $n/p$. Carrying out the
 sums over $n_1$ and $n_2$, we obtain

\begin{eqnarray*}
 \lefteqn{E(p-r,r)=}
 \\ &&
{n \over p} \int_q \left[ \sum_{n_1=0}^{p-r-1}\langle n_1|\rho (q)
\rho (-q)|n_1 \rangle
-\sum_{n_1,n_2=0}^{p-r-1}|\rho_{n_1n_2}|^2\right.\\ &&+ \left.
\sum_{n_1=0}^{r-1}\langle n_1|\rho (q) \rho (-q)|n_1 \rangle
-\sum_{n_1,n_2=0}^{r-1}|\rho_{n_1n_2}|^2\right]
 \end{eqnarray*}

It is now straightforward to compute the critical field for the
transition $|{\bf p-r,r}\rangle \to |{\bf p-r-1,r+1}\rangle$ by
invoking \beq E(p-r,r)-E(p-r-1, r+1) = g \left[ {e  \over
2m_e}\right]B^c {n\over p} \eeq
\subsection{Calculation of $1/T_1$}

As explained in the main text, our strategy for computing $1/T_1$
will be to  compute $K^{max}$,  the Knight shift at the center of
the well,  in a fully polarized sample, in terms of  an unknown
nuclear matrix element squared $|u(0)|^2$, and then to express
$1/T_1$ (which depends on $|u(0)|^4$) in terms of the measured
value of $K^{max}$.
\subsubsection{Calculation of Knight shift}

In first quantization, the hyperfine interaction of electrons with
a nucleus at the origin  is \beq H_{hf}= {8 \pi \over
3}\gamma_n\gamma_e {\bf S\cdot I}\ \delta^3(0)\eeq where
$\gamma_e$ and $\gamma_n$ are gyromagnetic ratios of the nucleus
and electron, ${\bf S}={1 \over 2}{\bf \sigma}$ and ${\bf I}$ is
the nuclear spin ($3/2$  in this case).

Our plan is to first express this operator in second-quantized
form (in the electron basis) and then  transform to the CF basis.

Consider now the quantum well, which we take to be infinite in the
$x-y$ plane and of width $w$ in the $z$-direction. In the absence
any nuclear potential the single-particle wavefunctions will be
given by \beq \Phi_{({\bf k},s)}({\bf r},z,s) = {e^{i{\bf k \cdot
r}}\over \sqrt{L^2}} \psi_{\perp}(z) \chi_s \eeq where, $\chi_s$
is a spinor with  $s=\pm 1/2$,  ${\bf r}$, like ${\bf k}$ lies in
the $x-y$ plane, $\psi_{\perp}(z)$ is a function like
$\sqrt{2/w}\sin (\pi z /w)$. Note that in the $z$-direction there
is essentially just one wavefunction, other  excitations being too
high in energy to interest us.  We now turn on  the nuclear
potential, whose effect will be to modify the states to \beq
\Phi_{({\bf k},s)}({\bf r},z,s)={e^{i{\bf k \cdot r}}}u_{({\bf
k},s)}({\bf r},z) \chi_s \label{200}, \eeq where, by Bloch's
theorem, $
 u_{{\bf k},s}({\bf
r},z)$ is invariant under translations by the unit cell {\em in
the $x-y$ plane}. The unit cell here extends into the z-direction
from one end of the well to the other and may contain many nuclei.
In the $z$-direction, $u_{{\bf k},s}({\bf r},z)$ will vary and
possibly vanish rapidly beyond the ends of the well. We will
assume $u_{({\bf k},s)}({\bf r},s)$ is normalized to unity over
the sample. (This means it contains all normalization factors like
$1/\sqrt{L}$ that we may usually associate with the plane wave.)
The only thing  of importance is that  the  states are still
labeled by $({\bf k},s)$. As we shall see,detailed knowledge of
$u_{({\bf k},s)}({\bf r},s)$ will not come into play.

In  second quantization we define a spinor field  operator \beq
\Psi ({\bf r},z) = \sum_{s,{\bf k}}d({\bf k},s)e^{i{\bf k \cdot
r}}u_{({\bf k},s
)}({\bf r},s)\chi_s\eeq

(Note that this operator is good for low energy physics in which
higher excited states in the $z$-direction are ignored. We could
add all these to  the sum over  states to obtain an operator that
had Dirac delta function anticommutation relations. However if we
limit ourselves to the low energy sector as defined above, this is
a waste.)

The spin density at the origin becomes
 \beq {\bf S}({\bf 0})
=\sum_{s,{\bf k}}\sum_{s',{\bf k'}}u_{ ({\bf k},s)}({\bf
0})u^{*}_{ ({\bf k'},s')}({\bf 0})d^{\dag}({\bf k'},s')
{\mbox{\boldmath $\sigma $}^{ss'}\over 2}d({\bf k},s) \eeq where
\beq u_{ ({\bf k},s)}^{*}({\bf 0})=u_{ ({\bf
k},s)}^{*}(0,0,0).\eeq

Now we make an assumption that is often made: $u_{ ({\bf
k},s)}({\bf 0})$ does not depend on $s$ or ${\bf k}$. We will
simply call it $u(0)$. Given  this assumption
 \begin{eqnarray} {\bf S}({\bf
0}) &=& |u(0)|^2 \sum_{{\bf k,k'}s,s'}d^{\dag}_{{\bf
k'},s'}{\mbox{\boldmath $\sigma $}^{s,s'}\over 2} d_{{\bf k}s}\\
&=& |u(0)|^2 \sum_{{\bf k,q}s,s'}d^{\dag}_{{\bf
k+q},s}{\mbox{\boldmath $\sigma $}^{s,s'}\over 2} d_{{\bf k}s}\\
&=&|u(0)|^2 \sum_{\bf q}\cal{S} ({\bf q})
\end{eqnarray}
where $\cal{S} ({\bf q})$ is familiar expression for  spin density
at momentum ${\bf q}$, the effect of the nuclear potential being
encoded in  $|u(0)|^2$. Note that $d^{\dag}_{{\bf k},s}$ still
creates particles in states that are solutions to the nuclear and
quantum well potential and not plane waves. However the
commutation rules of the $d$'s are canonical  and the spin density
operator will obey the usual commutation rules \beq \left[ {\cal
S}^{a}({\bf q}),{\cal S}^{b}({\bf q'})\right] =\i\varepsilon^{abc}
{\cal S}^{c}({\bf q+q'}). \eeq

When we go to the CF basis, we assume we will see the same change
as in the case of the charge density: \beq \sum_{\bf
k}d^{\dag}_{{\bf k+q}} d_{{\bf k}} \to \sum_{\bf k}d^{\dag}_{{\bf
k+q}} d_{{\bf k}} e^{i{\bf q \times k}l^2/2}e^{-q^2l^2/4} \eeq
where I have included the factor $e^{-q^2l^2/4}$ so as to work
with the projected density and not the magnetic translation
operator. Thus we write
\begin{eqnarray}
{\bf S}({\bf 0})^{CF} &=& |u(0)|^2 \sum_{{\bf
k,k'}s,s'}d^{\dag}_{{\bf k'},s'}{\mbox{\boldmath $\sigma
$}^{s's}\over 2} d_{{\bf k}s}\nonumber \\ &\times &e^{i{\bf k
\times k'}l^2/2}e^{-|{\bf k -k'}| ^2l^2/4}
\end{eqnarray}
The factor $e^{i{\bf k \times k'}l^2/2}e^{-|{\bf k -k'}|
^2l^2/4}$,  which is just $e^{i{\bf q \times k}}e^{-q^2l^2/4}$
ensures that the above  spin operators (without  the $|u(0)|^2$)
have the right commutation relations ( Eqn. (28) of Moon {\em et
al}\cite{moon}) among themselves  and the projected charge
density. (The same criterion was used by Murthy in Ref.\cite{GMMP}
in the small $q$ limit.)

We are now ready to eliminate $|u(0)|^2$ in terms of a measurable
quantity.  The hyperfine interaction takes the form of an average
field $\langle {\bf B}\rangle$ acting on the nuclei:
\begin{eqnarray}
H_{hf} &=& \gamma_n    \langle {\bf B}\rangle \\ \langle {\bf
B}\rangle &=&|u(0)|^2 {8 \pi \over 3} \gamma_e   \langle
\sum_{{\bf k}s} d^{\dag}_{{\bf k},s} d_{{\bf k},s}{1 \over
2}\sigma^{ss}_z\rangle
\end{eqnarray}
where I have used the fact that by symmetry, only  ${\bf k=k'}$
and $S_z$ can have mean values.

Assume we are in a fully polarized state. Then the Knight shift is
readily found to be \beq K^{max} = {2 \over 3}\gamma_e \gamma_n
  N |u(0)|^2 \eeq where $N$ is the total number of particles.
For future use we invert this to write \beq |u(0)|^2 = \left[ {3
K^{max}\over 2\gamma_e \gamma_n   N}\right]. \label{u0}\eeq

\subsubsection{Calculation of $1/T_1$}
Consider now the relaxation rate.  By  the standard procedure
\cite{slichter} one arrives at the following expression for
$1/T_1$:
\begin{eqnarray}
{1 \over T_1}&=& {1 \over 2}{\sum_{mn}W_{mn}(E_m-E_n)^2\over \sum
E^{2}_{n}}\\ W_{mn}&=& {2 \pi  } \overline{\sum_F \sum_I |\langle
m F|H_{hf}|n I\rangle |^2\delta (E_F-E_I)}
\end{eqnarray}
where the bar indicates a thermal average, $(m,\ n)$ label nuclear
spin states,  ($I,\ F$) denote  many-body fermionic states, the
energy difference, $E_n-E_m$, between nuclear spin states has been
neglected in the energy conserving delta function,  and $H_{hf}$
is the hyperfine interaction, which now know how to write in the
CF basis.  Thus we have
\begin{eqnarray}
W_{mn}&=& {2\pi  }\left({8\pi \gamma_e \gamma_n   |u(0)|^2 \over
3}\right)^2\nonumber \\ &\times & \sum_{F,I}|\sum_{\alpha}\langle
m|I_{\alpha}|n\rangle  \langle F|S_{\alpha}^{CF}(0)|I\rangle|^2
\delta (E_F-E_I)\label{prev} \end{eqnarray} and we arrive at
\begin{eqnarray}
1/T_1&=& \pi \left({8\pi \gamma_e \gamma_n |u(0)|^2 \over
3}\right)^2\nonumber
\\ & &\times \int_{-\infty}^{\infty}dt \sum_{I}\sum_{ \alpha \ne z
}|\langle I|S_{\alpha}^{CF}(0)S_{\alpha}^{CF}(t)|I\rangle
\label{1/t}
\end{eqnarray}
where $S(t)$ is the Heisenberg operator at time $t$.

Here is a brief explanation of some steps leading  to Eqn.
(\ref{1/t}). First  I have used \beq {\sum_{mn}\langle
m|I_{\alpha}|n\rangle \langle n|I_{\alpha'}|m\rangle
(E_m-E_n)^2\over \sum_n E_{n}^{2}} = \delta^{T}_{\alpha
\alpha'}\label{show} \eeq where the transverse delta function
$\delta^T$ means that  $\alpha \ne z$ and then done some standard
manipulations.

To establish Eqn. (\ref{show}), we need    to invoke the following
facts.
\begin{itemize}
  \item  The kets $|m\rangle$ and $|n\rangle$ are eigenstates of
  the nuclear hamiltonian
  $A I_z$ where $A$ is some constant.
  \item   The factors $(E_n-E_m) I_{\alpha}$ or $(E_n-E_m) I_{\alpha'}$
  can be traded for
   commutators  of
  some other  $I_{\beta}$ or $I_{\beta'}$ with $A\ I_z$. The result is zero if
  either $\alpha$ or $\alpha'$ equals $z$, which explains the
  transverse delta function that emerges.
  \item  The previous step allows one to invoke completeness and
  reduce the numerator to the trace of the hamiltonian $(A I_{z})^{2}$, which then
  cancels the denominator.
  \end{itemize}
  Next we recall that
\begin{eqnarray}
S_{x}^{CF}(t)&=& {1\over
2}\sum_{k,s,k'}d^{\dag}_{k',-s}(t)d_{k,s}(t) e^{i{\bf k\times
k'}l^2/2}e^{-|{\bf{k-k'}}|^2l^2/2}\nonumber \\ d_{ks} (t) &=&
d_{ks}(0)e^{-i{\cal E}_s({\bf k}) t}\nonumber \end{eqnarray} (and
similarly for $ S^{CF}_{y}$ ) in the HF approximation. In the
above, ${\cal E}_s({\bf k})$ is the HF energy of a fermion of spin
$s$ and momentum ${\bf k}$.

To  arrive at Eqns. (\ref{oneovert}-\ref{oneovert1}) in the text,
we do the integral over $t$ (obtaining a delta function), use the
explicit expressions for spin operators, resort to
 standard HF factorization
of the quartic operators and perform a standard change of
variables in the measure, and   finally   eliminate  $|u(0)|^2 $
via Eqn. (\ref{u0}).

\subsection{Symbols}



\begin{tabular}{|c|c|}\hline
  Symbol & Significance \\ \hline
  $\nu $ & $=p/(2ps+1)$=filling fraction\\
  $p$ & Number of CF LL's\\
  $2s$ & Number of vortices attached\\
  $c^2$ & $2ps/(2ps+1)$\\
 $ B^*$ & Reduced field seen by CF  $=B/(2ps+1)$\\
 $ A^*$ & Reduced potential by CF  $=A/(2ps+1)$\\
 $l$ & electron magnetic length \\
  $l^*$ & CF magnetic length $=l/\sqrt{1-c^2}$\\
  ${\bf \Pi}^*$& Velocity operator for CF\\
  $\bar{\rho} ({\bf q})$ & Electron density in CF basis, ${\cal
  O}(ql)$\\
$\bar{\chi} ({\bf q})$& Constraint in CF basis, ${\cal
  O}(ql)$\\
  $\bar{\bar{\rho}} ({\bf q})$ & Electron density in CF basis,\\
$\bar{\bar{\chi }}({\bf q})$ & Constraint in CF basis,\\
$\bar{\bar{\rho}}^p ({\bf q})$ &$\bar{\bar{\rho}} ({\bf q})- c^2 f
\bar{\bar{\chi}} ({\bf q})$\  (preferred charge)\\
 $f$ & Vortex form factor
($e^{-q^2l^2/(8ps)}$)\\ $\lambda$ & Defined by  $v(q) = 2\pi e^2
e^{-ql\lambda}/q$\\ $b$ & Defined by  $v(q) = 2\pi e^2
e^{(qlb)^2}\ Erfc \  (qlb)/q$\\
  $\Delta_{a,p}$& Activation or polarizarion gap\\
  $\delta$  & $ \Delta/(e^2/\varepsilon l)$\\
  $1/m_{a,p}^{(2s)}$& Defined by $\Delta_{a,p} =
  eB^*/(m_{a,p}^{(2s)}$\\
  $C_{a/p}^{(2s)} $ & Defined by ${1/ m^{(2s)}_{a,p}} = ({e^2 l /\varepsilon
  })
  C^{(2s)}_{a,p}$\\
  $m^{nor}_{a,p}$ & ${m_{a,p}/ (m_e \sqrt{B(T)}})$\ (normalized mass)\\
  $m_e$ & Electron mass in free space\\
  $P$ & Polarization\\
  $S$ & Number of spin up minus down CF\\
  $E(S)$ & Ground state energy density \\
  $g$ & $g$ -factor of CF, taken to be .44\\
  $|{\bf p-r,r}\rangle$ & Many-body  CF state with \\
  & $p-r$ spin LL's
  and $r$ spin-down LL's.\\
  ${\cal E}_{\pm}(k)$ & Hartree Fock energy for up/down spin\\
  $\theta$ & Tilt angle \\
  $\rho_{n_1n_2}$& One-particle Matrix element of
  $\bar{\bar{\rho}}^p$ \\ & between  LL
   $n_1$ and $n_2$\\ \hline
\end{tabular}


\begin{thebibliography}{99}

\bibitem{jain} J.K.Jain, Phys. Rev. Lett. {\bf 63}, 199, (1989);
Phys. Rev. {\bf B41}, 7653 (1990);  J.K.Jain and R.K.Kamilla,  in
``Composite Fermions'', Ref. (\ref{Review1}) Editor O. Heinonen,
World Scientific (1998).
\bibitem{review1}  For a review see {\em The Quantum Hall
Effect}, Edited by R.E. Prange and S.M Girvin, Springer-Verlag,
1990, T. Chakraborty and P. Pieti\"{a}\"{i}nen, {\em The
Fractional Quantum Hall Effect: Properties of an incompressible
quantum fluid}, Springer Series in Solid State Sciences, {\bf 85},
Springer-Verlag New York, 1988, A.H. MacDonald ed., {\em Quantum
Hall Effect: A Perspective}, Kluwer, Boston, 1989.\label{Review1}
 {\it Perspectives in
Quantum Hall Effects}, Edited by Sankar Das Sarma and Aron Pinczuk
( Wiley, New York, 1997), {\em Composite Fermions},
\label{review1} Editor O. Heinonen, World Scientific (1998).
\bibitem{gmrs} R.Shankar and G. Murthy,  Phys. Rev. Lett. {\bf 79},
4437, (1997).
 G.Murthy and R.Shankar,  in ``Composite
Fermions'', {\em ibid}.(cond-mat/9802244).
\bibitem{gaps} G. Murthy and R.Shankar,Phys. Rev. {\bf B  59}, 12260, (1999)
(gaps).
\bibitem{prl1} R.Shankar,  Phys. Rev. Lett. {\bf 83}, 2382,
(1999).
\bibitem{prl2} R.Shankar, Phys. Rev. Lett.  {\bf
84   }, 3946, (2000) cond-mat/9911288.
\bibitem{pmj} K. Park, N.Meskini,  and J.K. Jain, J. Phys. Condensed Matter, {\bf 11}, 7283, (1999).
This work also responds to some queries raised by R.Morf, Phys.
Rev. Lett. {\bf 83}, 1485, (1999).\label{pmj}
\bibitem{zds} F.C. Zhang
and S. Das Sarma, Phys. Rev. {\b B 33}, 2908, (1986). This paper
also considers the effect of LL mixing.
\bibitem{Morf} R.H. Morf, N. d'Ambrumenil and S. Das Sarma
(to be published).
\bibitem{du} R.R.Du, A.S. Yeh, H.L. St\"{o}rmer, D.C. Tsui, L. N. Pfeiffer, K.W.Baldwin
 and K.W. West, Phys. Rev. Lett., {\bf 70}, 2944, (1993).
 \bibitem{Pan} W. Pan, H.L. St\"{o}rmer, D.C.Tsui, L.N. Pfeiffer, K.W. Baldwin
and K.W. West,
 Phys. Rev. Lett. {\bf B61}, R5101, (2000).
 \bibitem{dis} F. C.  Zhang, V.Z. Vulovic, Y. Guo, and S. Das Sarma, Phys.
 Rev. {\bf  32}, 6920 (1985).
\bibitem{LDA} See for example the early work of F. Stern and S. Das Sarma,
Phys. Rev. {\bf B30}, 840, (1984), {\em ibid} {\bf B32}, 8442,
(1985) and
  M. Ortolana, S. He and S. Das Sarma,
{\em ibid}  {\bf B55}, 7702, (1997) and the recent work of
Ref.(\ref{pmj}).
 \bibitem{j24} X. Zu, K. Park and J.K. Jain,
 "Masses of Composite Fermions carrying two and four
 flux quanta Differences and similarities," cond-mat/9911331, Phys. Rev.
 {\bf B61}, R7850, (2000)..
 \bibitem{spin} B. I. Halperin, Helv. Phys. Acta., {\bf 56}, 75,
 (1983), F.C.Zhang and T. Chakraborthy, Phys. Rev. {\bf B29}, 7032,
 (1984), M. Rasolt, S. Perrot and A. MacDonald, {\bf 55}, 433,
 (1985).
\bibitem{parkjain} K. Park and J.K. Jain,  Phys. Rev. Lett. {\bf 80}, 4237, (1998.)
\bibitem{kuk} I.V.Kukushkin, K. v.
Klitzing and K. Eberl, Phys. Rev. Lett. {\bf 82}, 3665, (1999.)
\bibitem{dem} A.E. Dementyev, N.N. Kuzma, P. Khandelwal, S.E. Barrett,
 L.N. Pfeiffer, and K. W. West, Phys. Rev. Lett., {\bf 83}, 5074, (1999). cond-mat/9907280.
\bibitem{melinte} S.Melinte, N.Freytag, M. Horvatic, C. Berthier, L.P. Levy,
 V. Bayot and M. Shayegan, Phys. Rev. Lett., {\bf 84}, 354, (2000).
  cond-mat/9908098.
  \bibitem{cb} S.M. Girvin and A.H. MacDonald,
Phys. Rev. Lett. {\bf 58}, 1252, (1987). See also S. M. Girvin in
Ref. (\ref{review1}).\bibitem{zhk}S.C. Zhang, H. Hansson and S.
Kivelson, Phys. Rev. Lett. {\bf 62}, 82, (1989). N. Read, Phys.
Rev. Lett., {\bf 62}, 86, (1989).  See also D.-H. Lee and S.-C.
Zhang, Phys. Rev. Lett. {\bf 66}, 1220 (1991) and the
 excellent  review by S.C. Zhang,
Int. J. Mod. Phys. {\bf B6}, 25, (1992).
 C. L. Kane, S. Kivelson, D.H. Lee and S.C. Zhang, Phys.
Rev.  {\bf B 43 }, 3255 (1991).
\bibitem{halp} B.I. Halperin,Phys. Rev. Lett. {\bf 52}, 1583,
(1994.) See also Ref. \cite{spin}.
\bibitem{read1} N.Read, Semi. Sci. Tech. {\bf 9}, 1859 (1994);
Surf. Sci., {\bf 361/362}, 7 (1996).
 \bibitem{gj} S.M. Grivin and T. Jach,  Phys. Rev., {\bf
B29}, 5617, (1984).
\bibitem{Su} W.P. Su, Phys. Rev. {\bf B33}, 1031, (1986).
\bibitem{focus} V. J. Goldman, B.Su and J. K. Jain, Phys. Rev.
Lett., {\bf 72}, 2065, (1994), J. H. Smet, D. weiss, R.H. Blick,
K. von Klitzing, P.T. Coleridge, {\em ibid}, {\bf 77}, 2272,
(1996).
\bibitem{lopez} A. Lopez and E. Fradkin, Phys. Rev. {\bf B 44}, 5246,
(1991), {\em ibid} {\bf 47}, 7080, (1993), Phys. Rev. Lett., {\bf
69}, 2126, (1992).
\bibitem{leinaas} J.M. Leinaas and J. Myrheim, {\it Nuovo Cimento} {\bf
37B}, 1 (1977).
\bibitem{kalmeyer} V. Kalmeyer and S. C. Zhang, Phys. Rev. {\bf B46},
9889, (1992).
\bibitem{brad}H.J. Kwon, J.B. Marston and A. Houghton, Phys. Rev.
Lett., {\bf 73}, 284, (1994) and Phys. Rev. {\bf B52}, 8002,
(1995).
\bibitem{hlr} B. I. Halperin, P.A. Lee and N. Read, Phys.
Rev. {\bf B47}, 7312, (1993).
\bibitem{fetter} A.L. Fetter, C. B. Hannah and R.B. Laughlin,
Phys. Rev., {\bf B 39}, 9679, (1989).
\bibitem{rs}  R. Rajaraman and S. Sondhi, Int. J. Mod. Phys.
{\bf B 10}, 793 (1996). R. Rajaraman, Phys. Rev. {\bf B56}, 6788,
(1997).
\bibitem{BP} D. Bohm and D. Pines, Phys. Rev. {\bf 92}, 609,
(1953).
\bibitem{SSH} S.H. Simon, A. Stern, and B.I. Halperin, Phys. Rev. {\bf
B54}, R11114 (1996).
 \bibitem{gapsfour} G.Murthy,  K.Park, J.K.Jain, and R.Shankar,  Phys. Rev. {\bf
B58}, 13263, (1998) (scaling laws).
\bibitem{HS1} B.I.Halperin and A.Stern, \prl {\bf 80}, .5457
(1998); G. Murthy and R.Shankar, {\em ibid} , 5458 (1998). See
also  Ref. \ref{dh}.
\bibitem{simon} A. Stern, B.I. Halperin, F. von Oppen and S.Simon,
Phys. Rev., {\bf59},12547, (1999).
 cond-mat
9812135.
\bibitem{GMP} S.M. Girvin, A.H. MacDonald and P. Platzman, Phys. Rev.
{\bf B33}, 2481, (1986).
\bibitem{read2} N.Read, Phys. Rev.  {\bf B58}, 16262, (1998).
\bibitem{PH} V.Pasquier and F.D.M.Haldane, Nucl. Phys. {\bf B516}, 719, (1998).
\bibitem{GMMP} G. Murthy, Phys. Rev. {\bf
B60}, 13702 (1999).
\bibitem{LLmix} V. Melik-Alaverdian and N. Bonesteel, Phys. Rev.
{\bf B52}, R17032, (1995). This paper shows that LL mixing is
negligible at $\lambda \simeq 1$. Earlier works e.g., B. Yoshioka,
J. Phys. Soc., Jpn., {\bf 55}, 885, (1986) did not consider
$\lambda$ dependence.
\bibitem{DH} D.H.
Lee, Phys. Rev. Lett., {\bf 80}, 4745 (1998).\label{dh}
\bibitem{han} J.H. Han and S.R. Eric Yang,  Phys. Rev. {\bf B58}, R 10163,
(1998).
\bibitem{magp} N. d'Ambrumenil, and R. Morf, Phys. Rev. {\bf B40},
6108, (1989), P.M. Platzman and S. He, Phys. Rev. {\bf B49},
13674, (1989), S.He, S.Simon and B.I. Halperin, Phys. Rev. {\bf
B50}, 1823, (1994),

\bibitem{pseudo} F.D.M.Haldane, Phys. Rev. Lett., {\bf 51}, 605,
(1983).
\bibitem{scaling}  I am obliged to J.K. Jain, G.Murthy and H.
St\"{o}rmer for illuminating discussions on this question.
\bibitem{FH}F.F.Frank and W.E. Howard, Phys. Rev. Lett., {\bf 16}, 797,
(1966).
\bibitem{gmstep} G. Murthy, Phys. Rev. Lett., {\bf 84}, 350,
(2000).
\bibitem{cp} T. Chakroborthy and P. Pietilainen, cond-mat/9910019.
\bibitem{gmpol} G. Murthy, "Temperature dependence of
magnetization in fractional quantum Hall systems: Composite
fermion Hartree-Fock and beyond", cond-mat/0008259.
\bibitem{moon} K.Moon, H.Mori,K.Yang,S.M. Girvin, A.H. MacDonald,
L.Zheng, D.Yoshioka, and S.C.Zhang, Phys. Rev. {\bf B51}, 5138,
(1995).
\bibitem{slichter} C. P. Slichter, {\it Principles of Magnetic
Resonance}, Springer Verlag, 1990.



\end{thebibliography}
\end{document}